\documentstyle[harvard,epsfig]{mn}

\input{epsf}

\def\lsim{\mathrel{\hbox{\rlap{\hbox{\lower4pt\hbox{$\sim$}}}\hbox{$<$}}}}

\begin{document}

\title[Metallicity and the Cepheid Distance Scale]
{The Effects of Metallicity on the Cepheid P-L Relation and the 
Consequences for the Extragalactic Distance Scale}

\author[P.D. Allen and T. Shanks]
{
Paul D. Allen$^{1,2}$ and Tom Shanks$^{1}$
\\
1 Department of Physics, Science Laboratories, South Road, Durham DH1 3LE.
\\
2 Astrophysics, University of Oxford, Keble Road, Oxford OX1 3RH. \\
}
\maketitle

\begin{abstract}
If a Cepheid luminosity at given period  depends on metallicity, then the P-L
relation in galaxies with higher  metallicities may show a higher dispersion if
these galaxies also sample a wider range of intrinsic metallicities. Using
published HST Cepheid data from 25 galaxies, we have found such a correlation
between the the P-L dispersion and host galaxy metallicity which  is significant
at the $\approx3\sigma$ level in the V band.  In the I band the correlation is
less significant, although the tighter intrinsic dispersion of the P-L relation
in I makes it harder to detect such a correlation in the HST sample. We find that
 these results are unlikely to be explained by increased dust absorption in high
metallicity galaxies. The data support the suggestion of Hoyle et al. that the
metallicity dependence of the Cepheid P-L relation may be stronger than expected,
with $\Delta M/[O/H] \approx -0.66$ mag dex$^{-1}$ at fixed period.

The high observed dispersions in the HST Cepheid P-L relations have the further
consequence that the bias due to incompleteness in the P-L relation at faint
magnitudes is more significant than previously thought.  Using a maximum
likelihood technique which takes into account  the effect on the P-L relations of
truncation by consistently defined magnitude completeness limits, we re-derive the
Cepheid distances to the  25 galaxies and find that the average distance is
increased by $\approx$0.1mag. However, in the cases of two high metallicity
galaxies at large distances the effect is severe, with the published distance
modulus underestimating the true distance modulus by $\approx$0.5mag.

In the HST sample, galaxies at higher distance tend to have higher metallicity.
This means that when a full metallicity correction is made, a scale error in the
published Cepheid distances is seen in the sense that the published distance
moduli are increasingly underestimates at larger distances, with the average
difference now being $\approx$0.3mag. This results in the average distance
modulus to the four galaxies in the Virgo cluster core increasing from
$(m-M)_0=31.2\pm0.19$ to $(m-M)_0=31.8\pm0.17$ with similar increases for the
Fornax and Ursa Major clusters. For the 18 HST galaxies with good Tully-Fisher
distances and $m-M_0>29.5$ the Cepheid-TF distance modulus average residual
increases from 0.44$\pm$0.09 mag to 0.82$\pm$0.1mag indicating a significant
scale error in TF distances and resulting in the previous Pierce \& Tully TF
estimate of H$_0$=85$\pm$10 kms$^{-1}$Mpc$^{-1}$ reducing to H$_0$=58$\pm$7
kms$^{-1}$Mpc$^{-1}$, assuming a still uncertain Virgo infall model. Finally,
for the 8 HST galaxies with SNIa, the metallicity corrected Cepheid distances now
imply a metallicity dependence of SNIa peak luminosity in the sense that
metal-poor hosts have lower luminosity SNIa. Thus SNIa Hubble diagram estimates
of both  $H_0$ and $q_0$ may also require significant metallicity corrections.

\end{abstract}

\begin{keywords}
Cepheids: Metallicity, Distance Scale, Hubble's Constant 
\end{keywords}

\section{Introduction}
\label{sec:intro}
Over fifty years after the discovery of the expansion of the universe, the value
of the expansion rate, H$_{0}$ is yet to be determined to high accuracy. Although
there is a general consensus amongst astronomers that it lies somewhere between
40 and 100 kms$^{-1}$ Mpc$^{-1}$  and probably between 50 and 70 kms$^{-1}$
Mpc$^{-1}$ , no conclusive result has yet been obtained. The Hubble constant is a
fundamental cosmological parameter, and its value is important for the
determination of the age, density and ultimate fate of the universe. H$_{0}$ is
usually measured by calculating the gradient of redshift against distance for a
sample of galaxies. However, the Hubble redshift of galaxies can only be accurately
determined at higher redshifts where the contribution of  peculiar motions are
proportionately smaller and here the measurement of extragalactic distances becomes 
increasingly difficult.

The primary extragalactic distance indicators are the Cepheid variables. Although
there are several standard candle techniques used to measure extragalactic 
distances, most are calibrated using the Cepheid distances to nearby galaxies.
Accurate Cepheid distances, which are free from systematic errors are 
therefore vitally important to determinations of H$_{0}$.

The Hubble Space Telescope has allowed new measurements of Cepheid distances to
many galaxies outside the Local Group and several groups have attempted to
measure H$_{0}$ using these distances. The aim of the largest group, (the HST Key
Project on the Extragalactic Distance Scale) is to calculate H$_{0}$ with an
accuracy of 10\%. If this is to be achieved, then several possible sources of
systematic error need to be looked at in  detail. These include the Cepheid P-L
calibration in the LMC. A recent recalibration \cite{feastcatch} could cause
H$_{0}$ to decrease by $\approx$ 10\% \cite{webb}.

In this paper we first check the effects of metallicity on the dispersion of the
Cepheid P-L relation, using data from the H$_{0}$ Key Project. Then we
systematically apply cuts to the data in period, in order to investigate the
effects on the dispersion/metallicity relationship. We also discuss the possible
causes of what is observed. We then investigate the consequences of both a global
metallicity correction  and the effects of bias due to incompleteness at faint
magnitudes. We use a maximum likelihood technique to fit P-L relations truncated
due to magnitude incompleteness and hence  derive new Cepheid distances.

\section{Data}
\label{sec:data}
The data analysed in this work is from sources that have been 
either published or accepted for publication. In total, results for 25 
galaxies were examined. Of these, 18 were from the `HST Key Project on the 
extragalactic distance scale' team. The photometry and Cepheid periods are 
used as published. Most of this data was taken from the Key Project archive 
website (http://www.ipac.caltech.edu/H0kp). This gives the photometry/period 
data for the Cepheids in each of the galaxies. This was used along with 
published papers to obtain exactly the same data as was used by the key 
project team in their distance determinations. The references are as follows: 
NGC 925 \cite{silbermann96}; NGC 1326A \cite{prosser}; NGC 1365 
\cite{silbermann99}; NGC 1425 \cite{mould00}; NGC2090 \cite{phelps}; NGC 2541 
\cite{fer98}; NGC 3031 \cite{freed94}; NGC 3198 \cite{kelson99}; NGC 3319 
\cite{sakai}; NGC 3351 \cite{graham97}; NGC 3621 \cite{rawson}; NGC 4321 
\cite{fer96}; NGC 4414 \cite{turner}; NGC 4535 \cite{macri}; NGC 4548 
\cite{graham99}; NGC 4725 \cite{gibsonb}; NGC 5457 \cite{kelson96}; NGC 7331 
\cite{hughes}. Details about how the data was obtained and reduced can be 
found in these papers and the references therein. 

The other 7 galaxies were from two sources: \citeasnoun{tanvir95},
\citeasnoun{tanvir99} for NGC 3368 and the supernova calibration team of 
\citeasnoun{sand96} for the galaxies NGC 3627, NGC 4639, NGC 4496A, NGC 4536, NGC 5253 and IC
4182. Once the data became available on the HST archive, the images for these
galaxies were reanalysed by the Key Project team in \citeasnoun{gibson}, using
the same software and reduction procedure as for the other 18 galaxies. For
consistency, the Cepheids in the HST images selected by \citeasnoun{gibson},
along with their photometry and period determination are used here, rather than
the originally published data.

\section{P-L Dispersion}
\label{plrels}

Cepheid data was taken from the Key Project website and the references given 
in Section \ref{sec:data}. An unweighted least squares fit was made to the 
magnitude versus period data for the Cepheids in each galaxy in both the 
V-band and the I-band. In this case, the line is the Cepheid P-L relation, 
which in the V-band is given by

\begin{equation}
m_{V} = -2.76log(P)-b
\label{eq:pl}
\end{equation}
    
A similar relationship is used in the I-band with a P-L slope of -3.06 now
assumed. The slope is held constant to that of the LMC P-L Relation \cite{mf91}.
This is done to minimise the effects of magnitude limited bias, since this would
tend to make the slope shallower. The Key Project team also follows this
procedure in fitting their P-L Relations.

The mean r.m.s. dispersion about the P-L relation was also calculated, this is 
generally considered to be due to the small variation in magnitude across the 
instability strip and any statistical noise/experimental errors. 

Plots of the P-L relations are given in Figure \ref{fig:pl1}. The mean 
magnitude found from the Key Project photometry for each Cepheid is plotted 
against the period as calculated by the Key Project team. This is done in both
the V- band and the I-band for all 25 galaxies. The plots show the best 
fitting line (solid black) to the data, along with the 2$\sigma$ dispersion 
lines (dashed) for the LMC, within which most of the data should lie. Also 
shown are magnitude limits (horizontal broken lines). These were calculated 
from the exposure times given in the relevant papers (see section 
\ref{sec:data}). The Key Project team do not calculate any magnitude limits, 
so we calculated consistent magnitude limits using the results of 
\citeasnoun{tanvir99} as a basis. Here magnitude limits of 26 (in the 
V-band), and 25 (in the I-band), could be imposed. This, whilst somewhat 
arbitrary, is justified by \citeasnoun{tanvir95}, as the region where 
statistical noise starts to become significant. The magnitude limits are 
therefore not absolute limits of non-detection, and many of the P-L relations 
in Figure \ref{fig:pl1} contain Cepheids that lie below their respective 
magnitude limits. Table \ref{tab:cephmag} lists the magnitude limits that 
have been calculated. The limits should be seen as the region where noise 
starts to become more significant and where there may be incompleteness in the
number of Cepheids. The magnitude limits for the other galaxies were
calculated by scaling the exposure times (under the assumption that observing 
conditions are  stable for the Hubble Space Telescope) relative to 
the exposure times given in \citeasnoun{tanvir99}, using signal-to-noise 
ratio considerations. However, at this stage the limits are not used in the 
calculation of the dispersion.




Also shown on the P-L plots in Figure \ref{fig:pl1}, are possible limits in
period for the P-L relations. This is due to the fact that short period 
Cepheids may not be detected in crowded stellar fields, since they are both
dimmer and more difficult to identify as Cepheids. Using artificial star 
tests, \citeasnoun{fer}, suggest that at Virgo and Fornax distances, P-L 
relations may be up to 50\% incomplete below 20-25 days. They then applied 
period cuts at 20 and 25 days to all their P-L relations, and found a small 
increase in distance modulus for a small number of galaxies. 
\citeasnoun{gibson} report a similar effect for some of the non-Key Project 
galaxies. The P-L relation best fits, and hence distance moduli, were then 
revised, (by the Key Project team) for these galaxies. Following this, those 
galaxies recommended for period cuts by \citeasnoun{fer} and 
\citeasnoun{gibson} were re-examined by ourselves, and the same period cuts 
applied. Cepheids with periods below 20 days (Log P = 1.3) are not included in
the best fit for the galaxies NGC 1326A, NGC 1425, NGC 4725, NGC 4536 and NGC 
3368. Cuts at 25 days were done for NGC 1365, NGC 4535, NGC 4496A and NGC 
3627. The intercept and dispersion shown on the P-L relations for these 
galaxies are for period cut samples.

\begin{figure*} 
\begin{tabular}{ccc} 
{\epsfxsize=5.5truecm \epsfysize=5.5truecm \epsfbox[17 144 590 715]{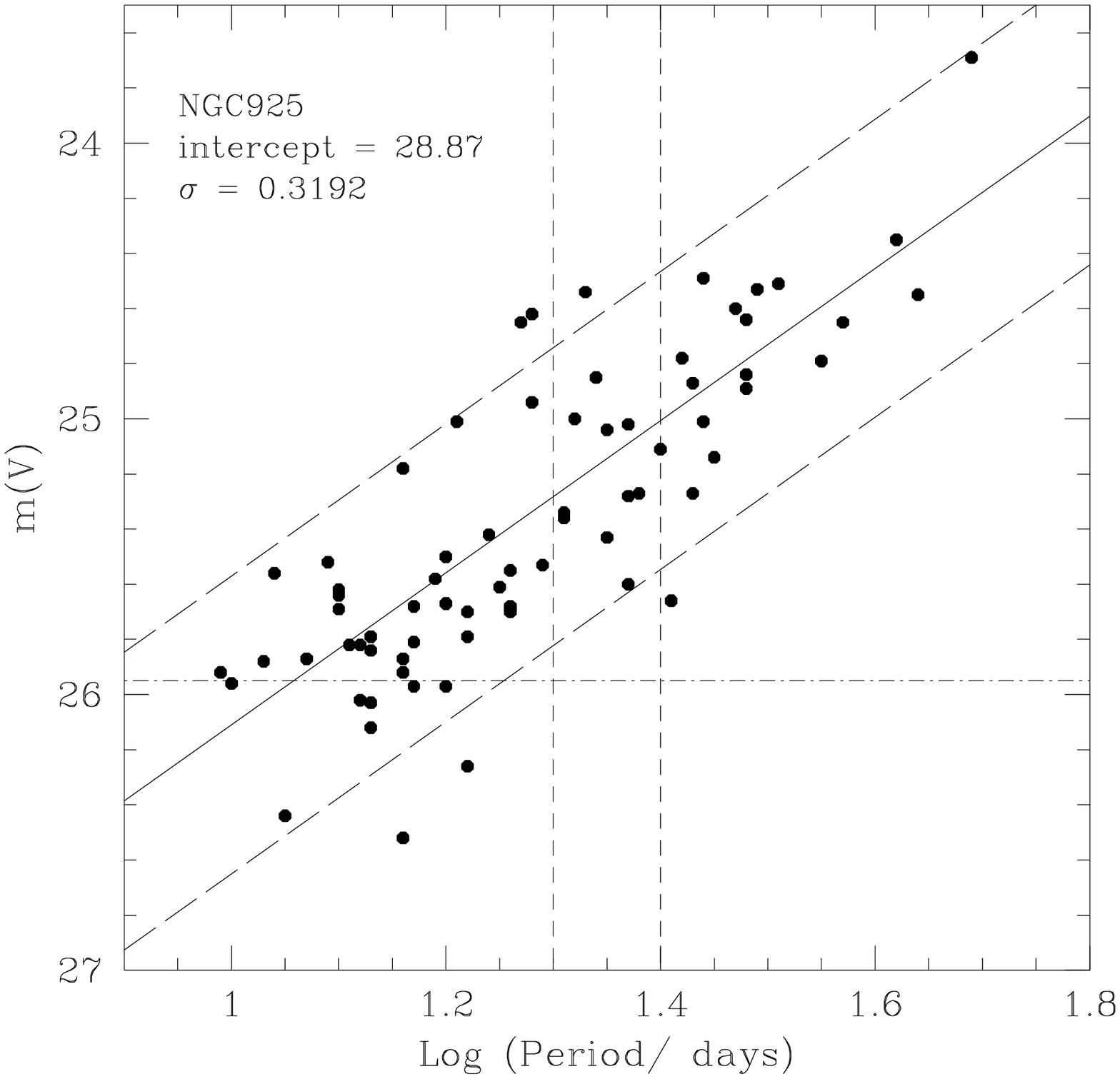}} &
{\epsfxsize=5.5truecm \epsfysize=5.5truecm \epsfbox[17 144 590 715]{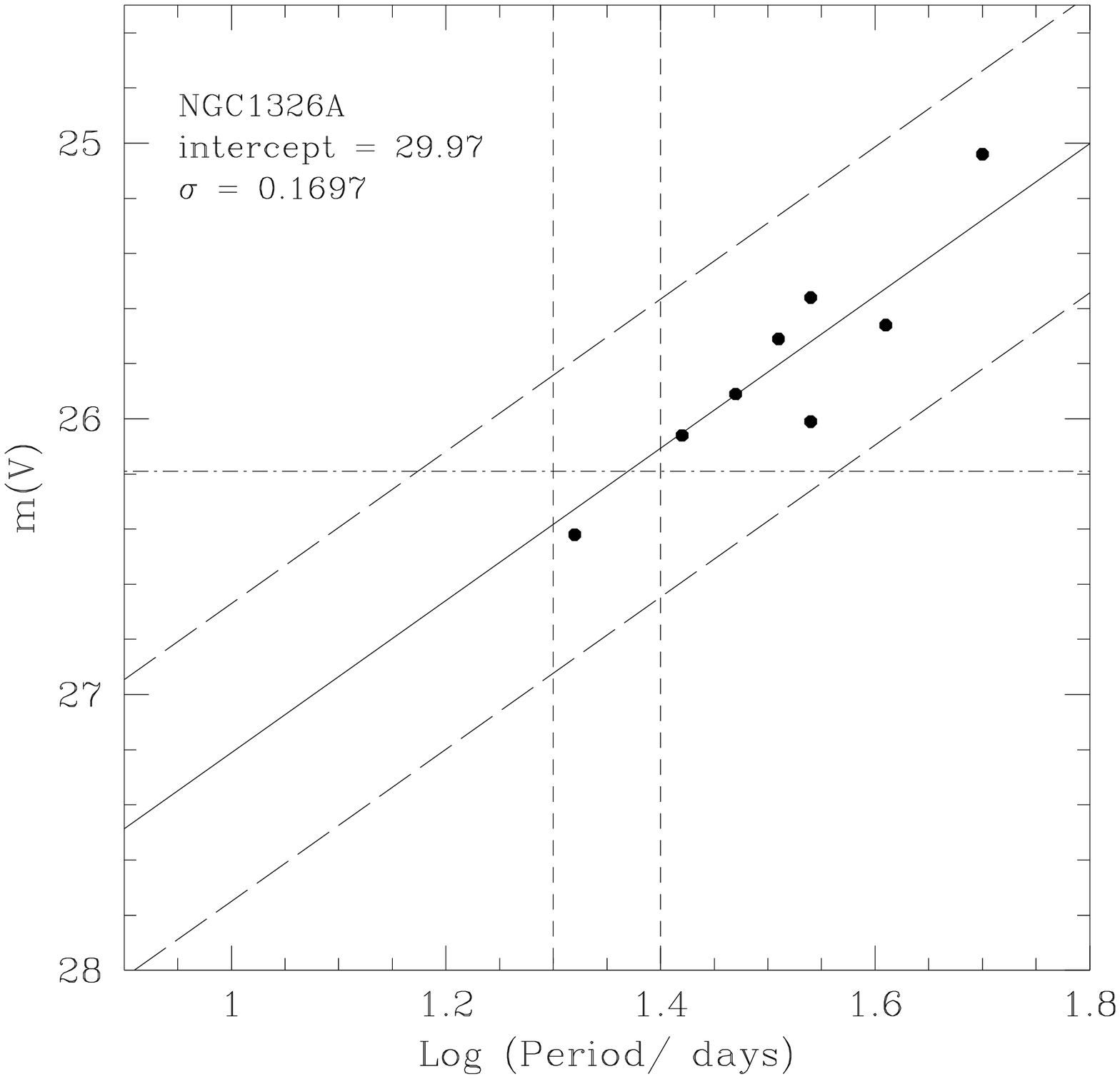}} &
{\epsfxsize=5.5truecm \epsfysize=5.5truecm \epsfbox[17 144 590 715]{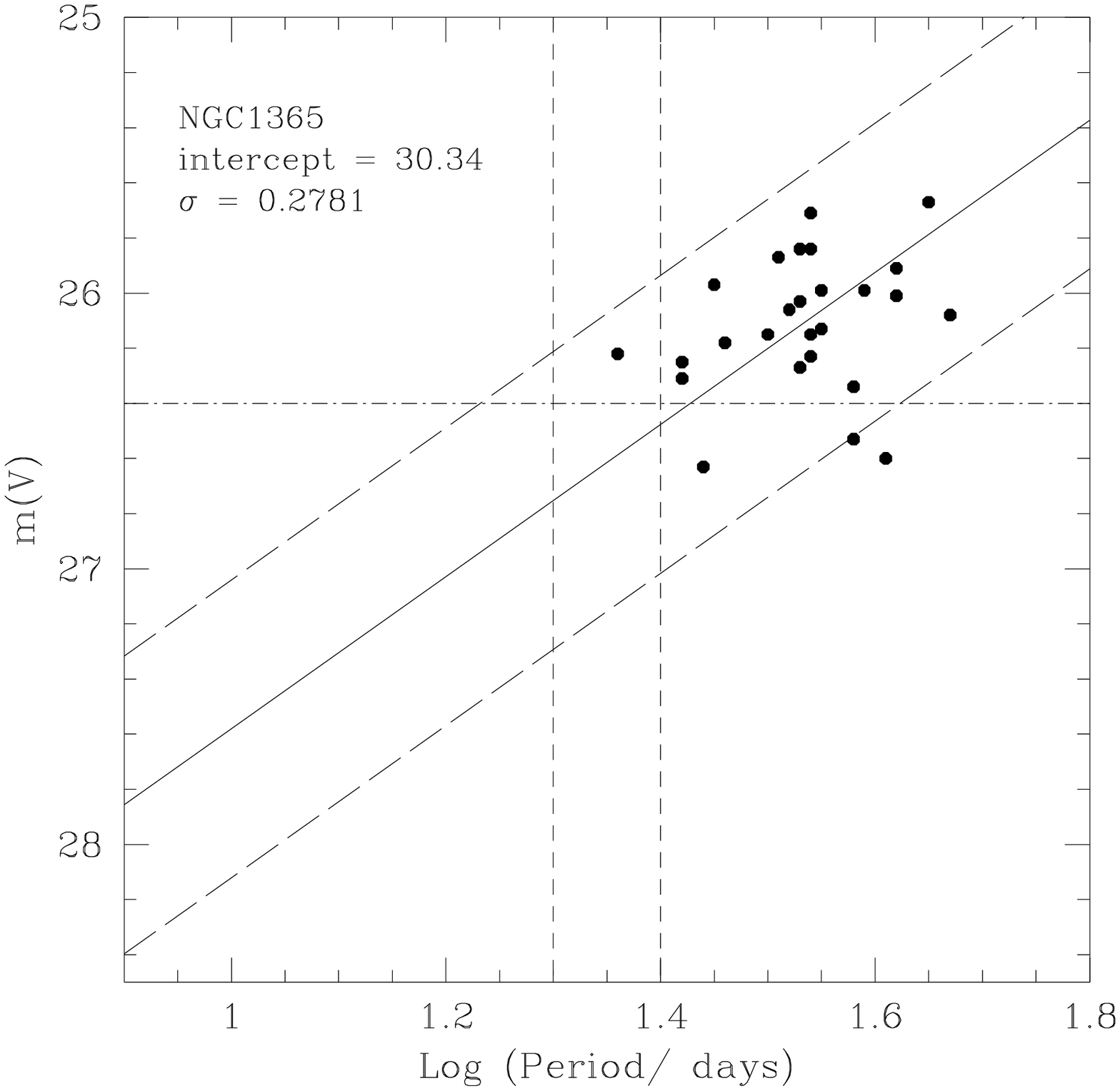}} \\
{\epsfxsize=5.5truecm \epsfysize=5.5truecm \epsfbox[17 144 590 715]{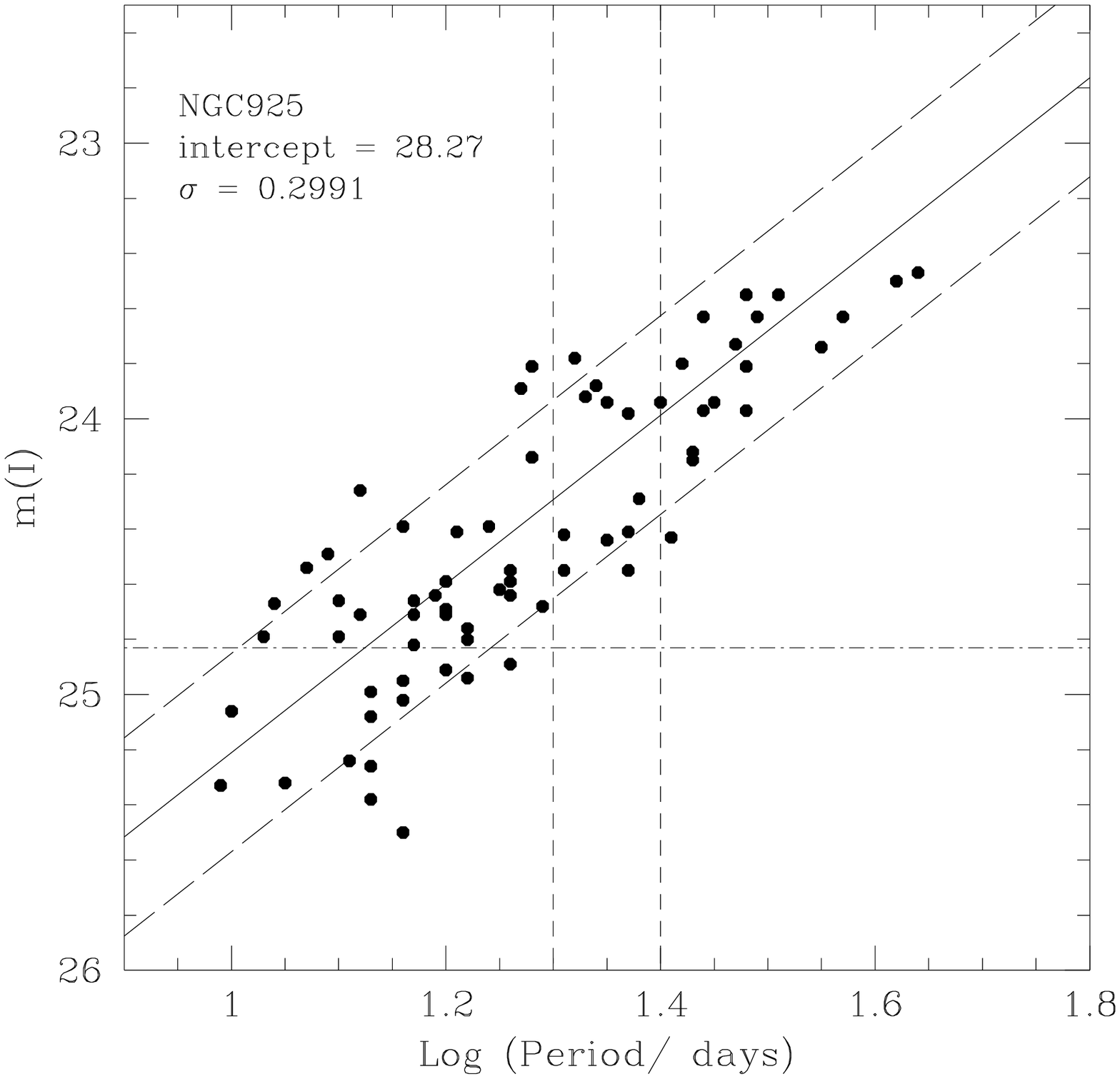}} &
{\epsfxsize=5.5truecm \epsfysize=5.5truecm \epsfbox[17 144 590 715]{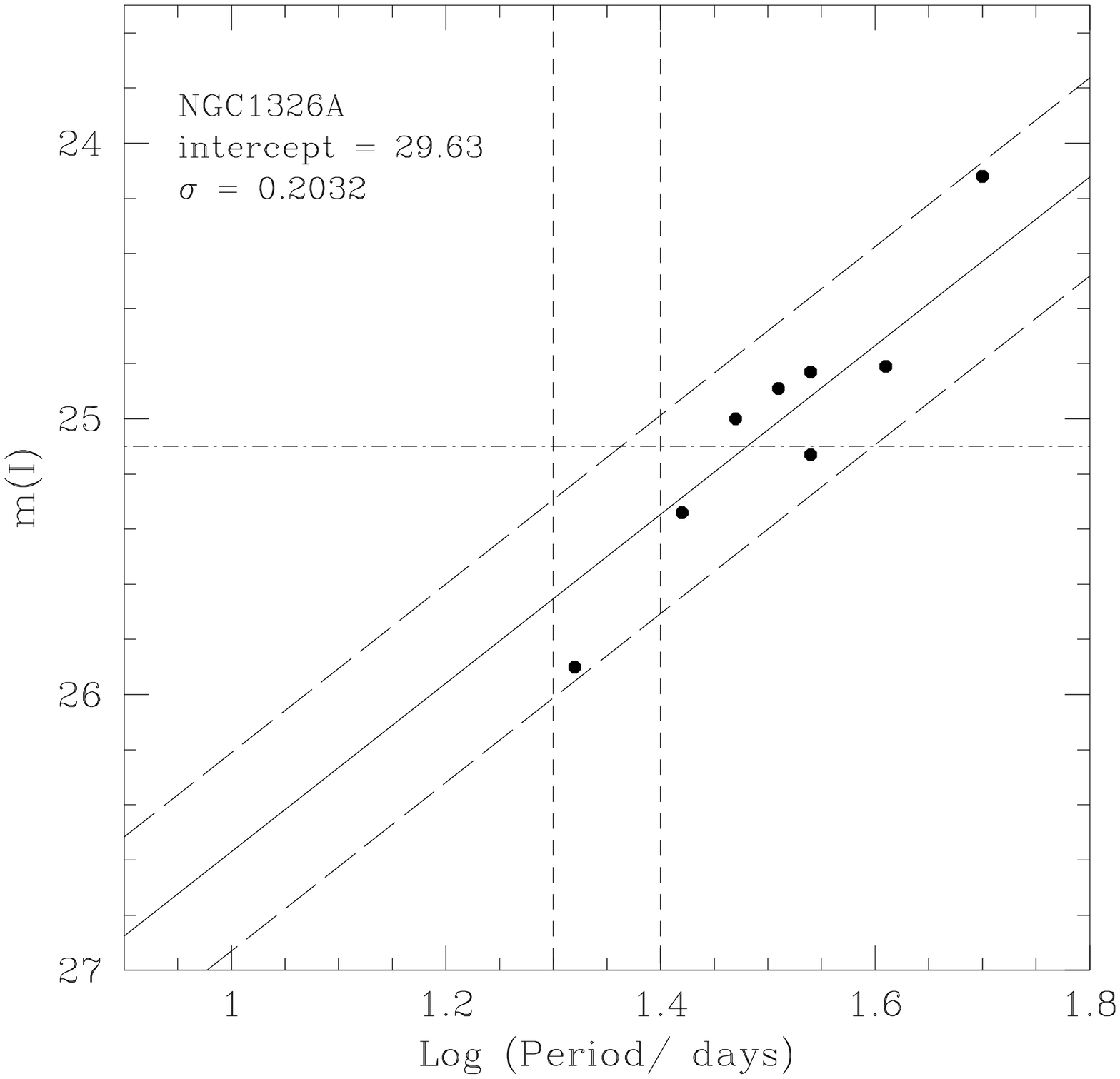}} &
{\epsfxsize=5.5truecm \epsfysize=5.5truecm \epsfbox[17 144 590 715]{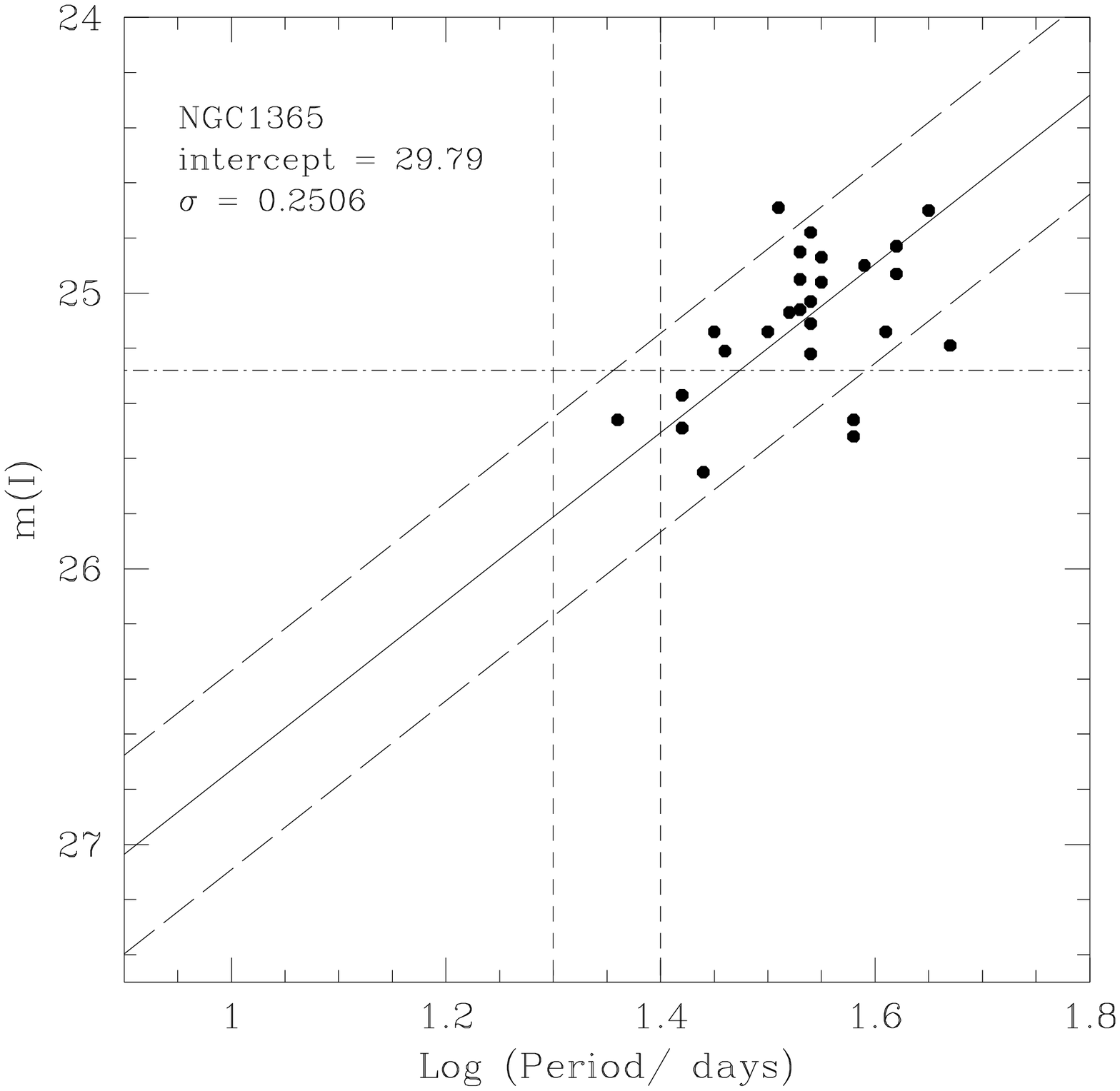}} \\
{\epsfxsize=5.5truecm \epsfysize=5.5truecm \epsfbox[17 144 590 715]{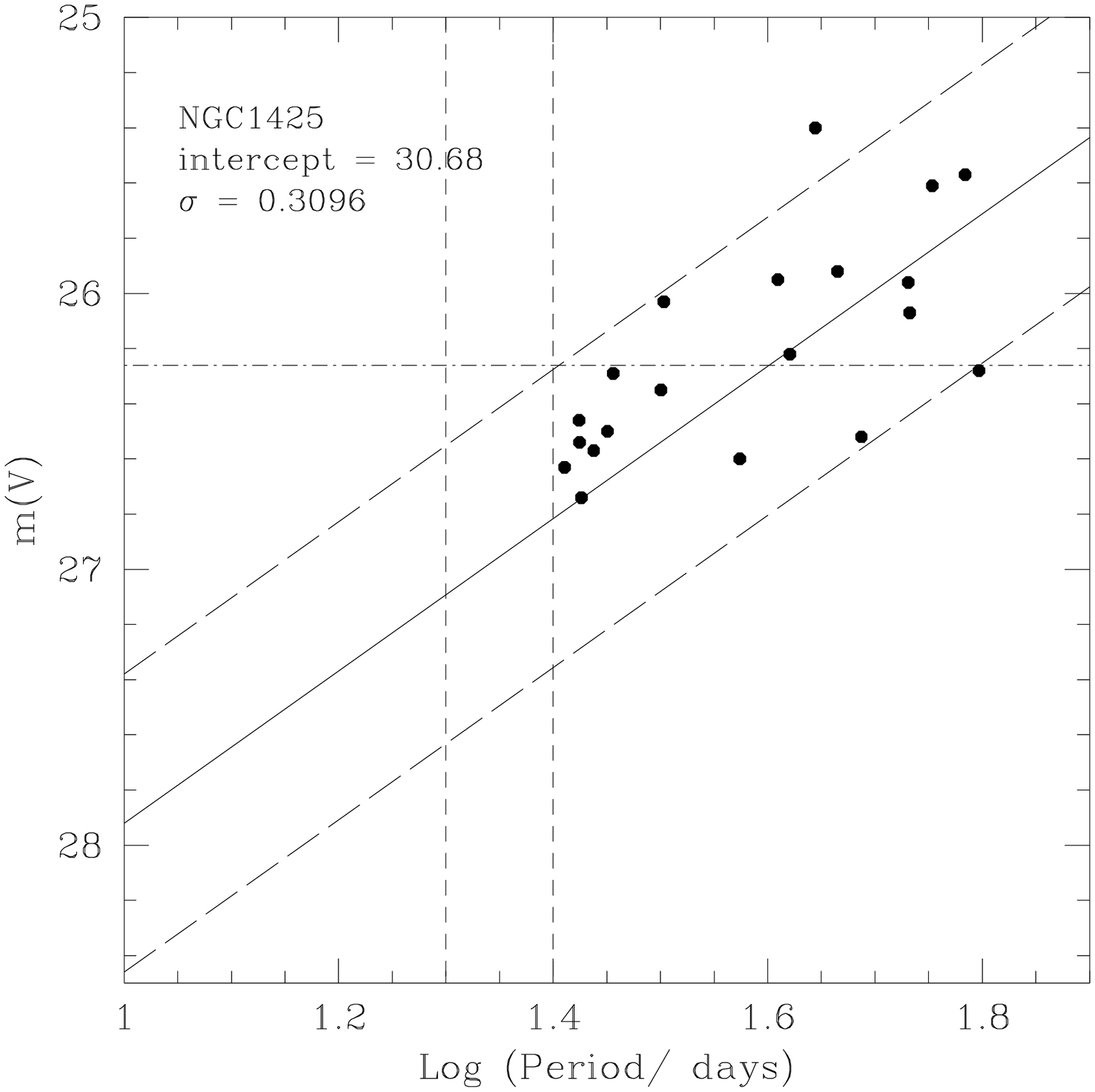}} &
{\epsfxsize=5.5truecm \epsfysize=5.5truecm \epsfbox[17 144 590 715]{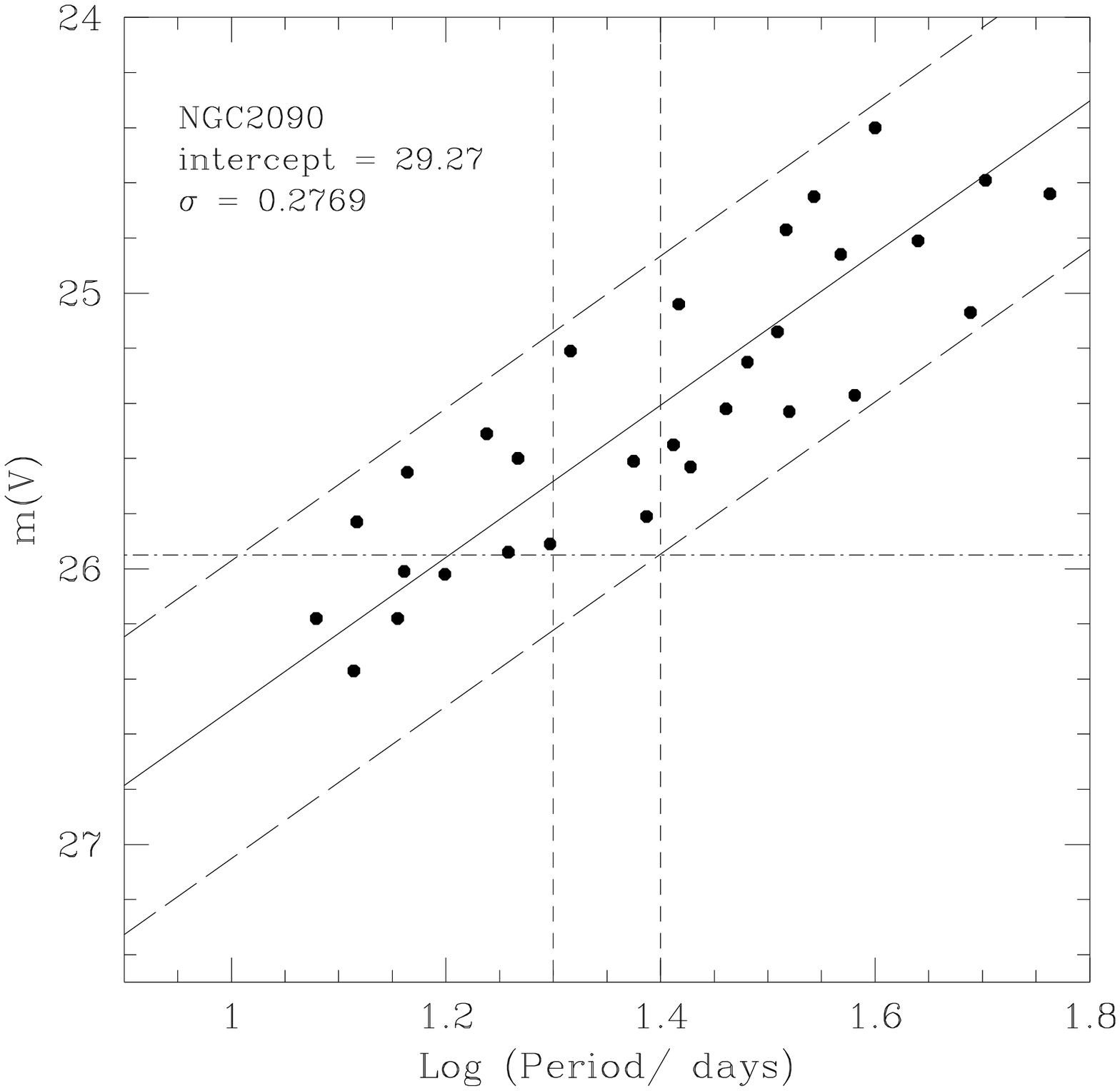}} &
{\epsfxsize=5.5truecm \epsfysize=5.5truecm \epsfbox[17 144 590 715]{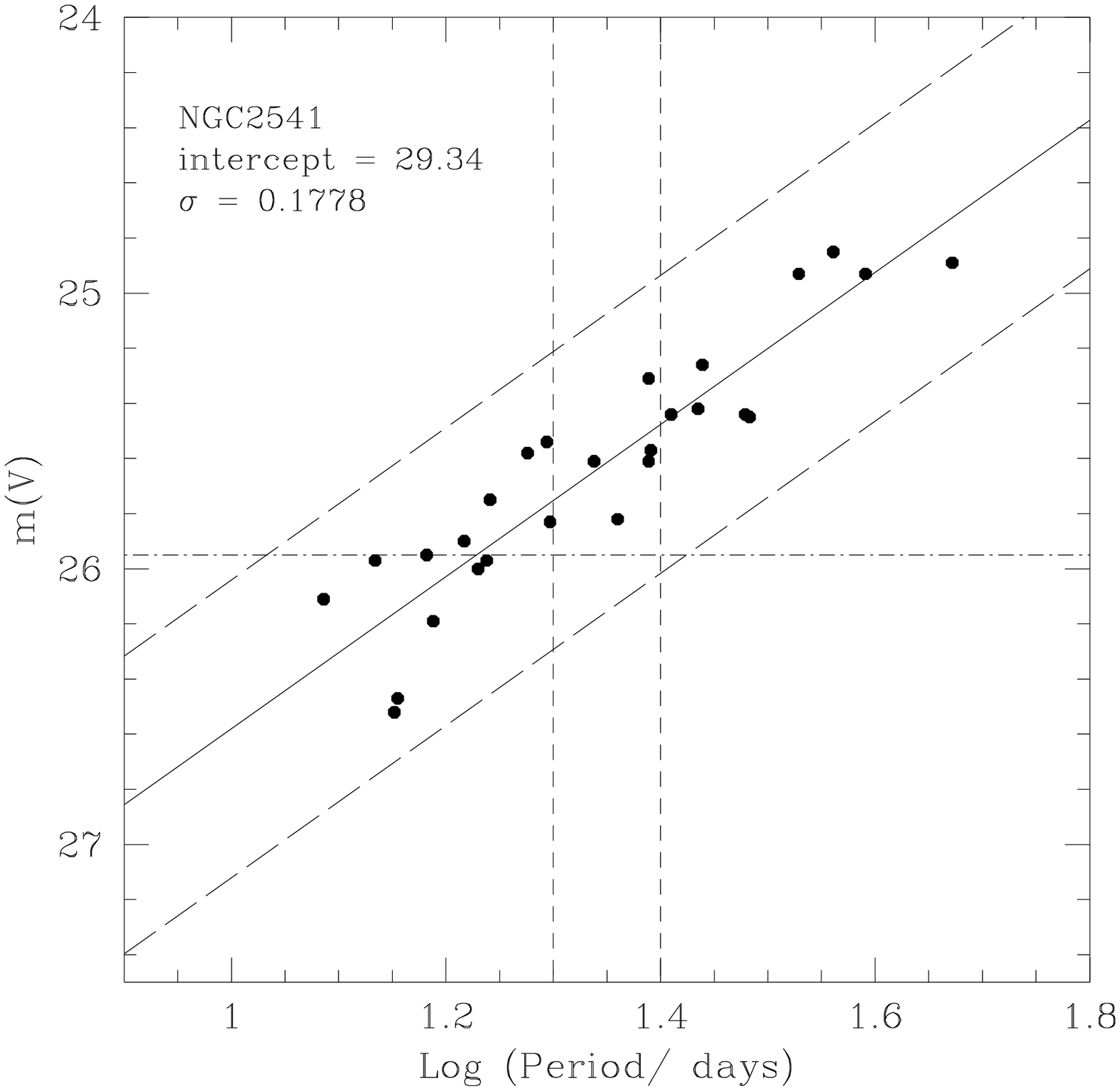}} \\
{\epsfxsize=5.5truecm \epsfysize=5.5truecm \epsfbox[17 144 590 715]{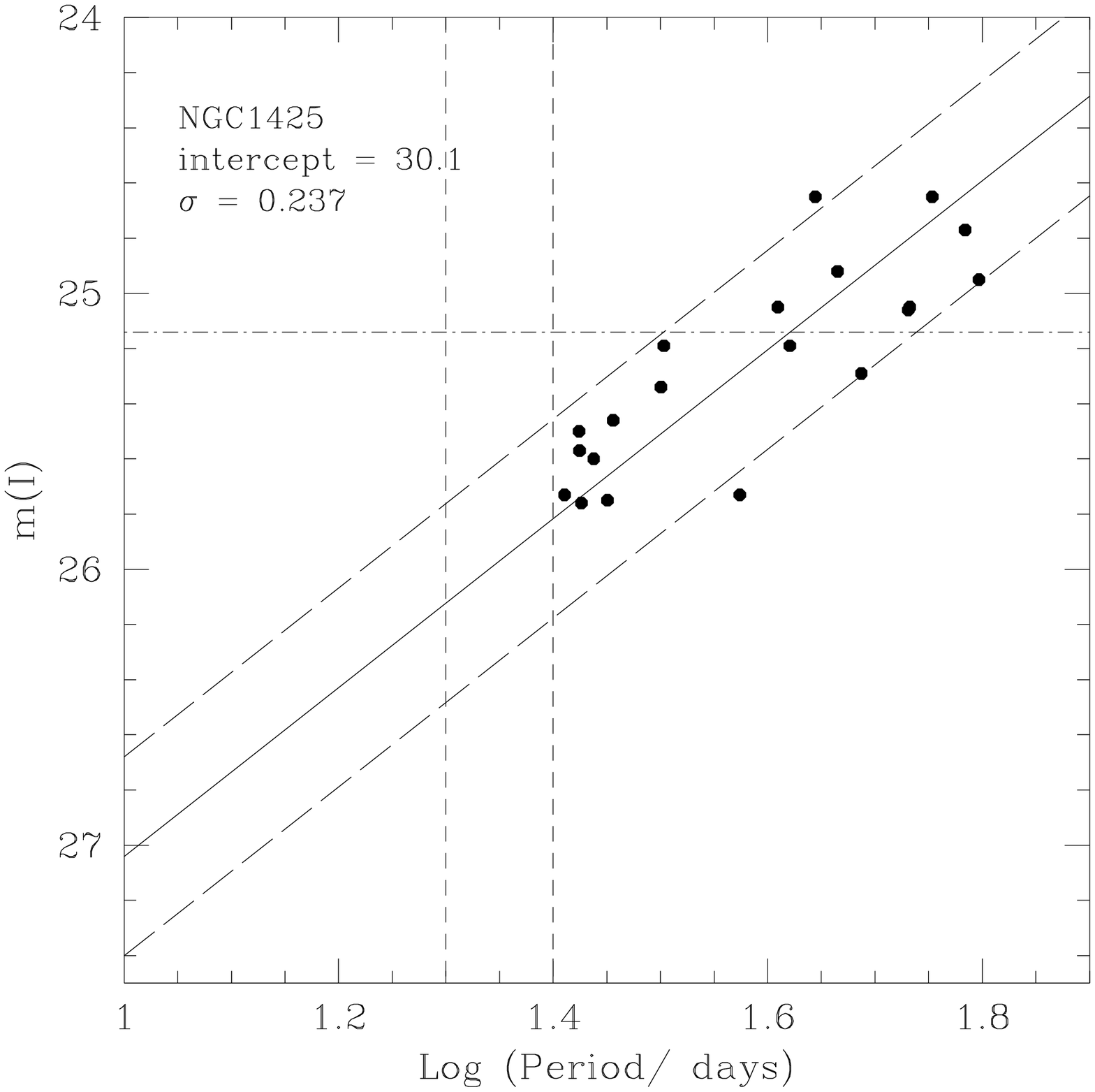}} &
{\epsfxsize=5.5truecm \epsfysize=5.5truecm \epsfbox[17 144 590 715]{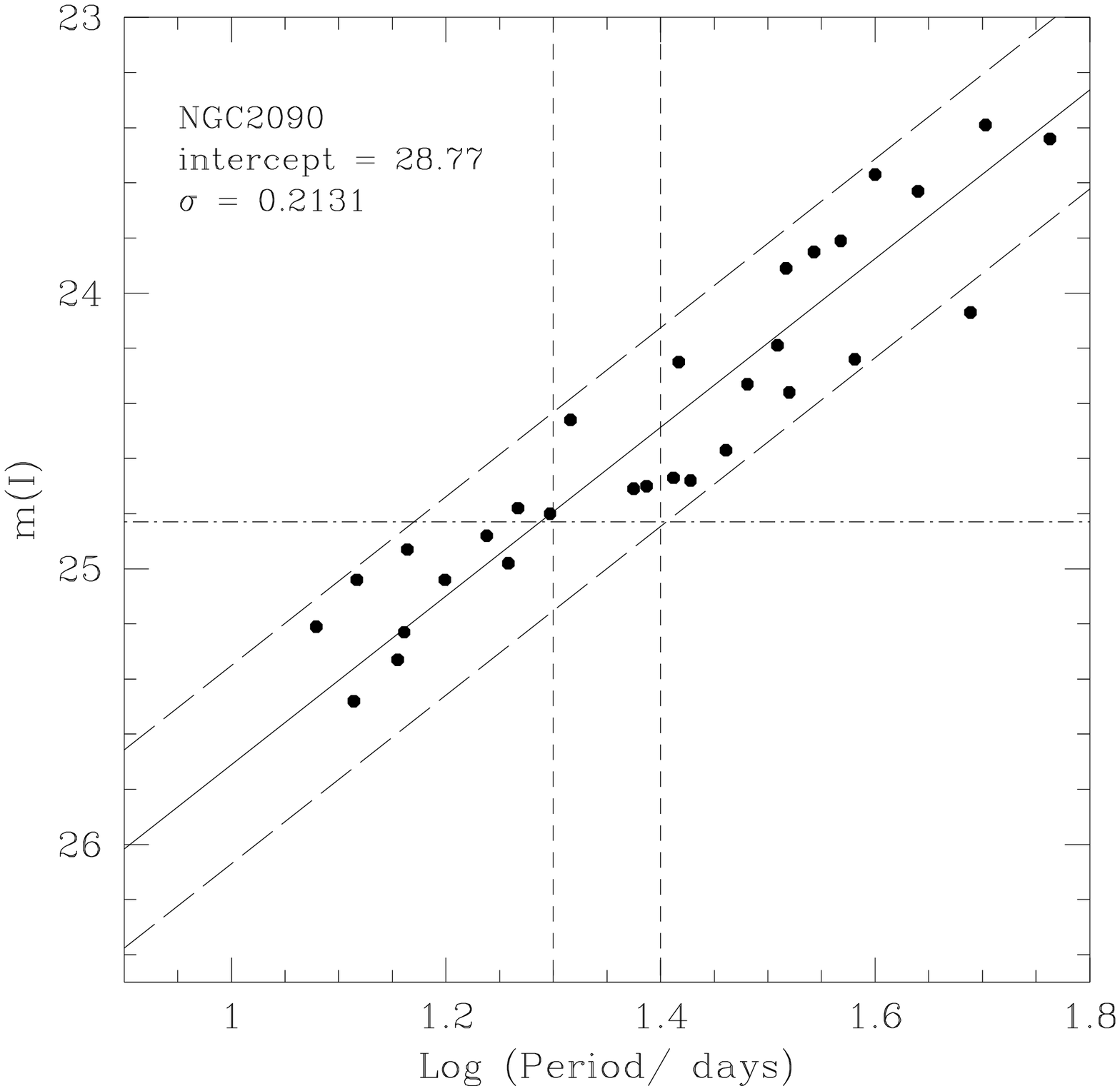}} &
{\epsfxsize=5.5truecm \epsfysize=5.5truecm \epsfbox[17 144 590 715]{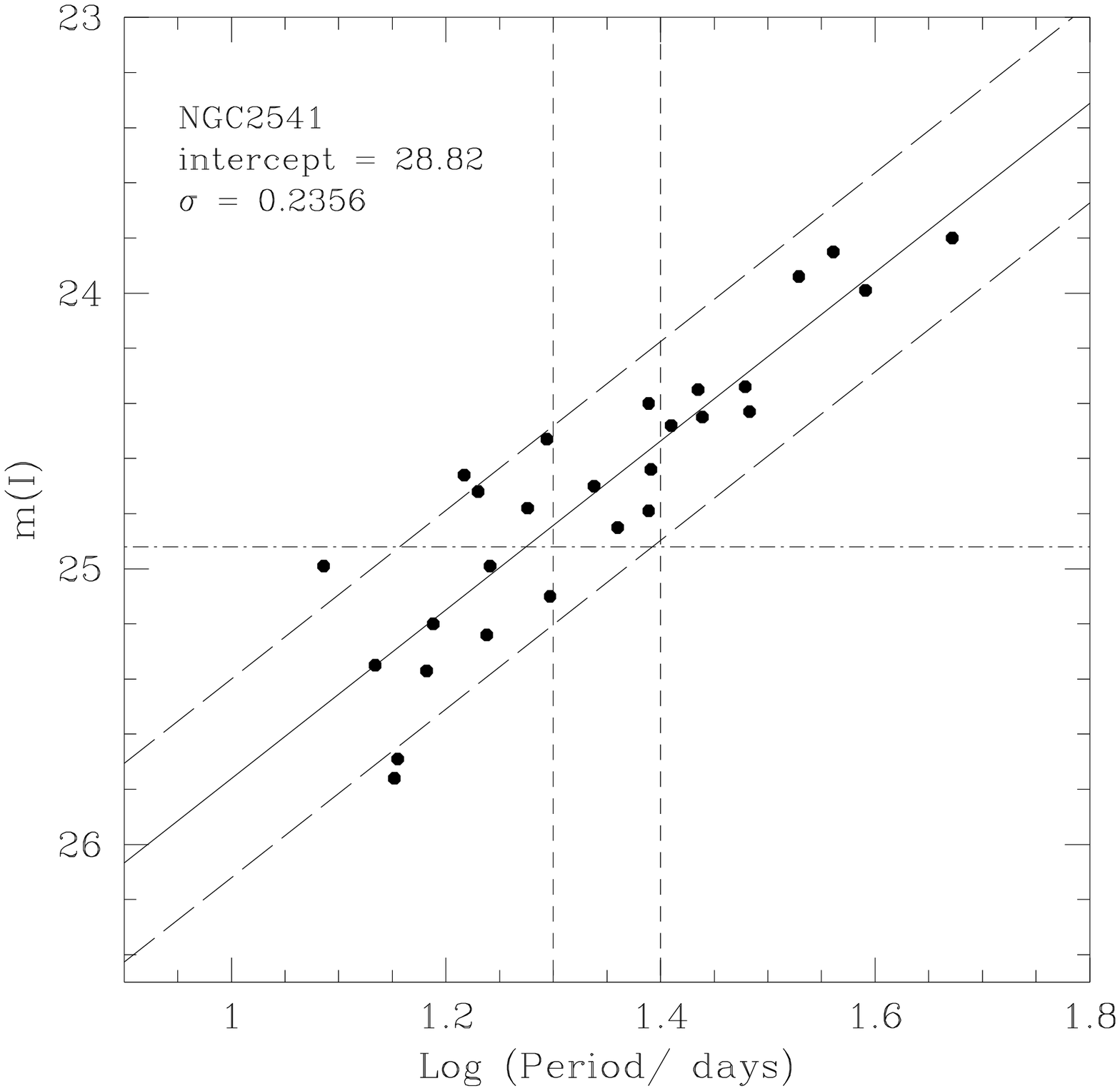}} \\
\end{tabular}
\caption{P-L Relation in V and I for 25 HST Cepheid Galaxies. The best fit to the data is shown as a solid black line along with 2$\sigma$ LMC envelopes. Magnitude limits are shown as horizontal broken lines, and possible period limits are shown as vertical dashed lines \emph{continued}.}
\label{fig:pl1} 
\end{figure*}

\begin{figure*} 
\begin{tabular}{ccc} 
{\epsfxsize=5.5truecm \epsfysize=5.5truecm \epsfbox[17 144 590 715]{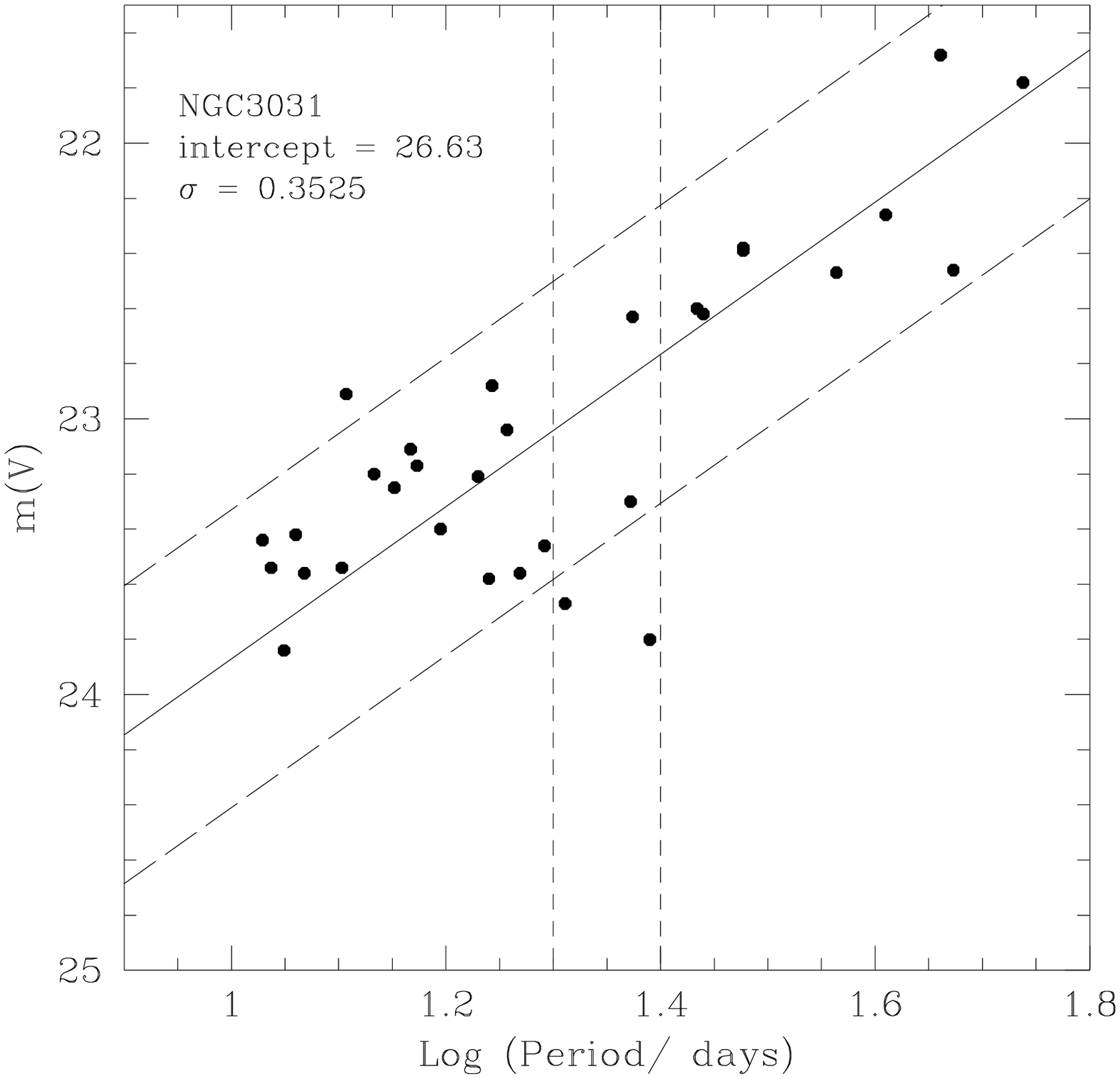}} &
{\epsfxsize=5.5truecm \epsfysize=5.5truecm \epsfbox[17 144 590 715]{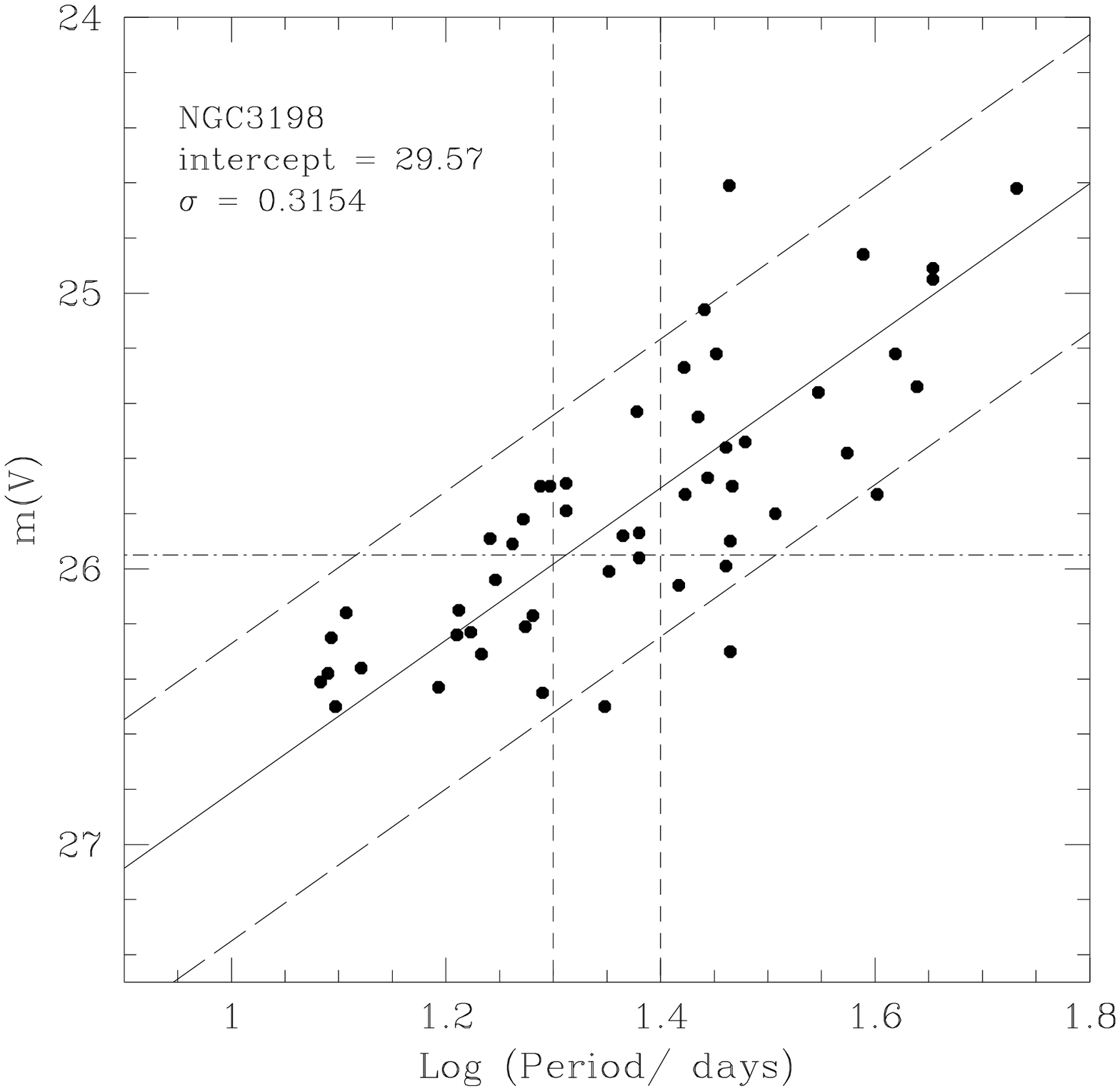}} &
{\epsfxsize=5.5truecm \epsfysize=5.5truecm \epsfbox[17 144 590 715]{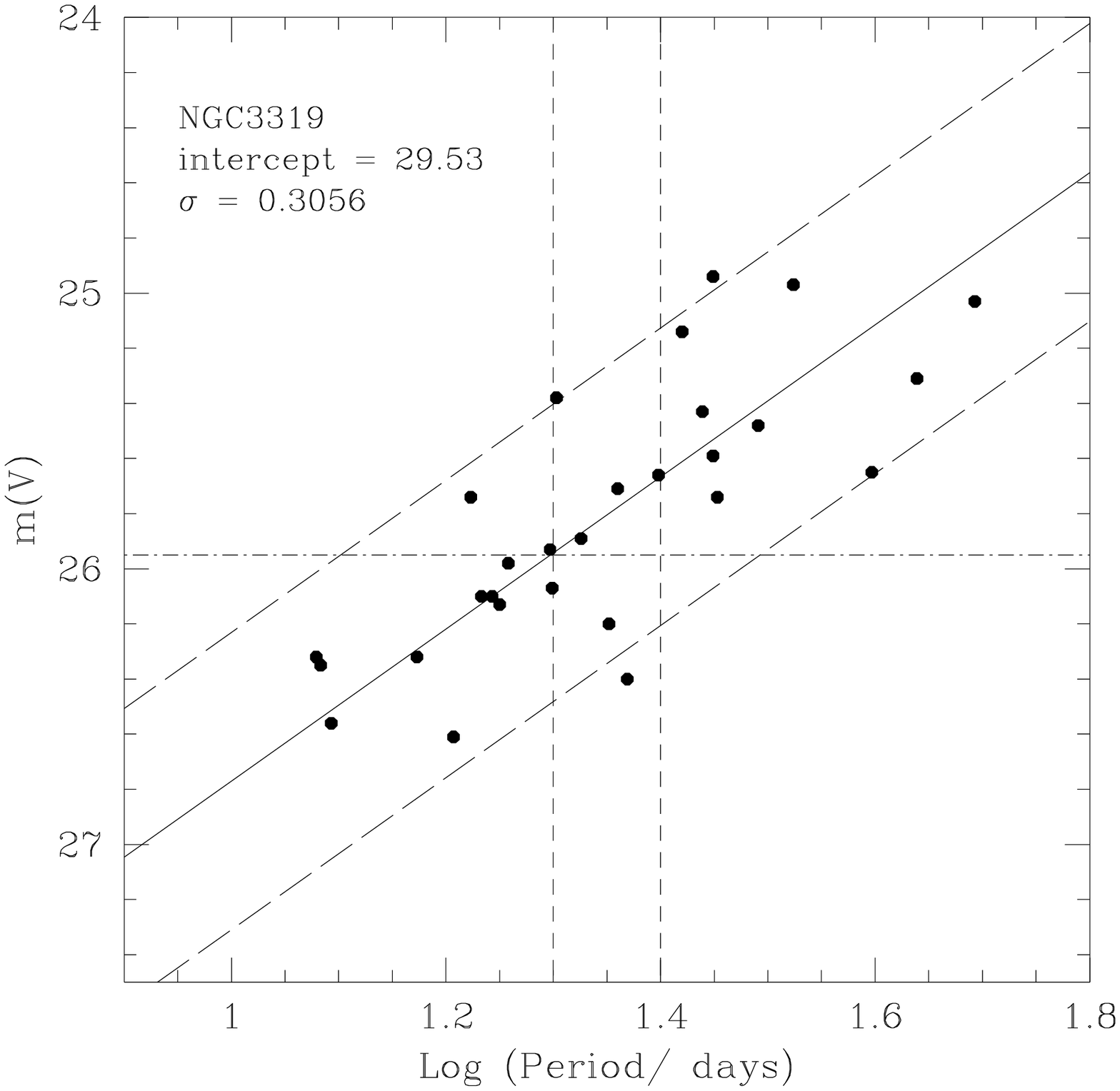}} \\
{\epsfxsize=5.5truecm \epsfysize=5.5truecm \epsfbox[17 144 590 715]{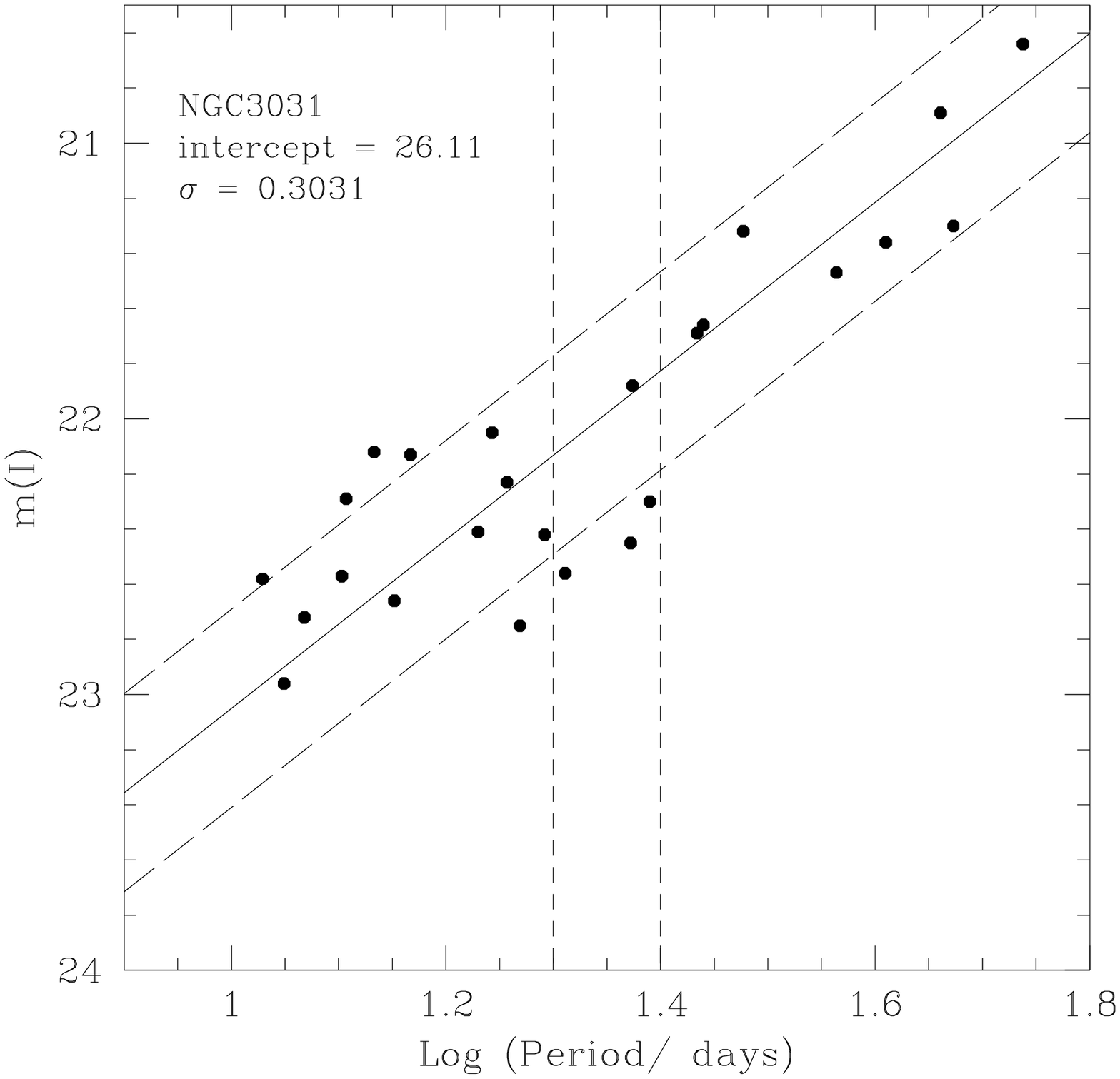}} &
{\epsfxsize=5.5truecm \epsfysize=5.5truecm \epsfbox[17 144 590 715]{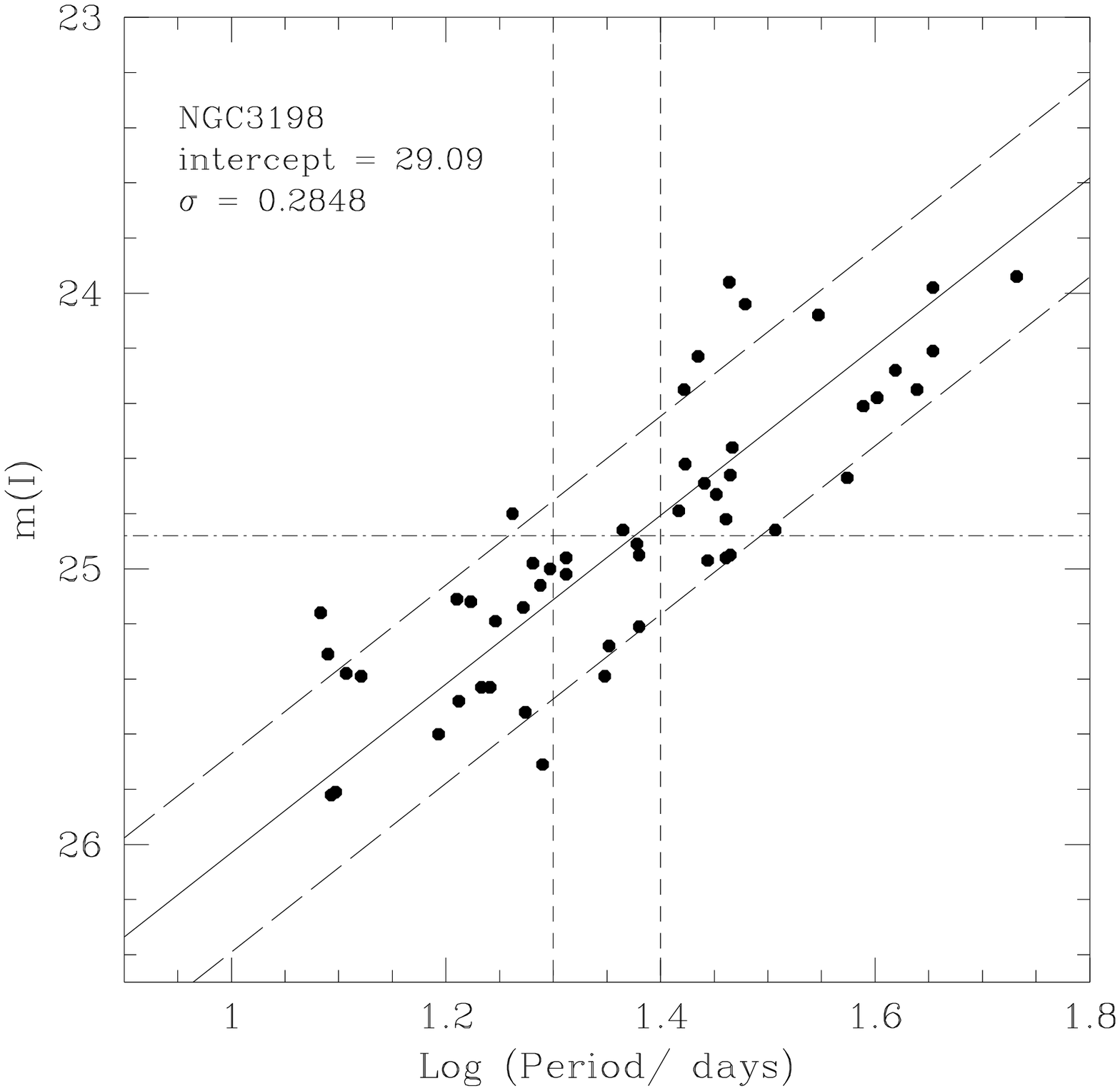}} &
{\epsfxsize=5.5truecm \epsfysize=5.5truecm \epsfbox[17 144 590 715]{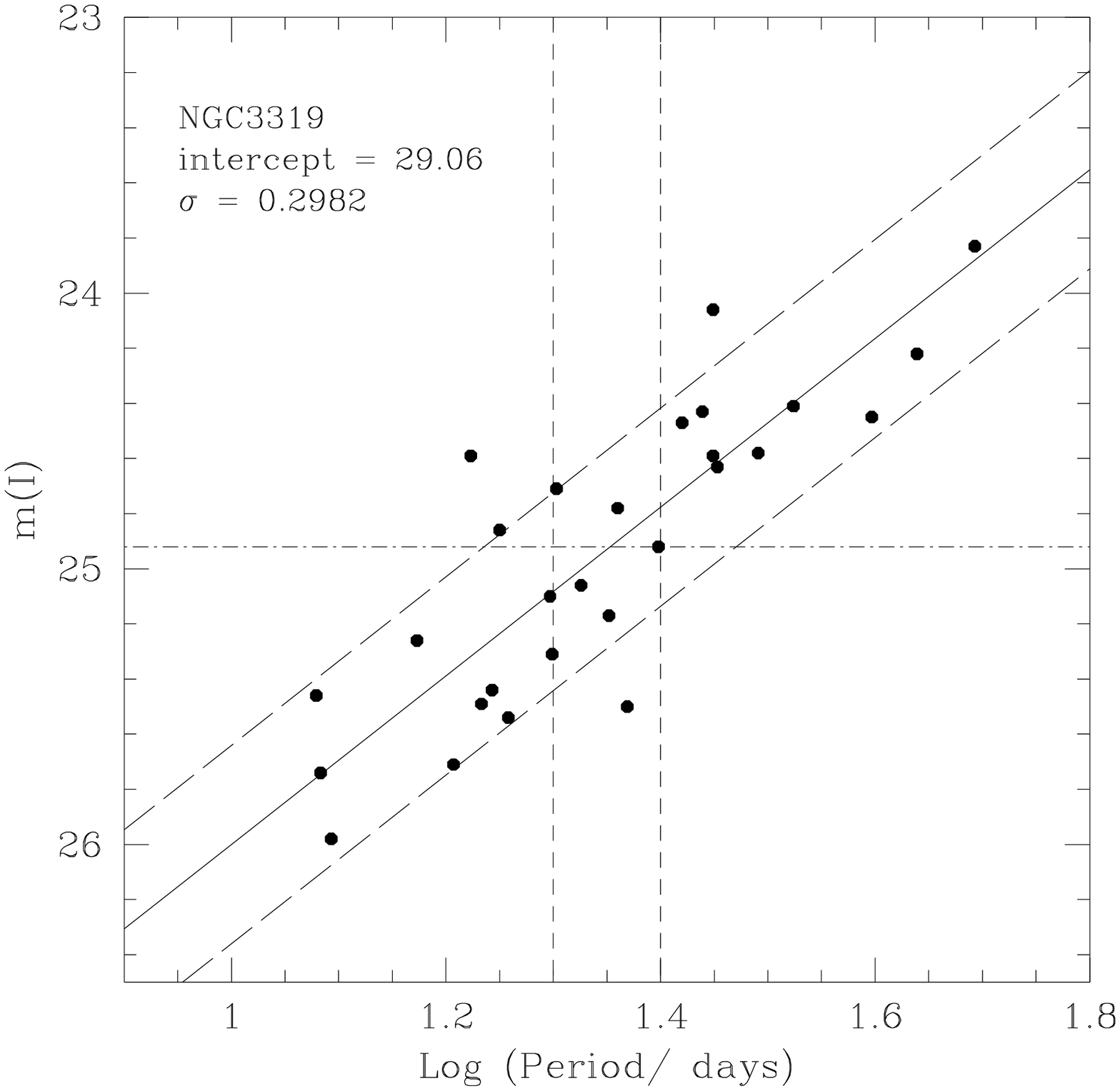}} \\
{\epsfxsize=5.5truecm \epsfysize=5.5truecm \epsfbox[17 144 590 715]{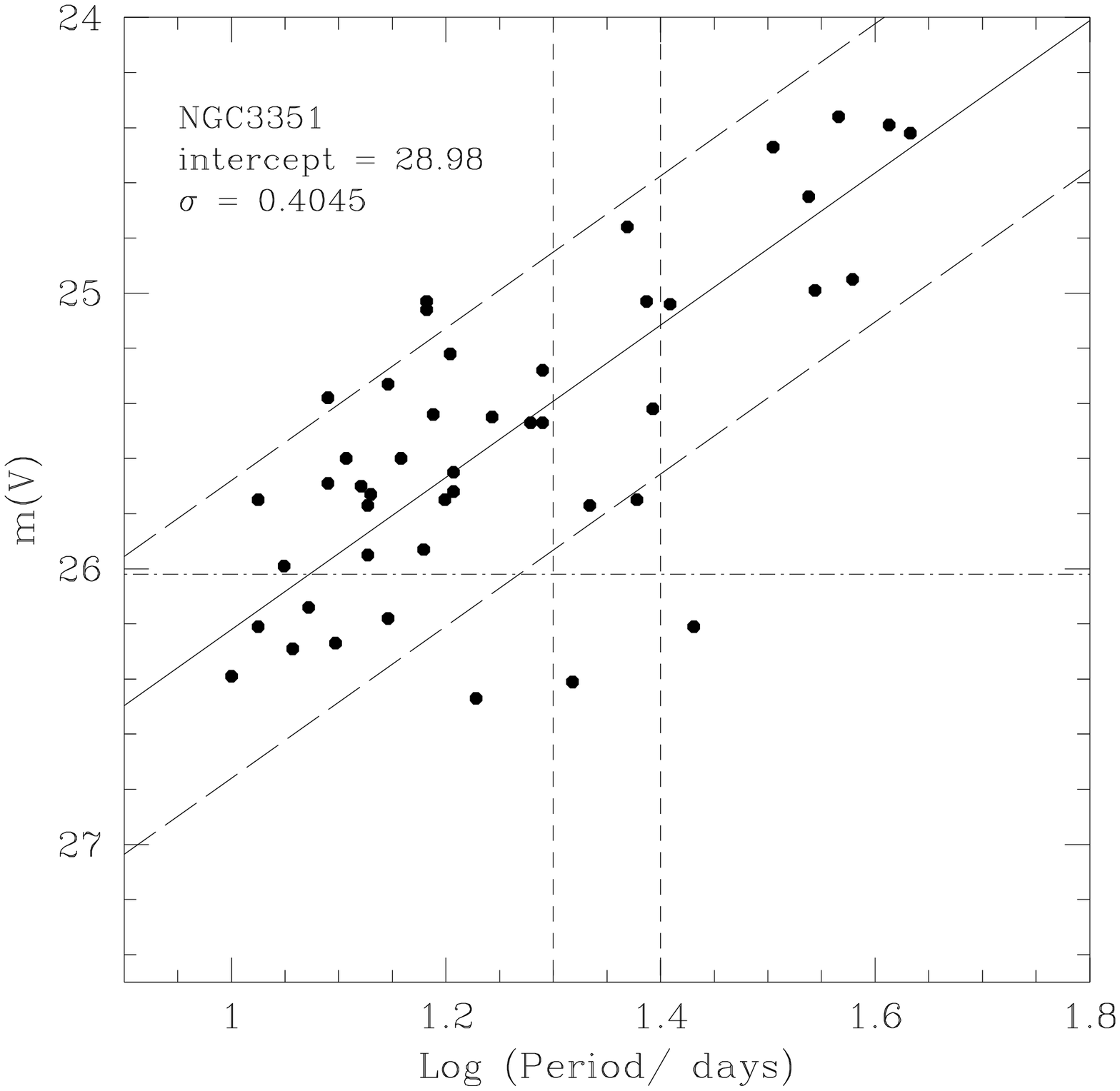}} &
{\epsfxsize=5.5truecm \epsfysize=5.5truecm \epsfbox[17 144 590 715]{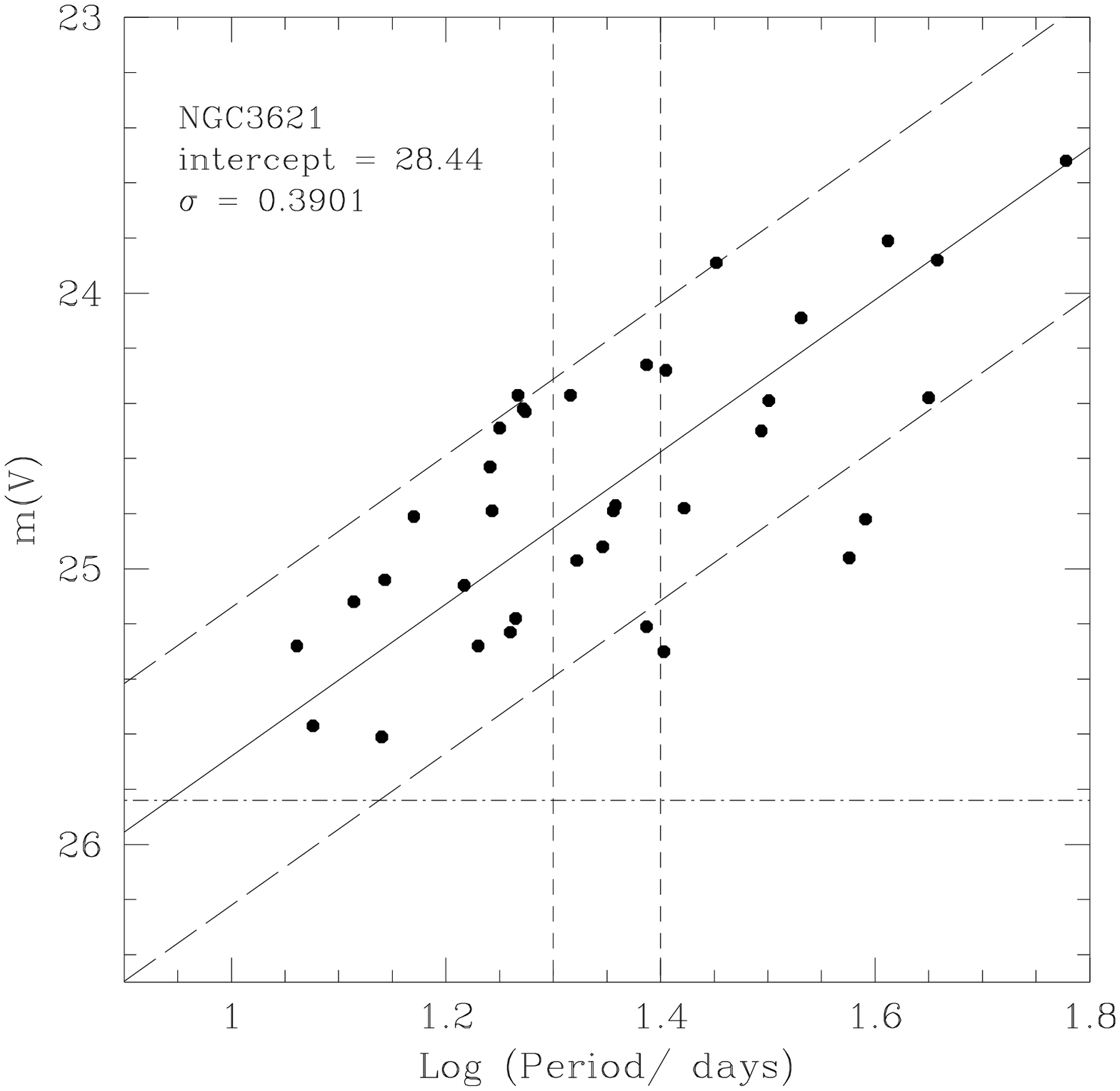}} &
{\epsfxsize=5.5truecm \epsfysize=5.5truecm \epsfbox[17 144 590 715]{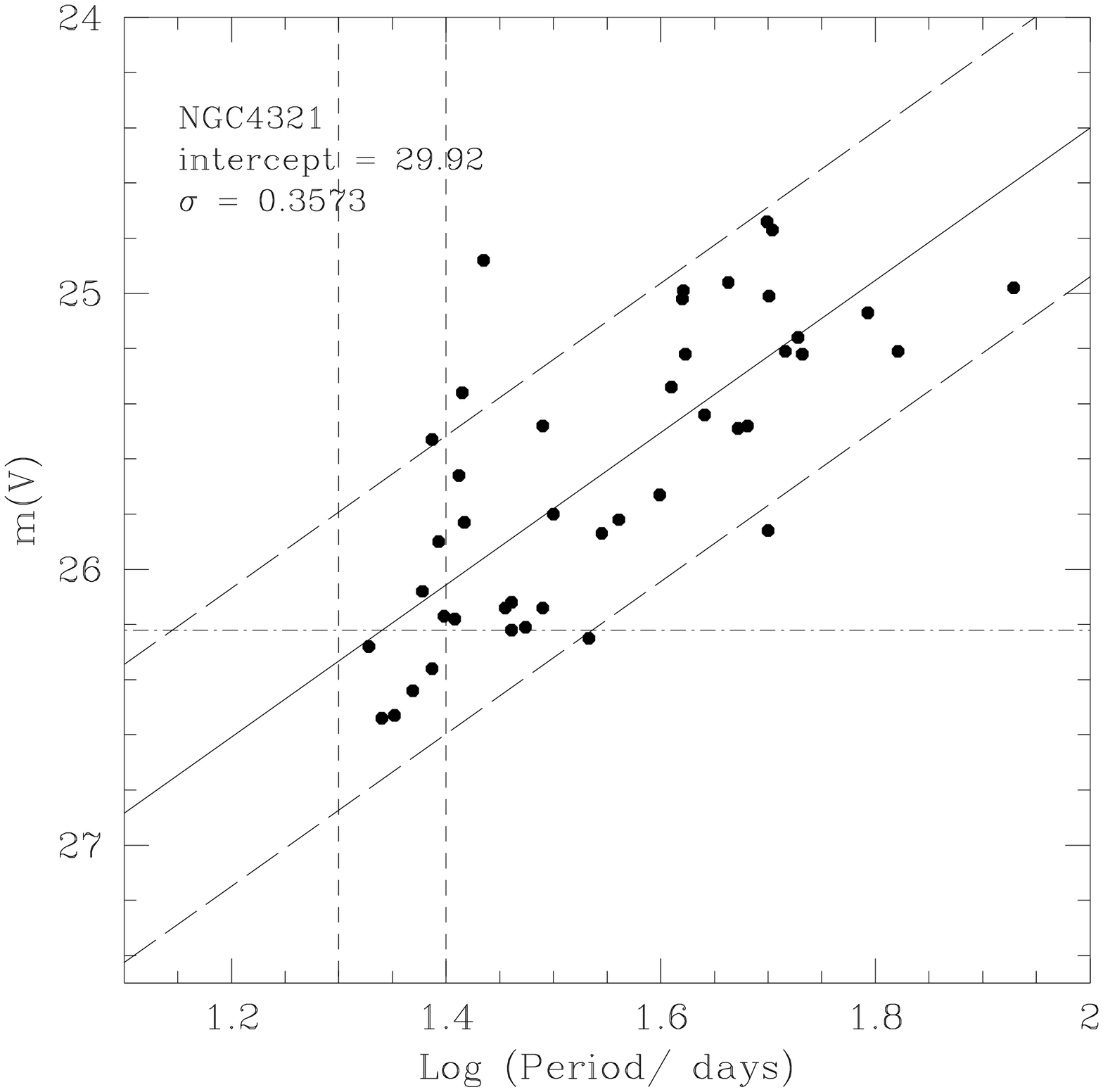}} \\
{\epsfxsize=5.5truecm \epsfysize=5.5truecm \epsfbox[17 144 590 715]{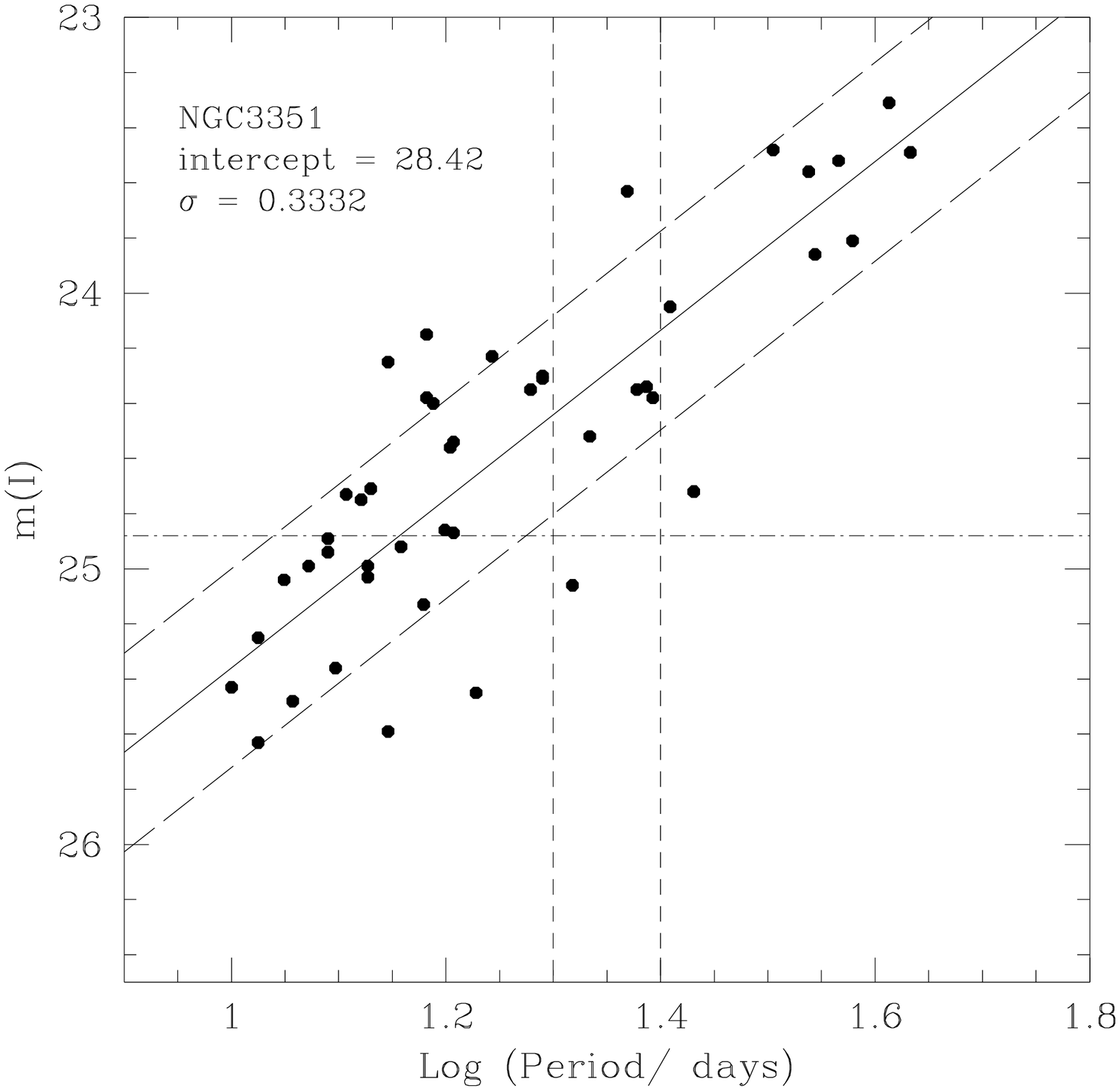}} &
{\epsfxsize=5.5truecm \epsfysize=5.5truecm \epsfbox[17 144 590 715]{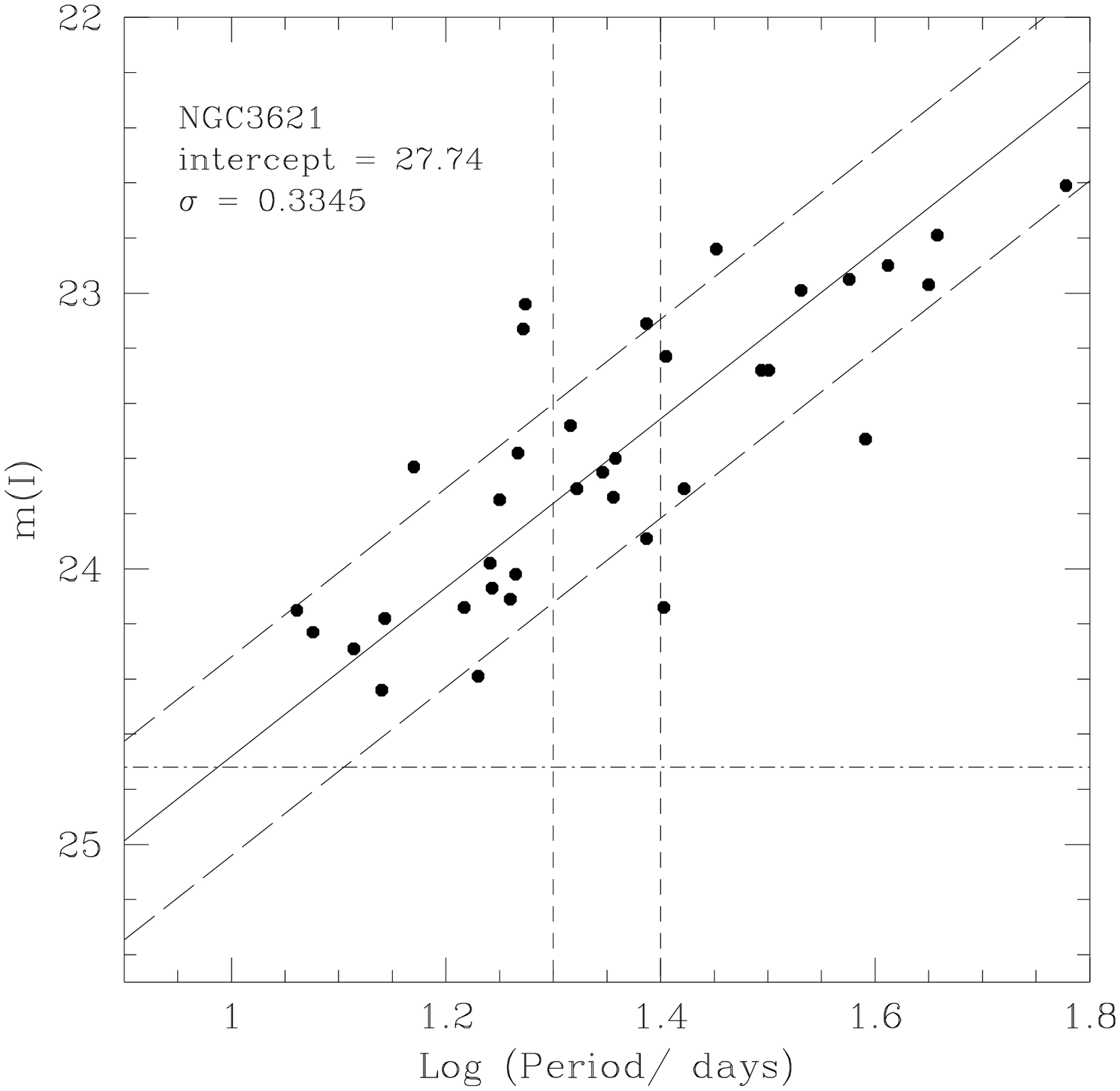}} &
{\epsfxsize=5.5truecm \epsfysize=5.5truecm \epsfbox[17 144 590 715]{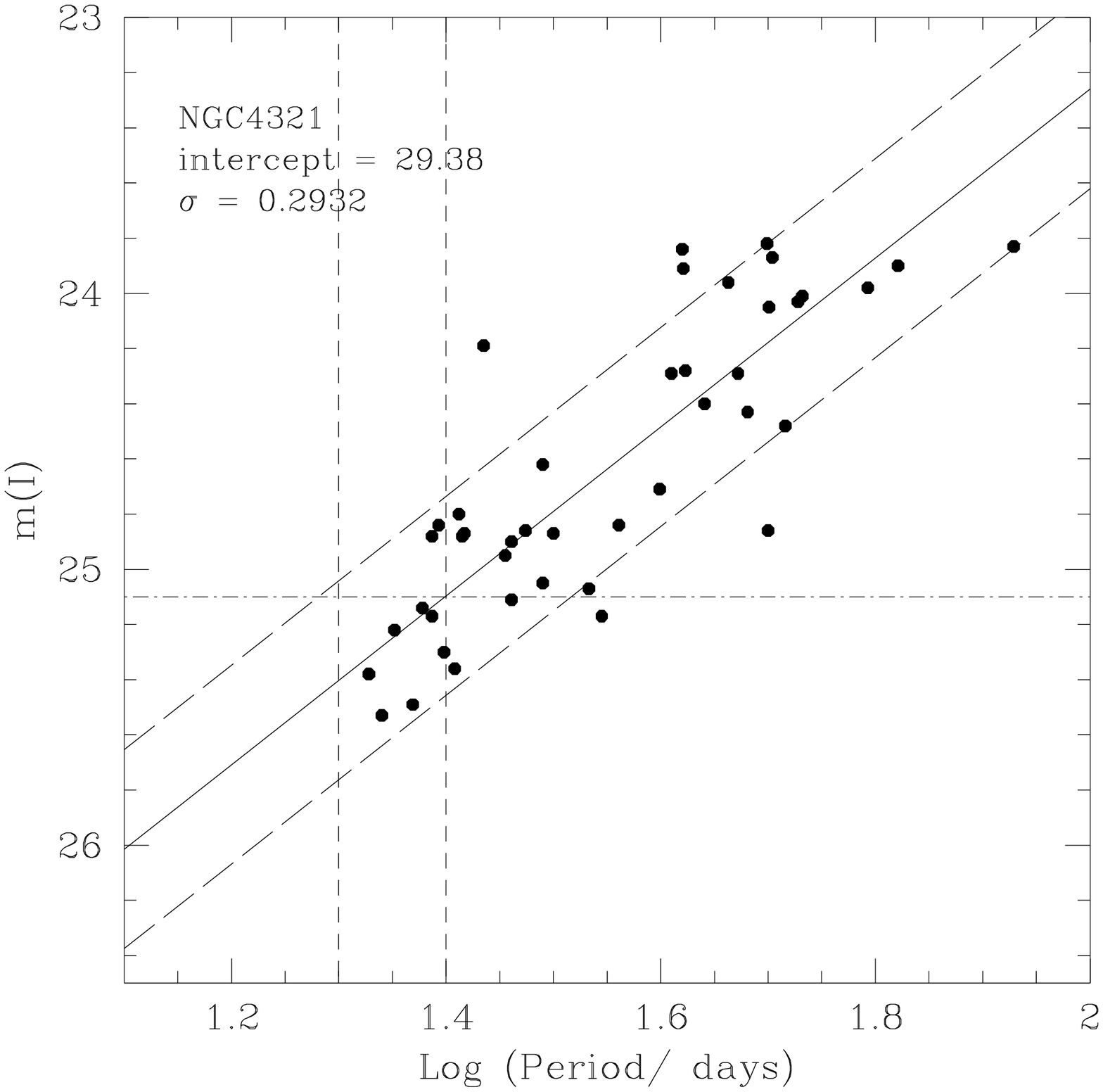}} \\
\end{tabular}
\addtocounter{figure}{-1}
\caption{\emph{continued}}
\label{fig:pl2} 
\end{figure*}

\begin{figure*} 
\begin{tabular}{ccc} 
{\epsfxsize=5.5truecm \epsfysize=5.5truecm \epsfbox[17 144 590 715]{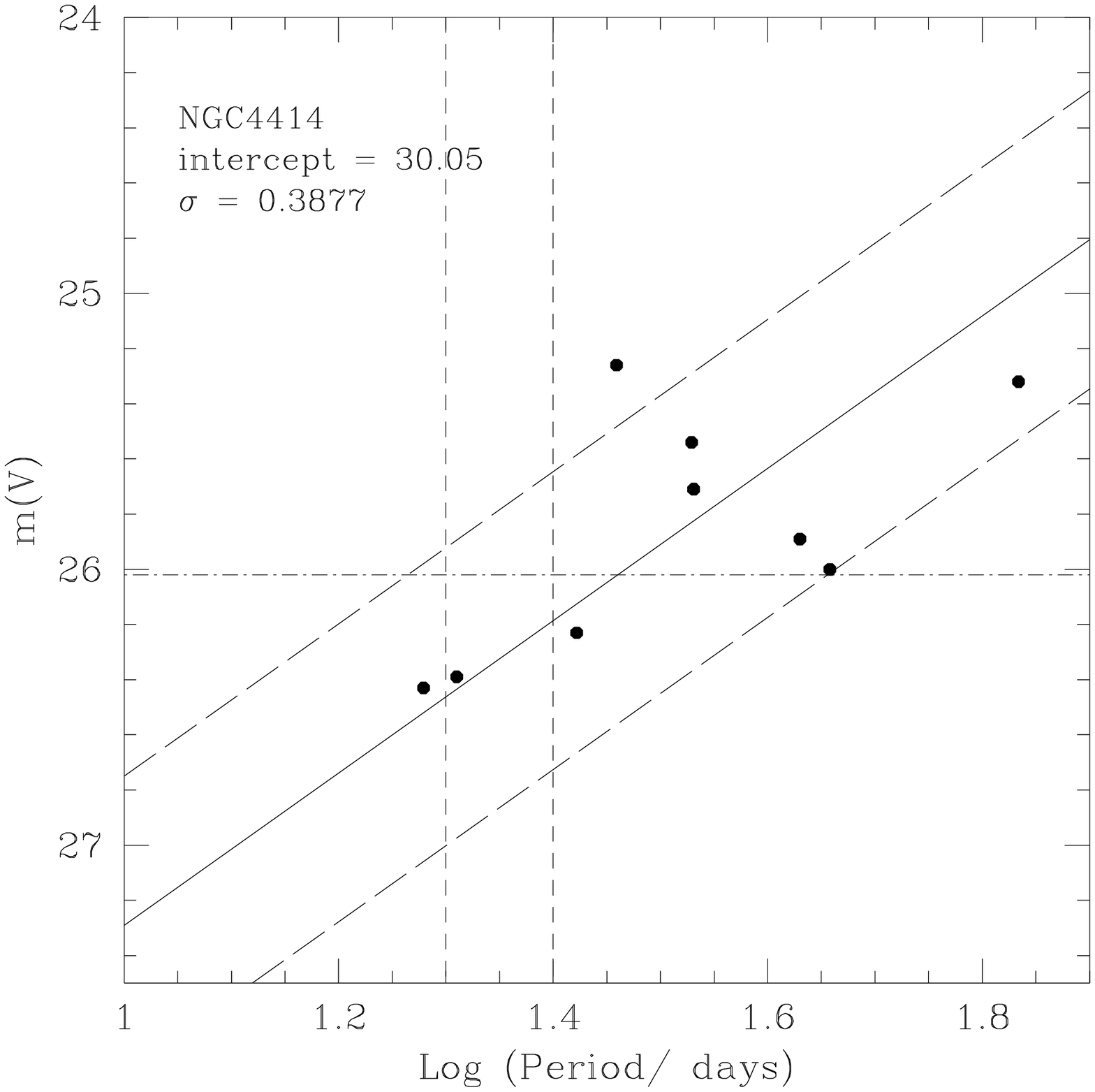}} &
{\epsfxsize=5.5truecm \epsfysize=5.5truecm \epsfbox[17 144 590 715]{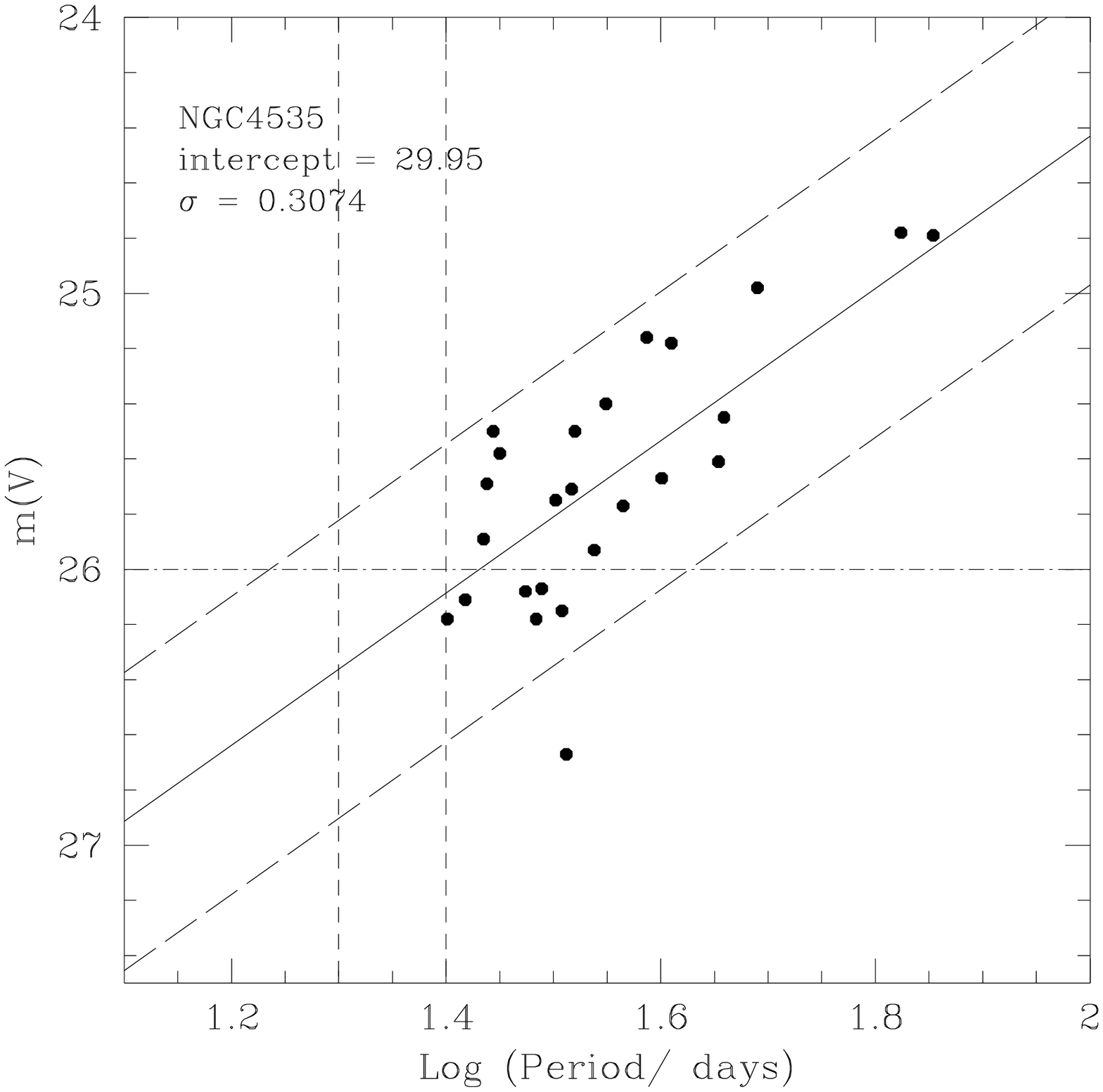}} &
{\epsfxsize=5.5truecm \epsfysize=5.5truecm \epsfbox[17 144 590 715]{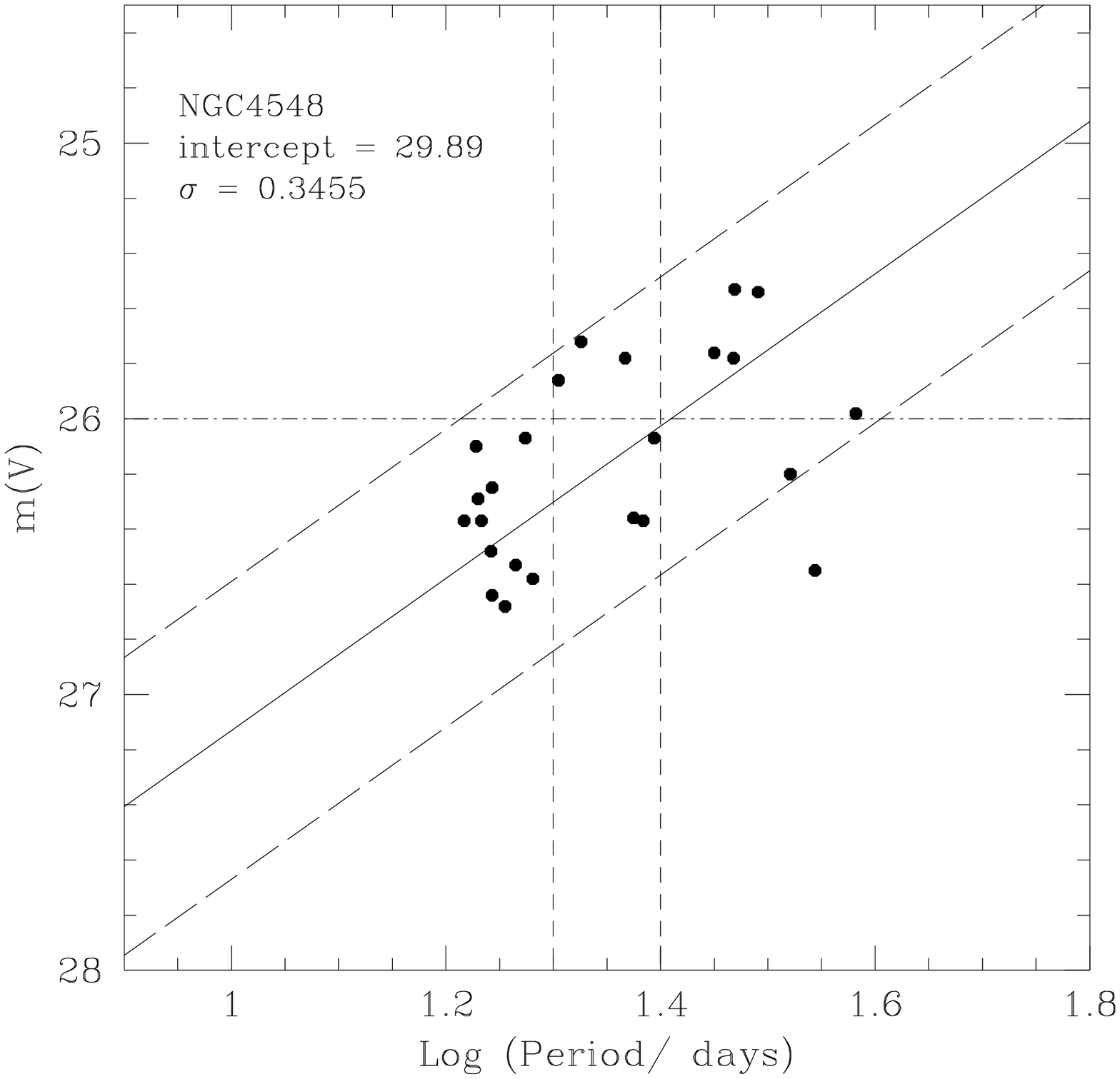}} \\
{\epsfxsize=5.5truecm \epsfysize=5.5truecm \epsfbox[17 144 590 715]{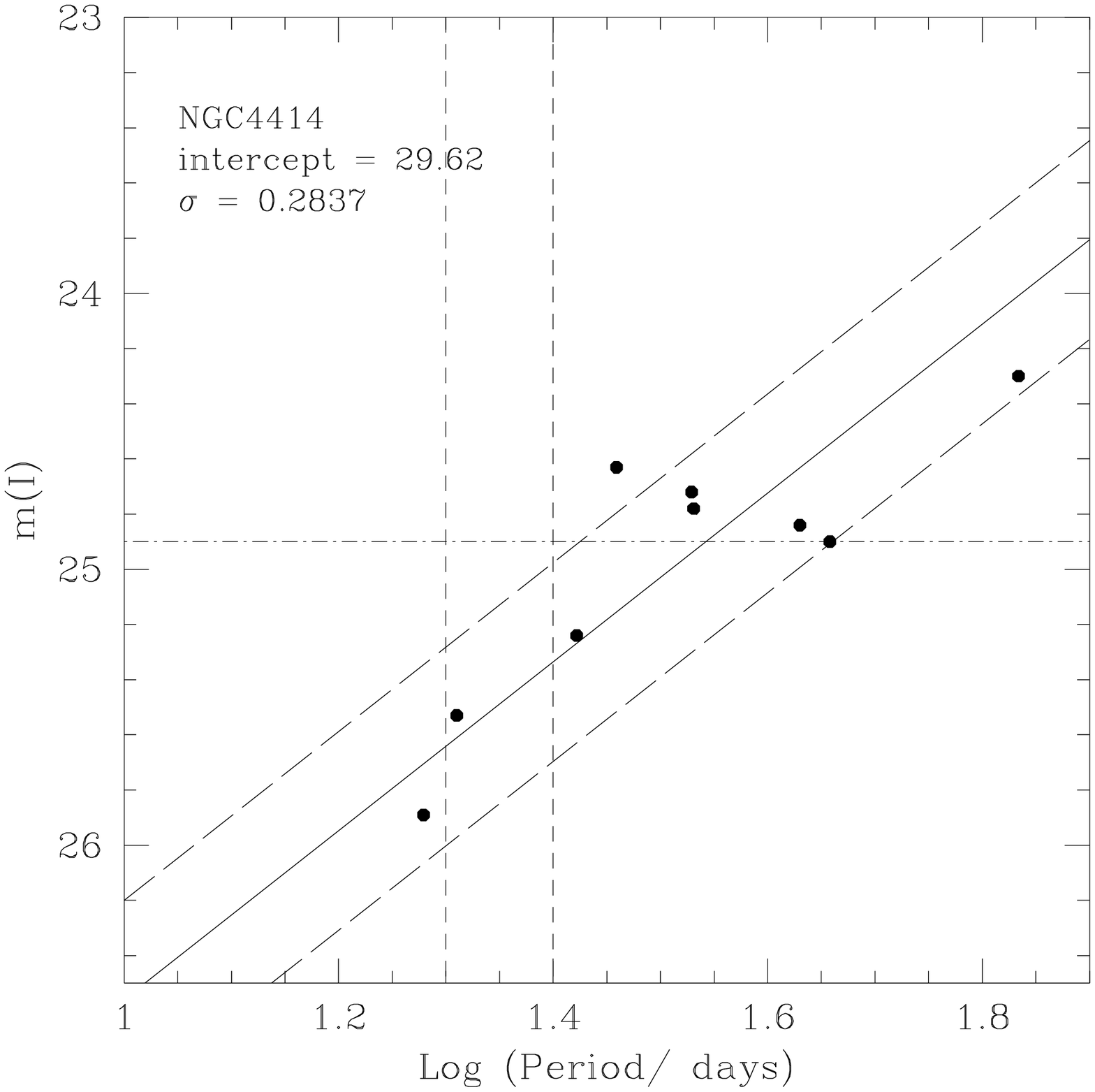}} &
{\epsfxsize=5.5truecm \epsfysize=5.5truecm \epsfbox[17 144 590 715]{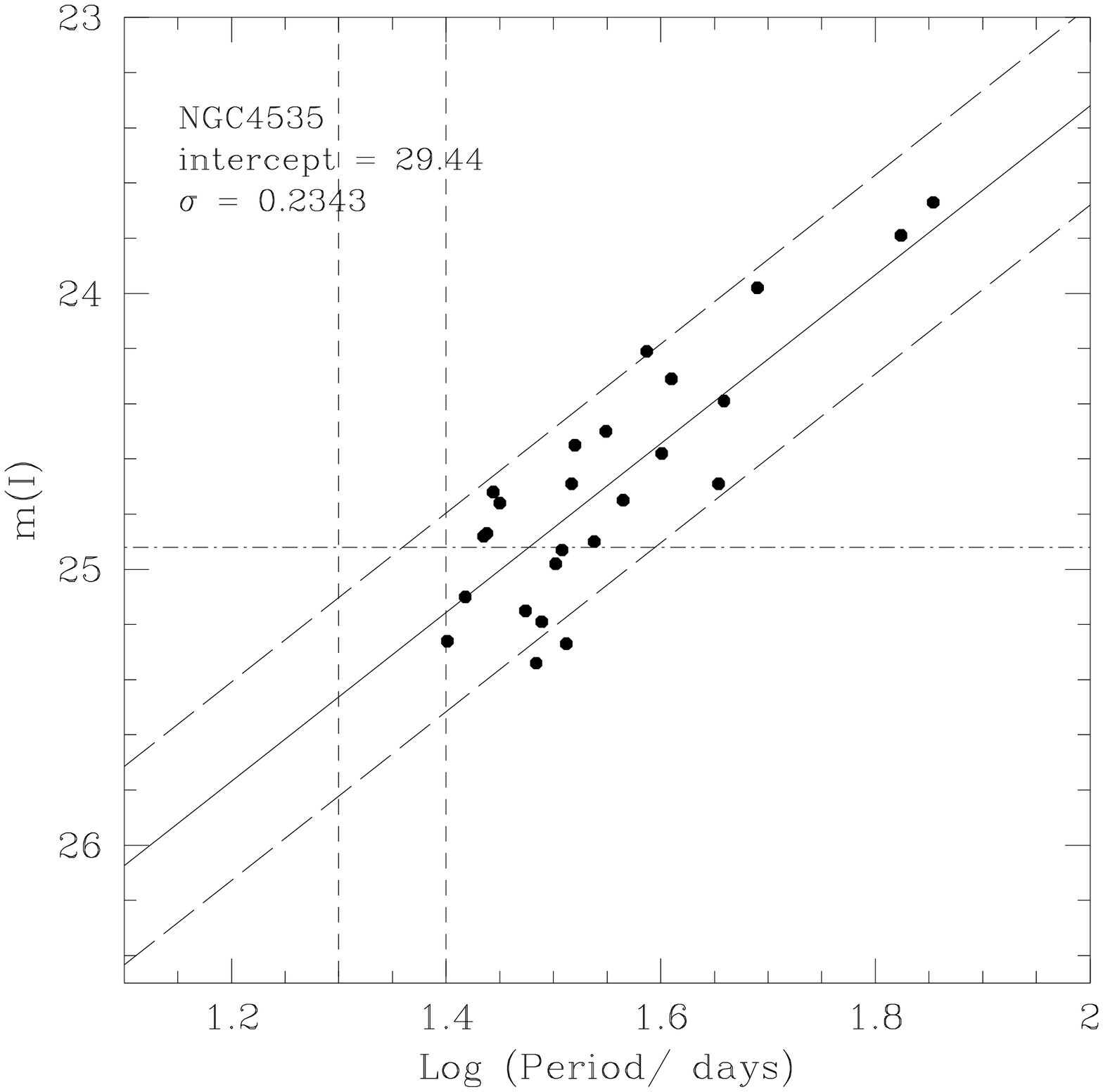}} &
{\epsfxsize=5.5truecm \epsfysize=5.5truecm \epsfbox[17 144 590 715]{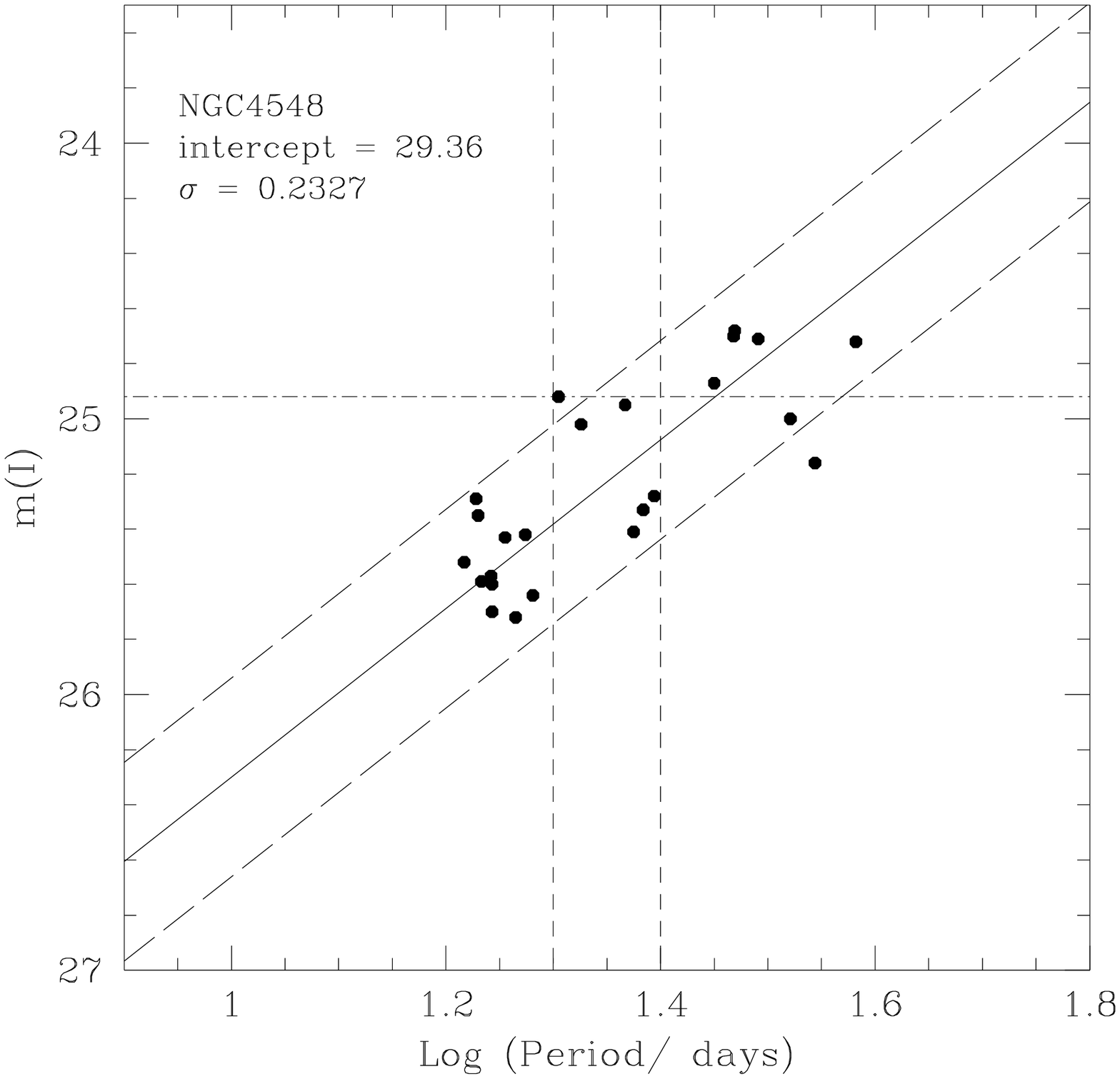}} \\
{\epsfxsize=5.5truecm \epsfysize=5.5truecm \epsfbox[17 144 590 715]{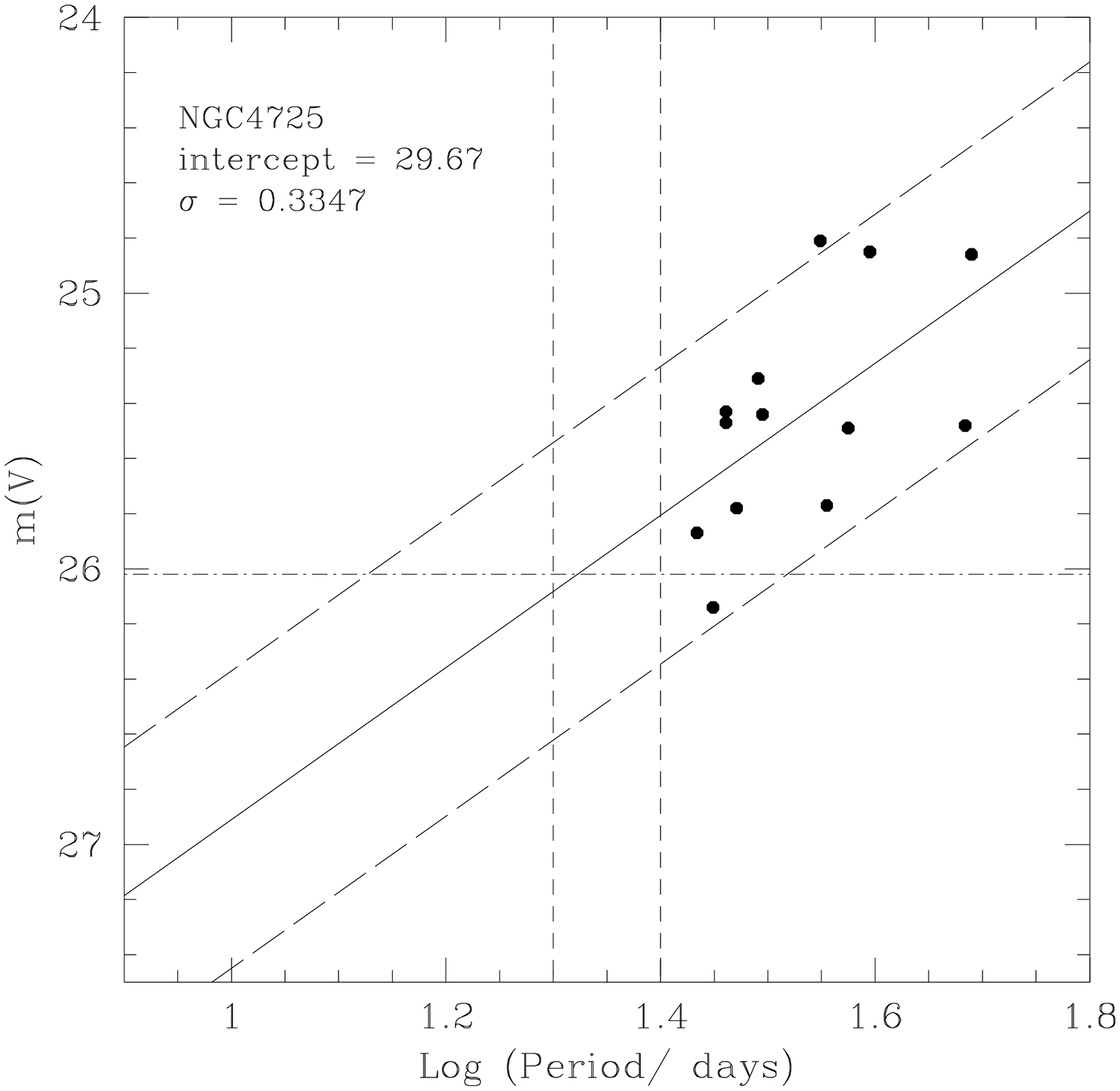}} &
{\epsfxsize=5.5truecm \epsfysize=5.5truecm \epsfbox[17 144 590 715]{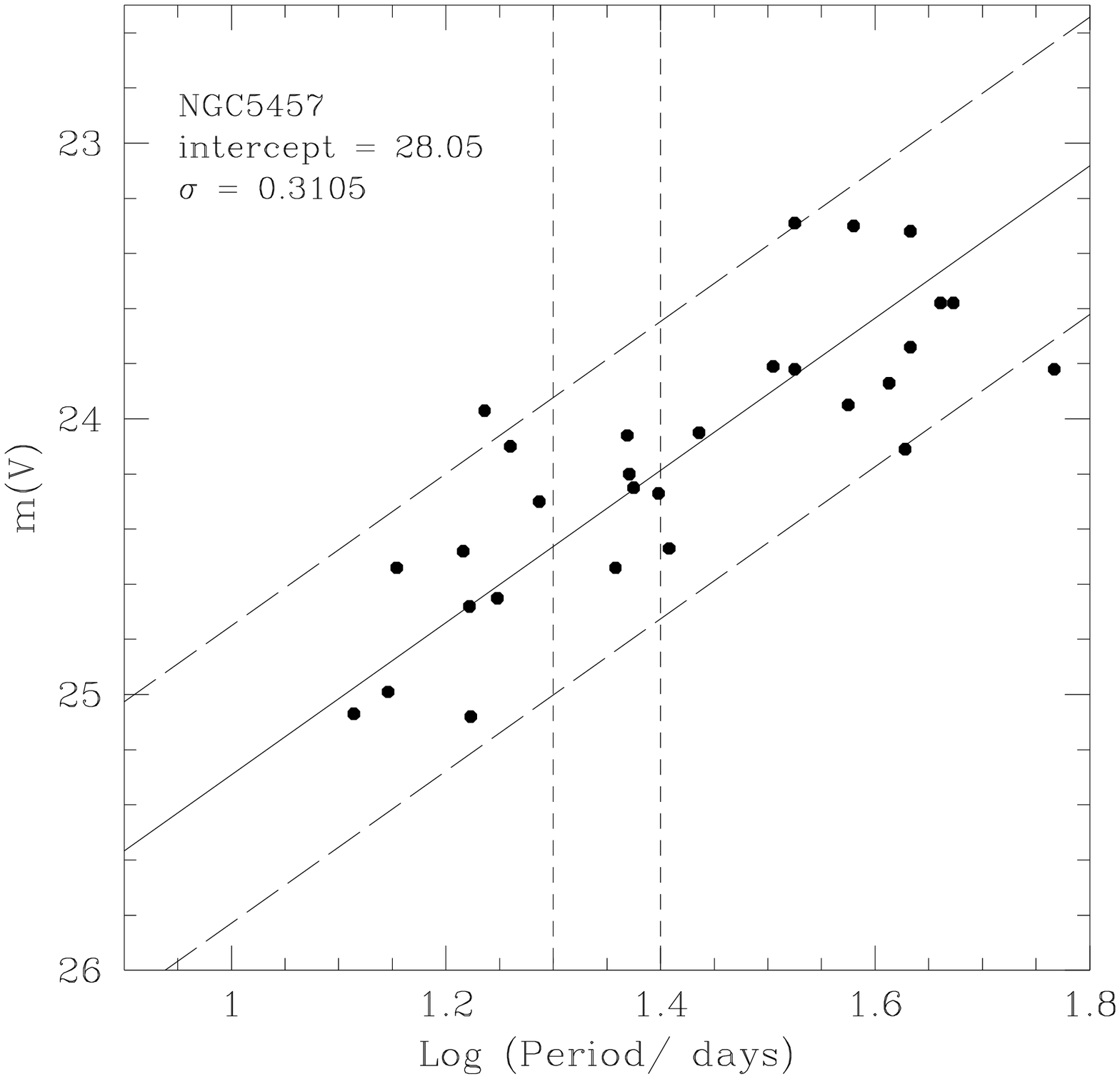}} &
{\epsfxsize=5.5truecm \epsfysize=5.5truecm \epsfbox[17 144 590 715]{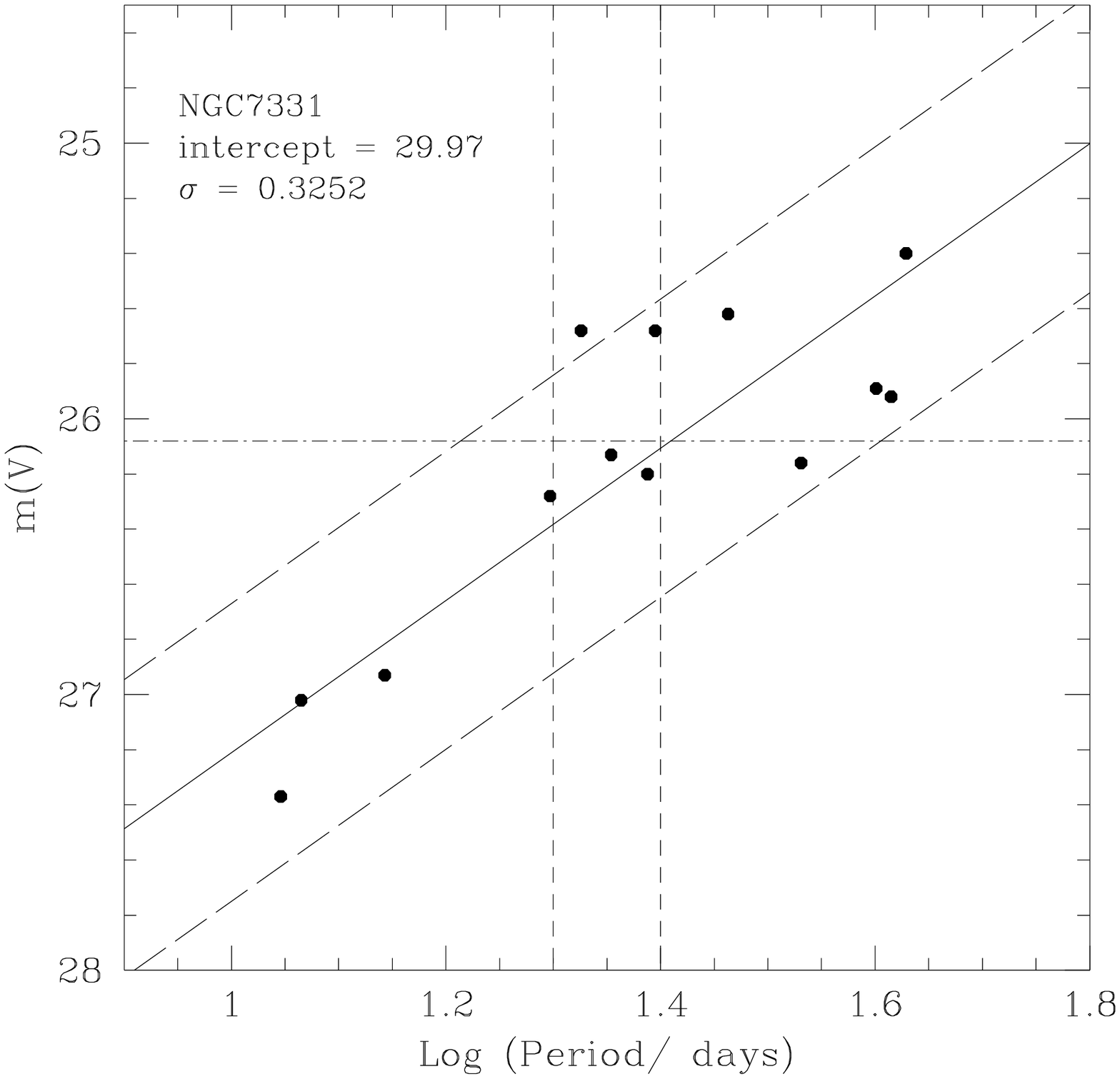}} \\
{\epsfxsize=5.5truecm \epsfysize=5.5truecm \epsfbox[17 144 590 715]{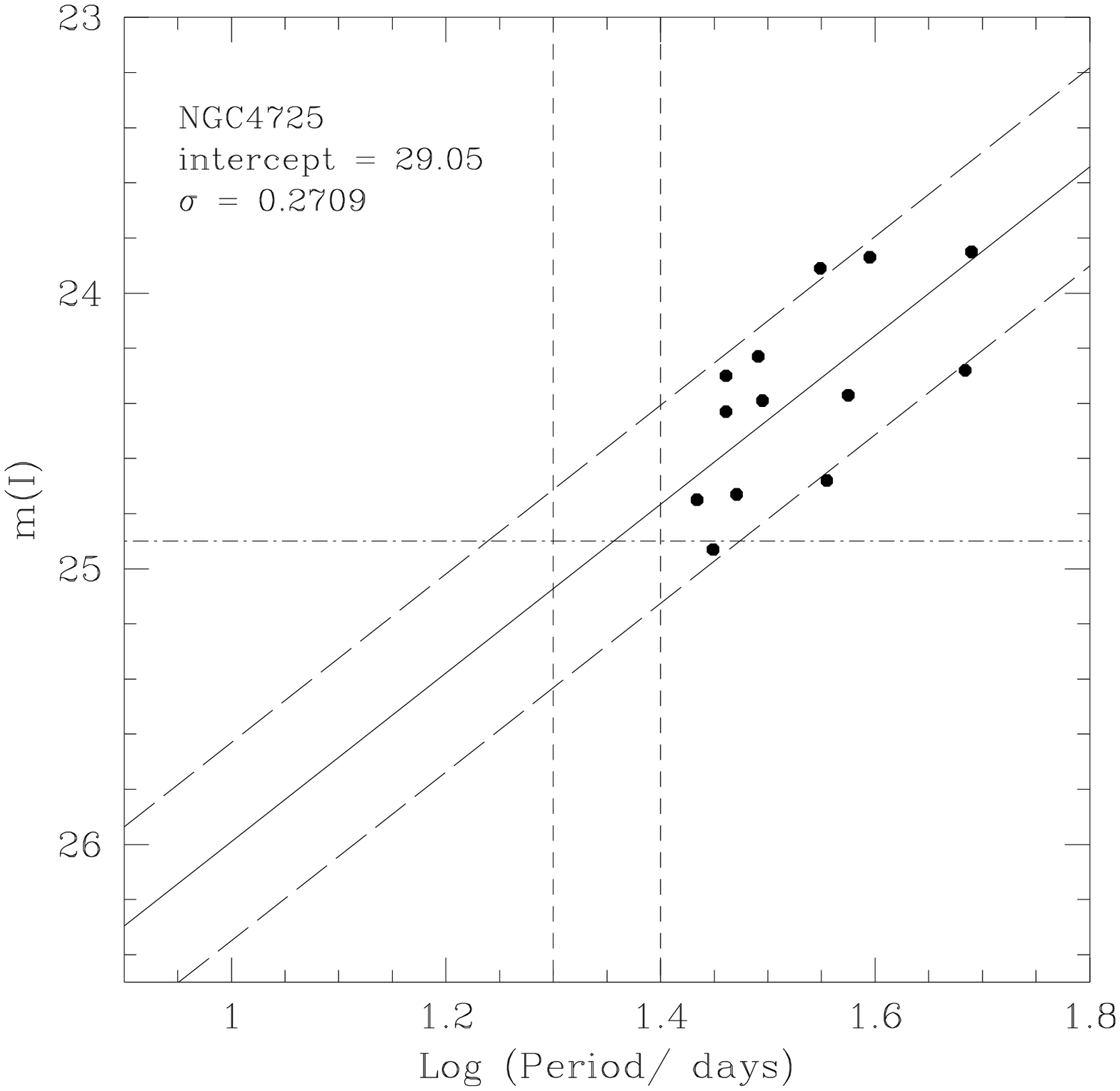}} &
{\epsfxsize=5.5truecm \epsfysize=5.5truecm \epsfbox[17 144 590 715]{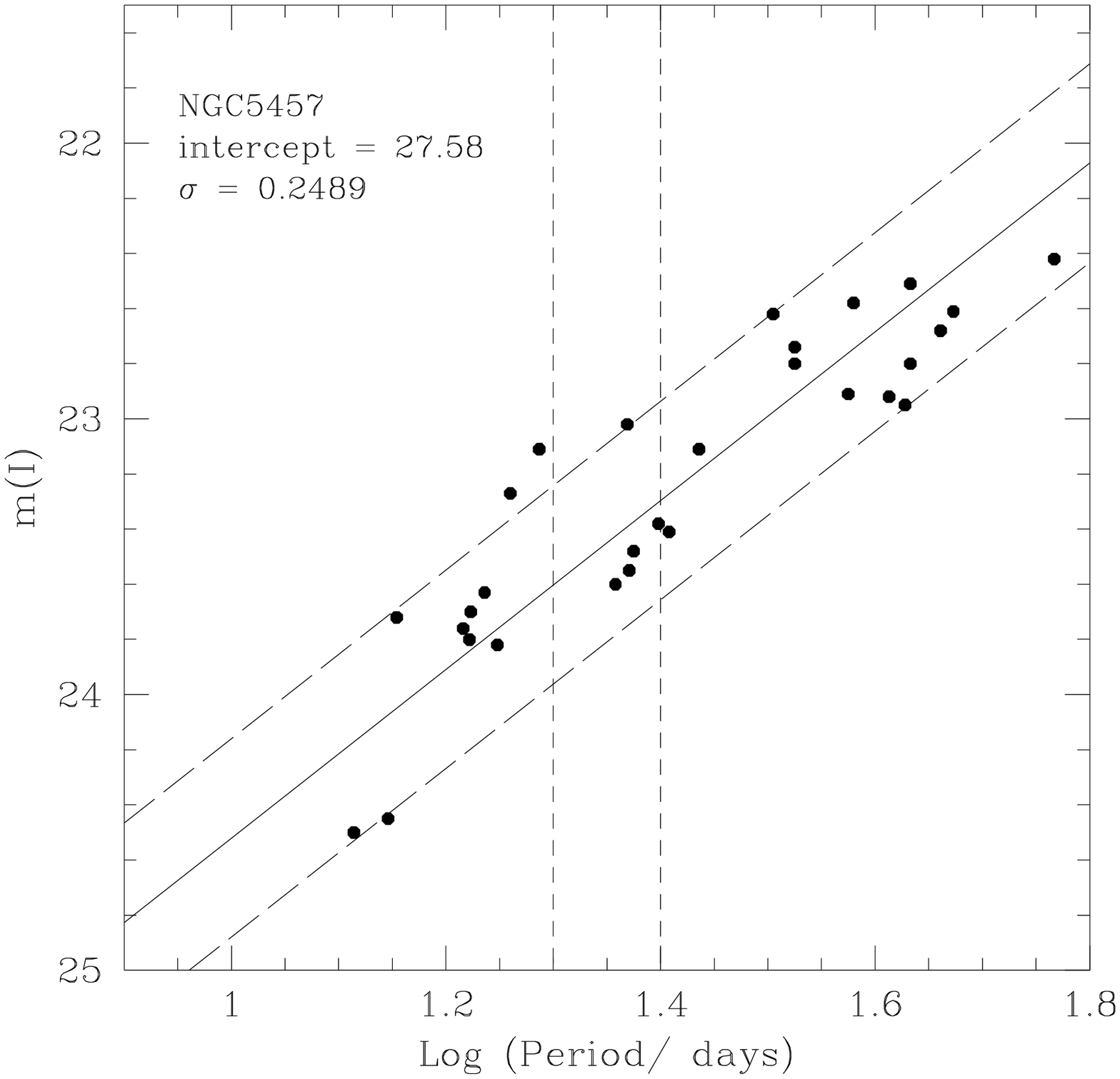}} &
{\epsfxsize=5.5truecm \epsfysize=5.5truecm \epsfbox[17 144 590 715]{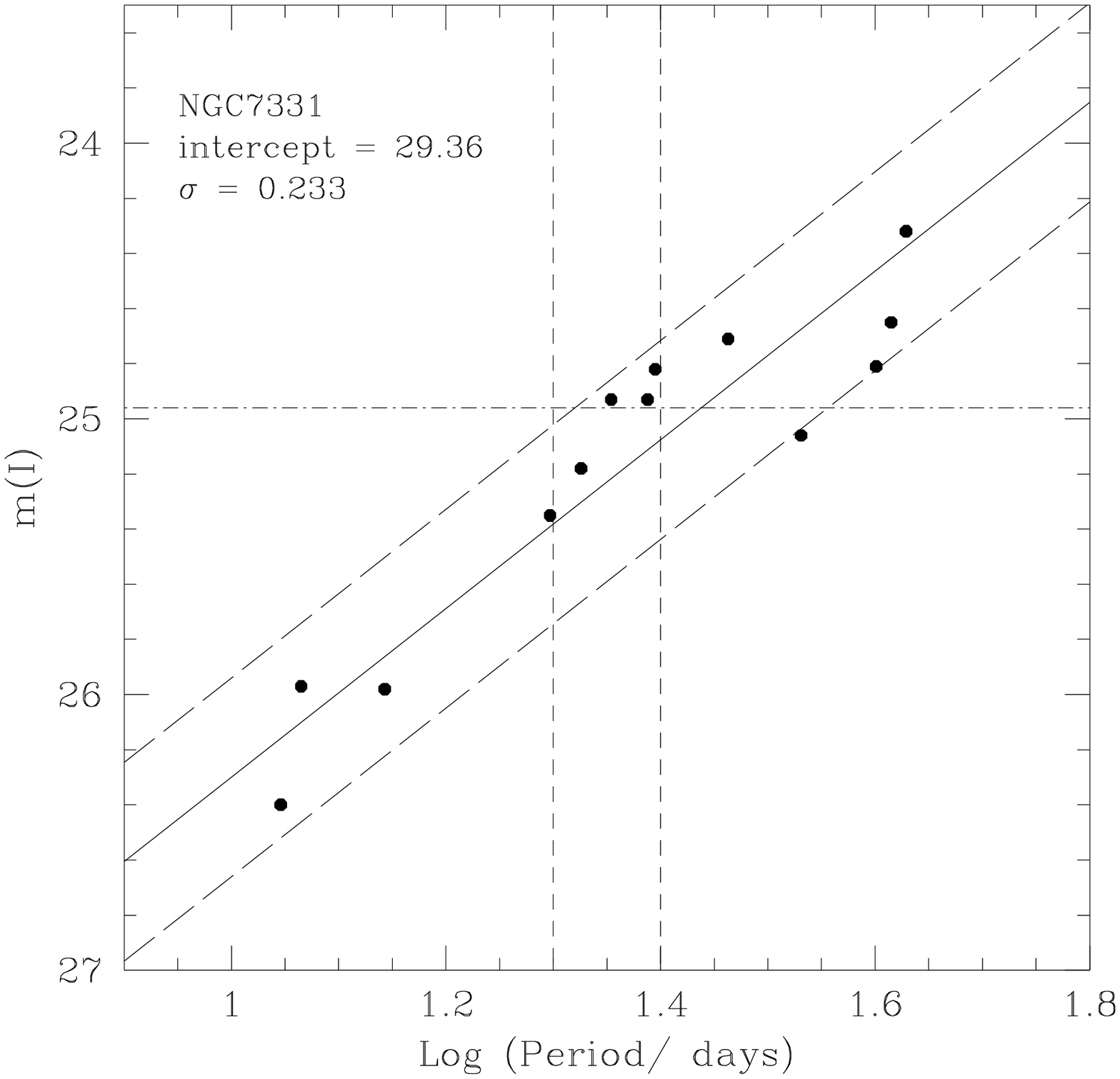}} \\
\end{tabular}
\addtocounter{figure}{-1}
\caption{\emph{continued}}
\label{fig:pl3} 
\end{figure*}

\begin{figure*} 
\begin{tabular}{ccc} 
{\epsfxsize=5.5truecm \epsfysize=5.5truecm \epsfbox[17 144 590 715]{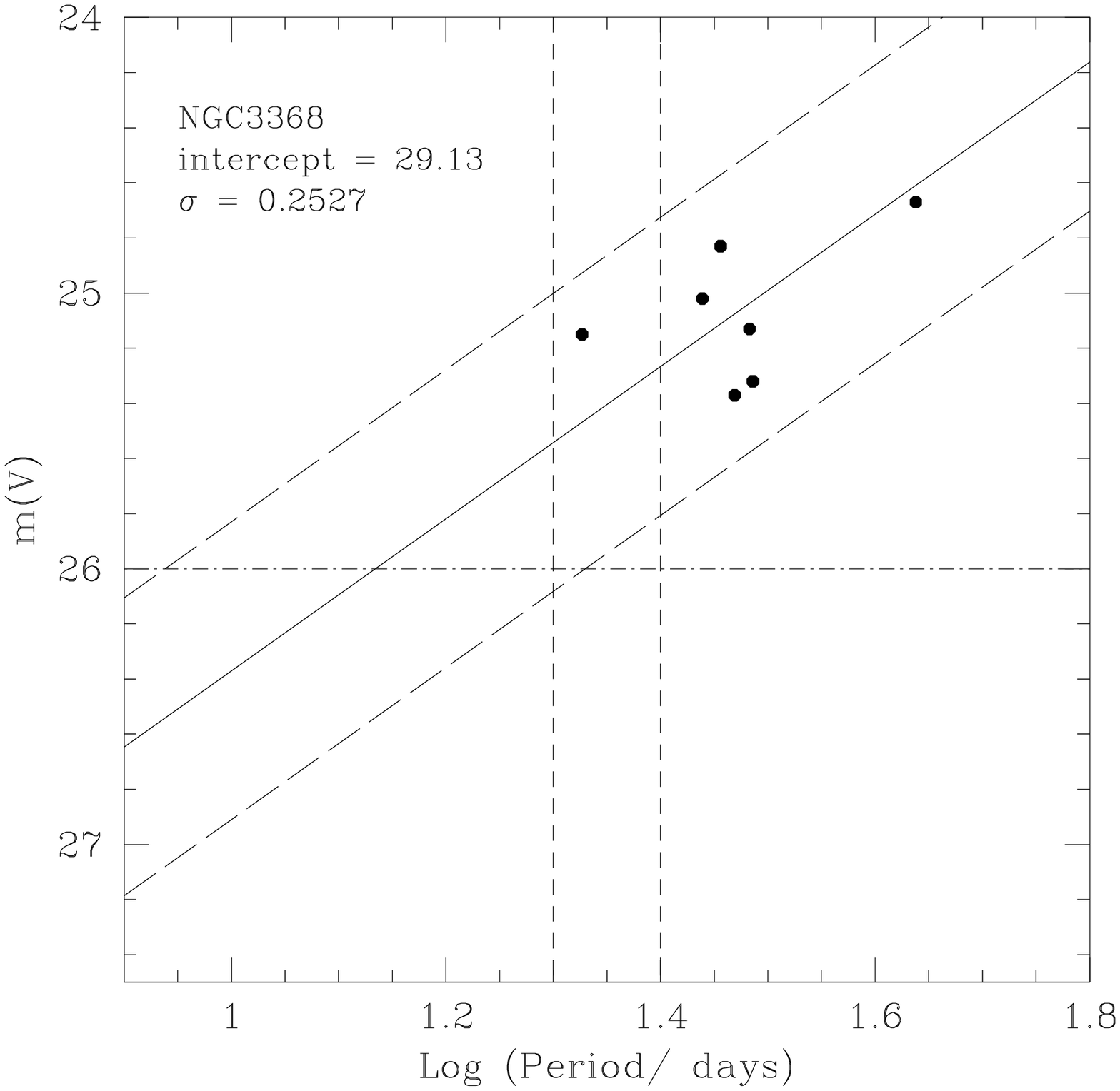}} &
{\epsfxsize=5.5truecm \epsfysize=5.5truecm \epsfbox[17 144 590 715]{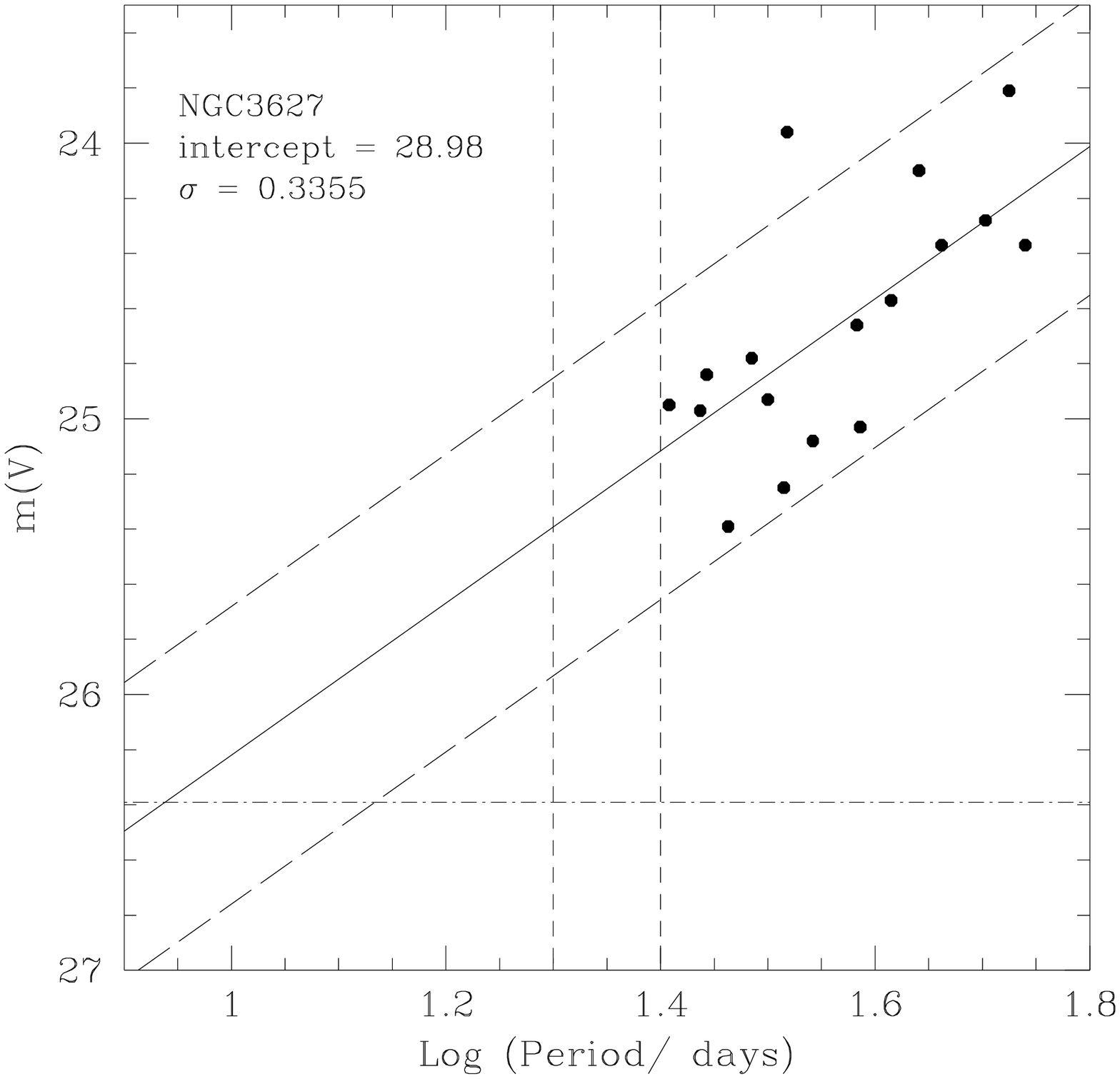}} &
{\epsfxsize=5.5truecm \epsfysize=5.5truecm \epsfbox[17 144 590 715]{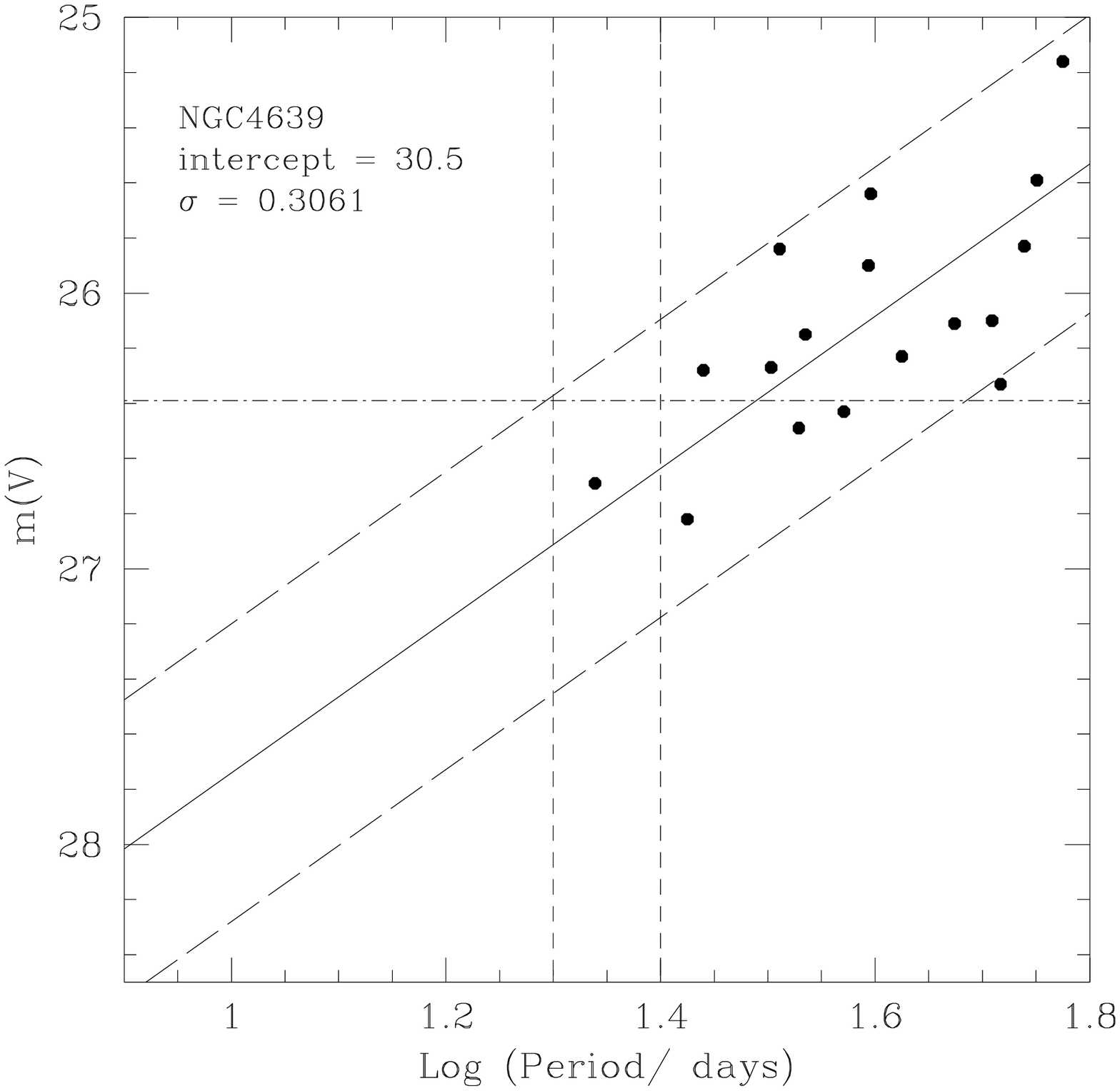}} \\
{\epsfxsize=5.5truecm \epsfysize=5.5truecm \epsfbox[17 144 590 715]{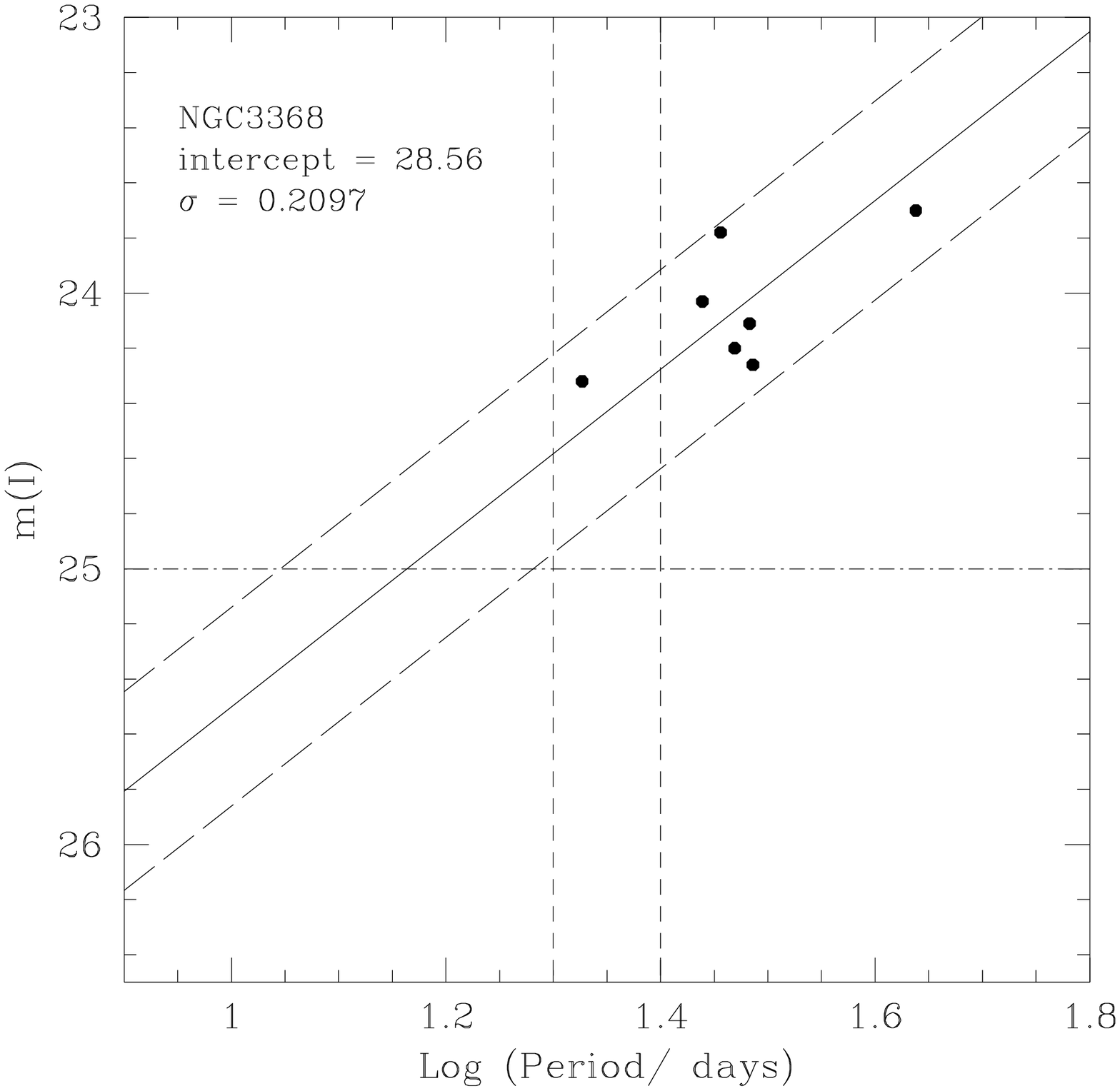}} &
{\epsfxsize=5.5truecm \epsfysize=5.5truecm \epsfbox[17 144 590 715]{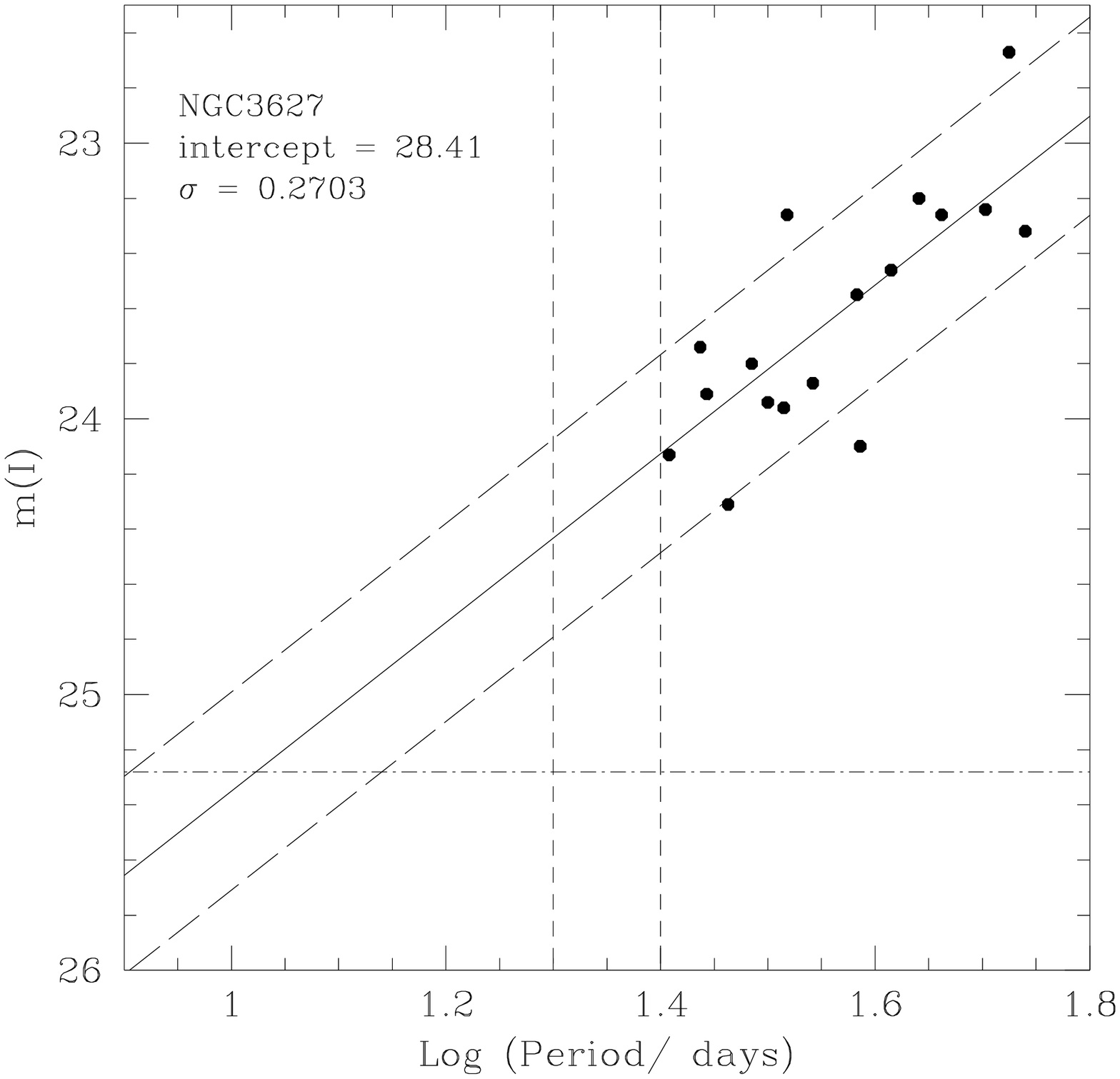}} &
{\epsfxsize=5.5truecm \epsfysize=5.5truecm \epsfbox[17 144 590 715]{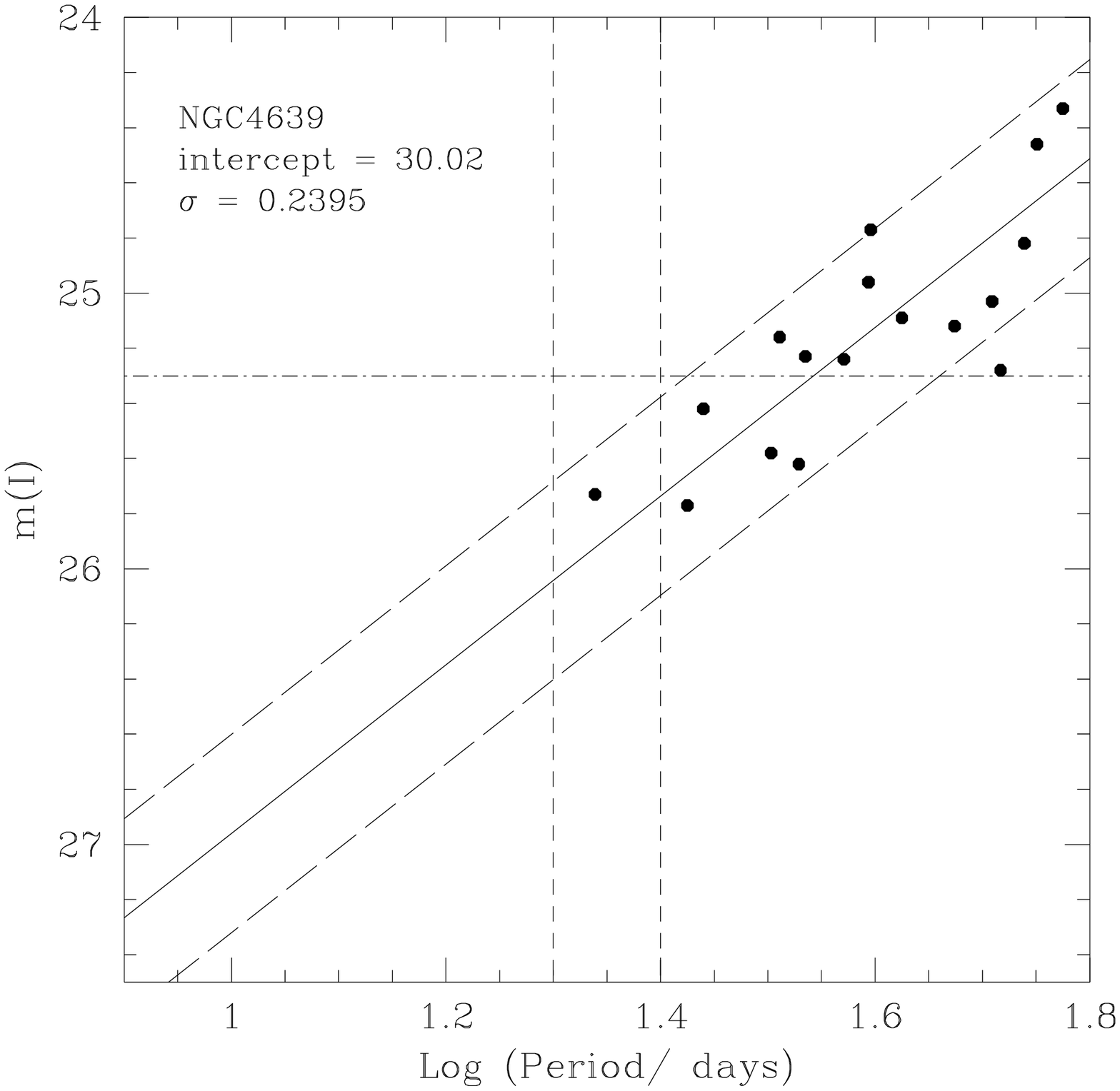}} \\
{\epsfxsize=5.5truecm \epsfysize=5.5truecm \epsfbox[17 144 590 715]{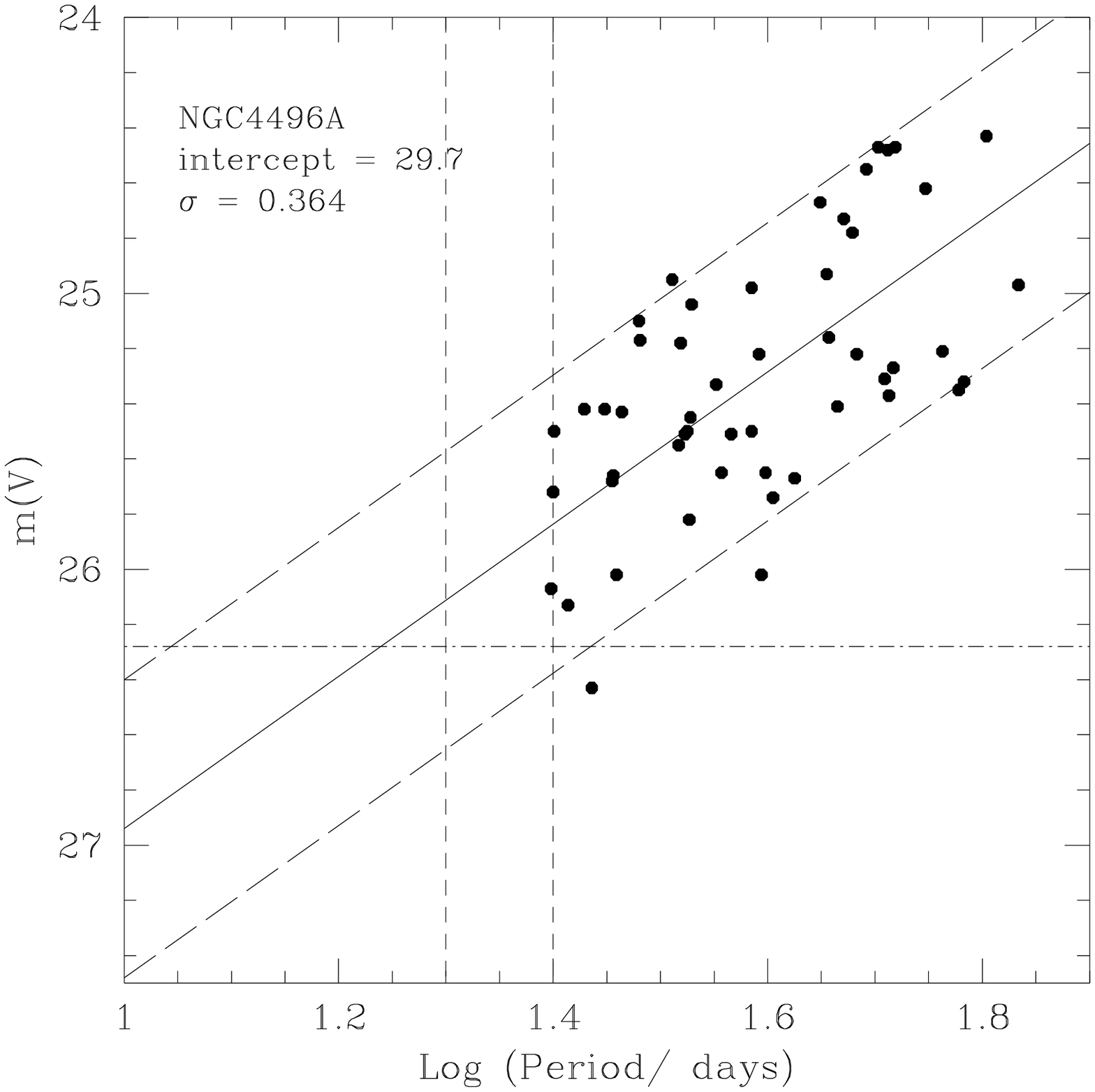}} &
{\epsfxsize=5.5truecm \epsfysize=5.5truecm \epsfbox[17 144 590 715]{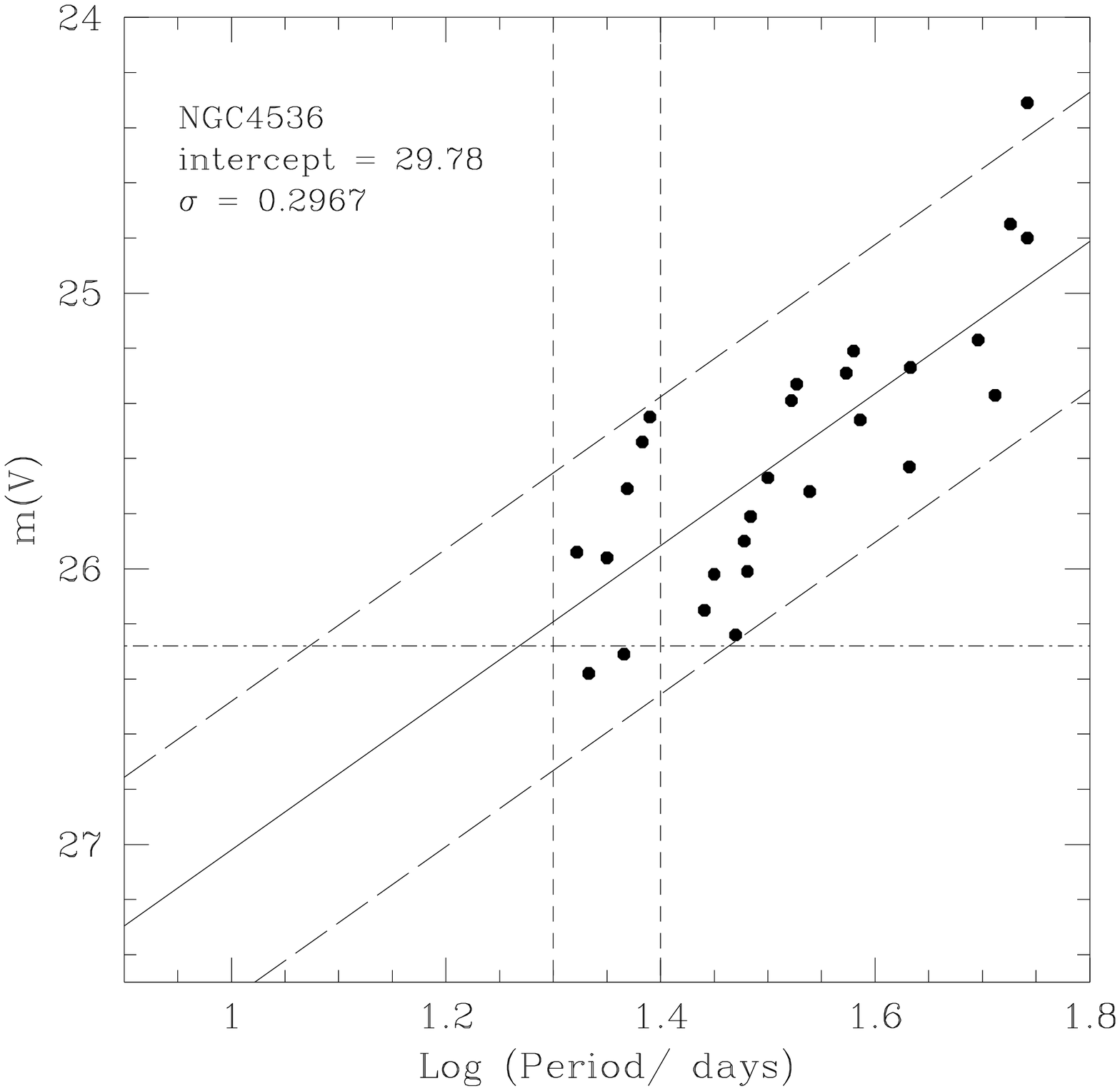}} &
{\epsfxsize=5.5truecm \epsfysize=5.5truecm \epsfbox[17 144 590 715]{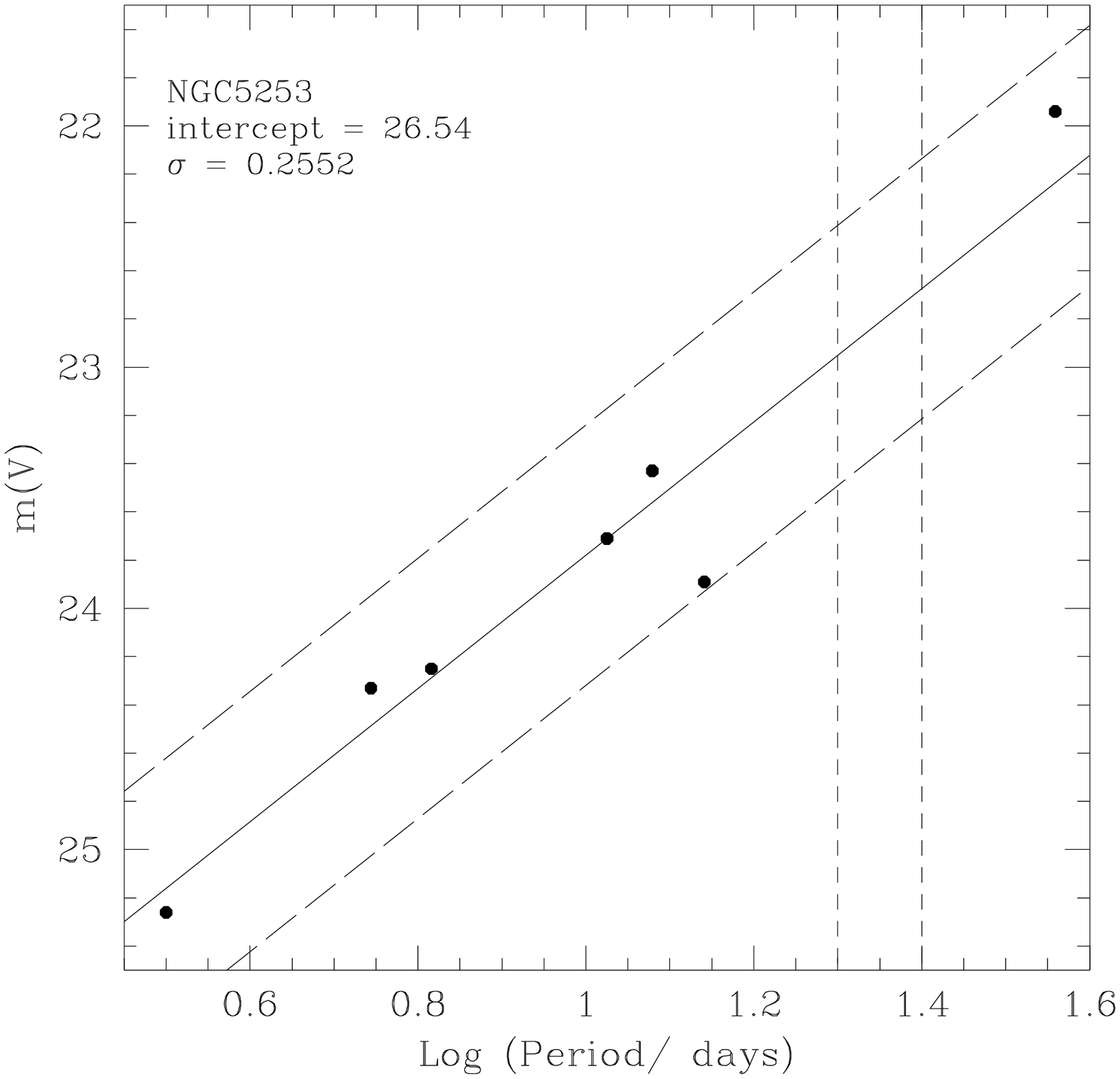}} \\
{\epsfxsize=5.5truecm \epsfysize=5.5truecm \epsfbox[17 144 590 715]{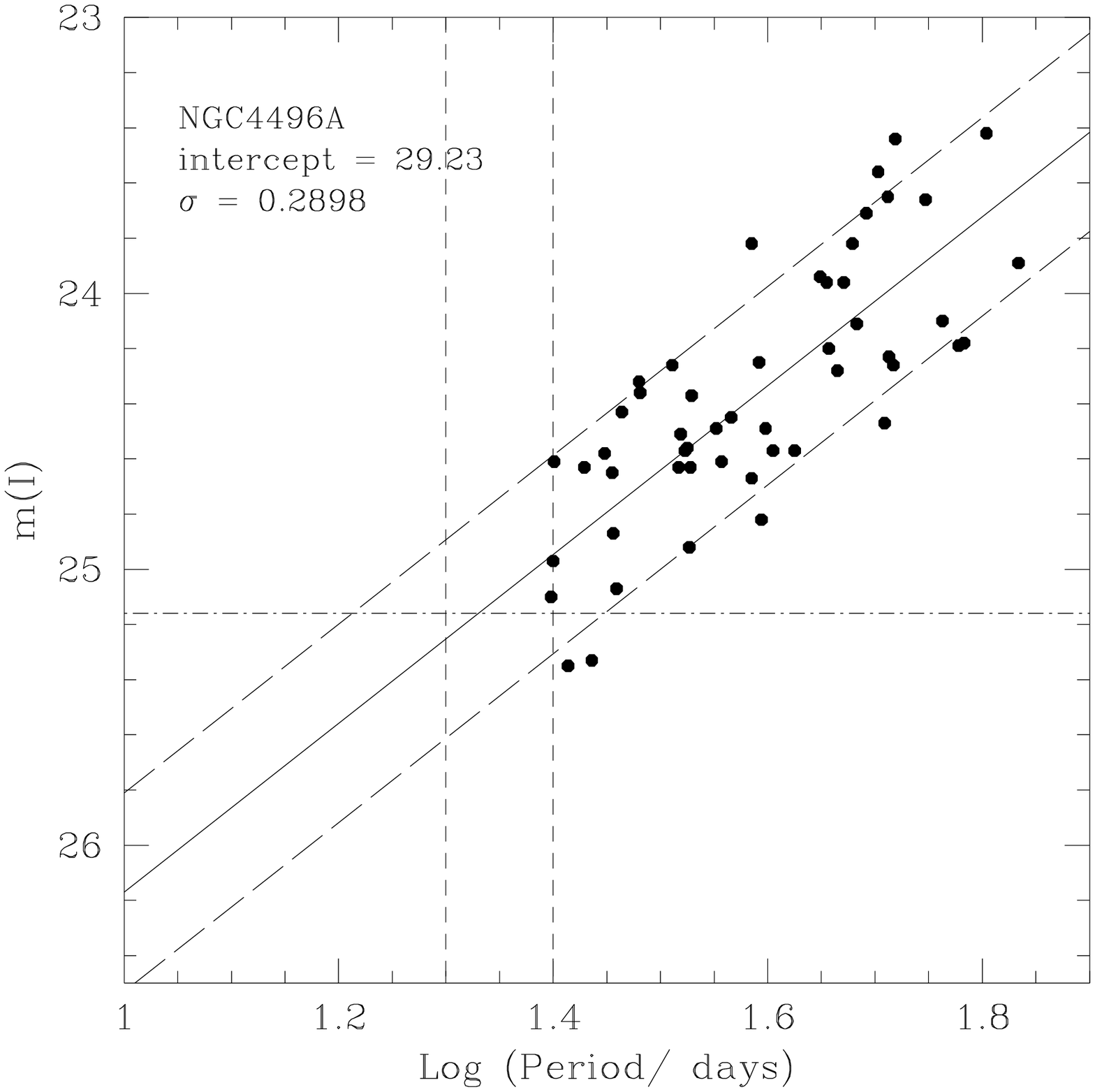}} &
{\epsfxsize=5.5truecm \epsfysize=5.5truecm \epsfbox[17 144 590 715]{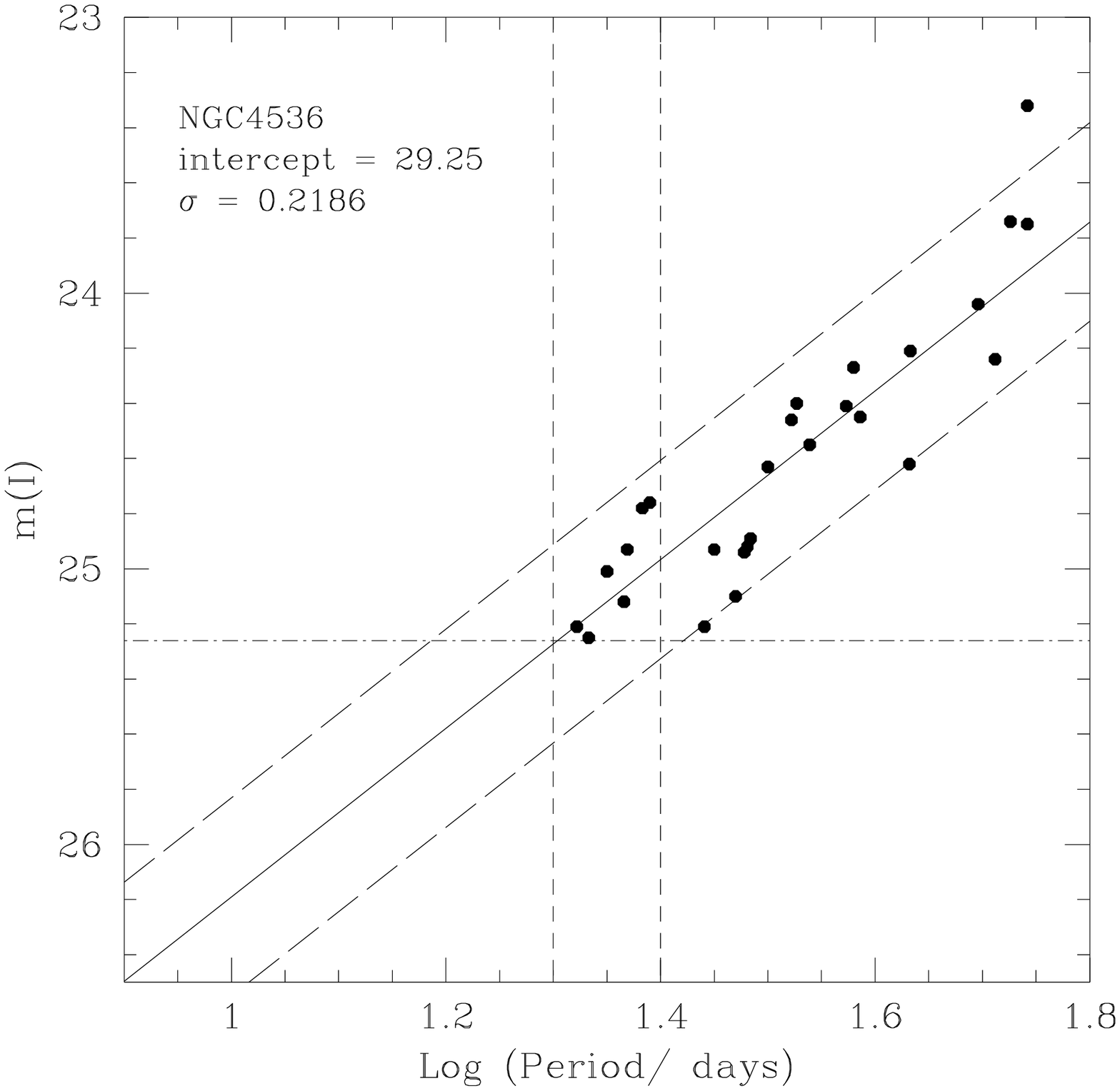}} &
{\epsfxsize=5.5truecm \epsfysize=5.5truecm \epsfbox[17 144 590 715]{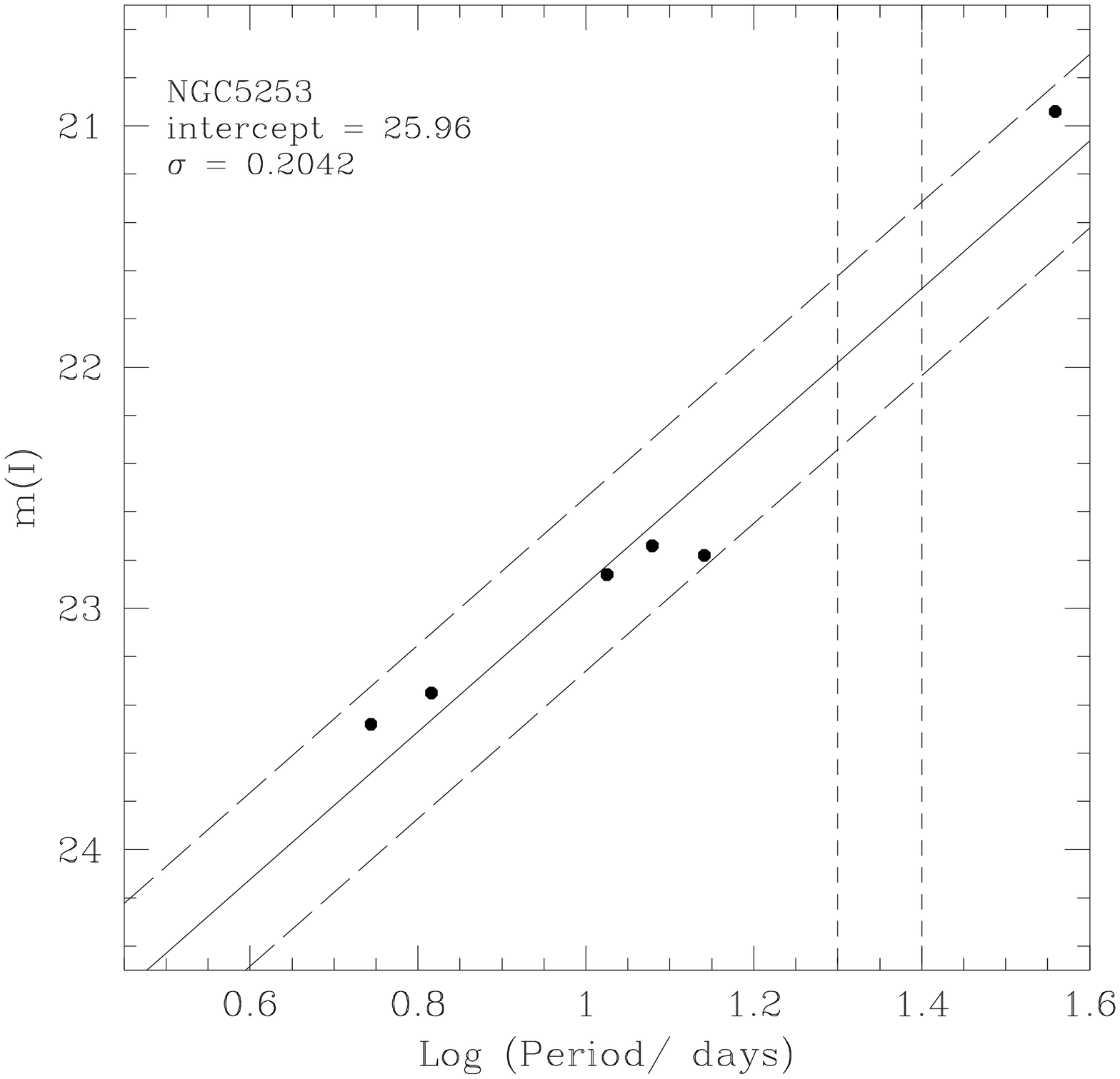}} \\
\end{tabular}
\addtocounter{figure}{-1}
\caption{\emph{continued}}
\label{fig:pl4} 
\end{figure*}

\begin{figure*} 
\begin{tabular}{ccc} 
{\epsfxsize=5.5truecm \epsfysize=5.5truecm \epsfbox[17 144 590 715]{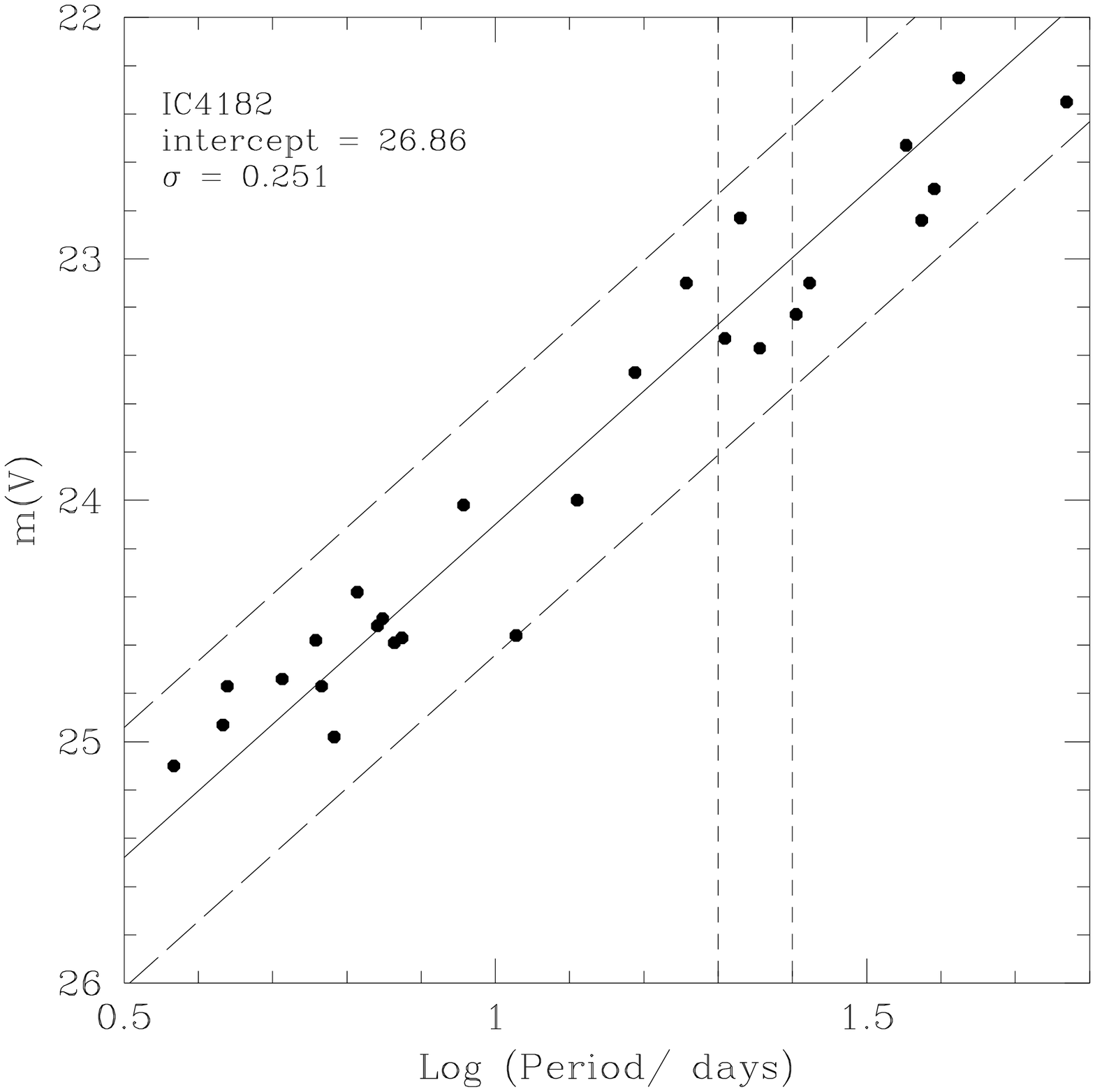}} &
{\epsfxsize=5.5truecm \epsfysize=5.5truecm \epsfbox[17 144 590 715]{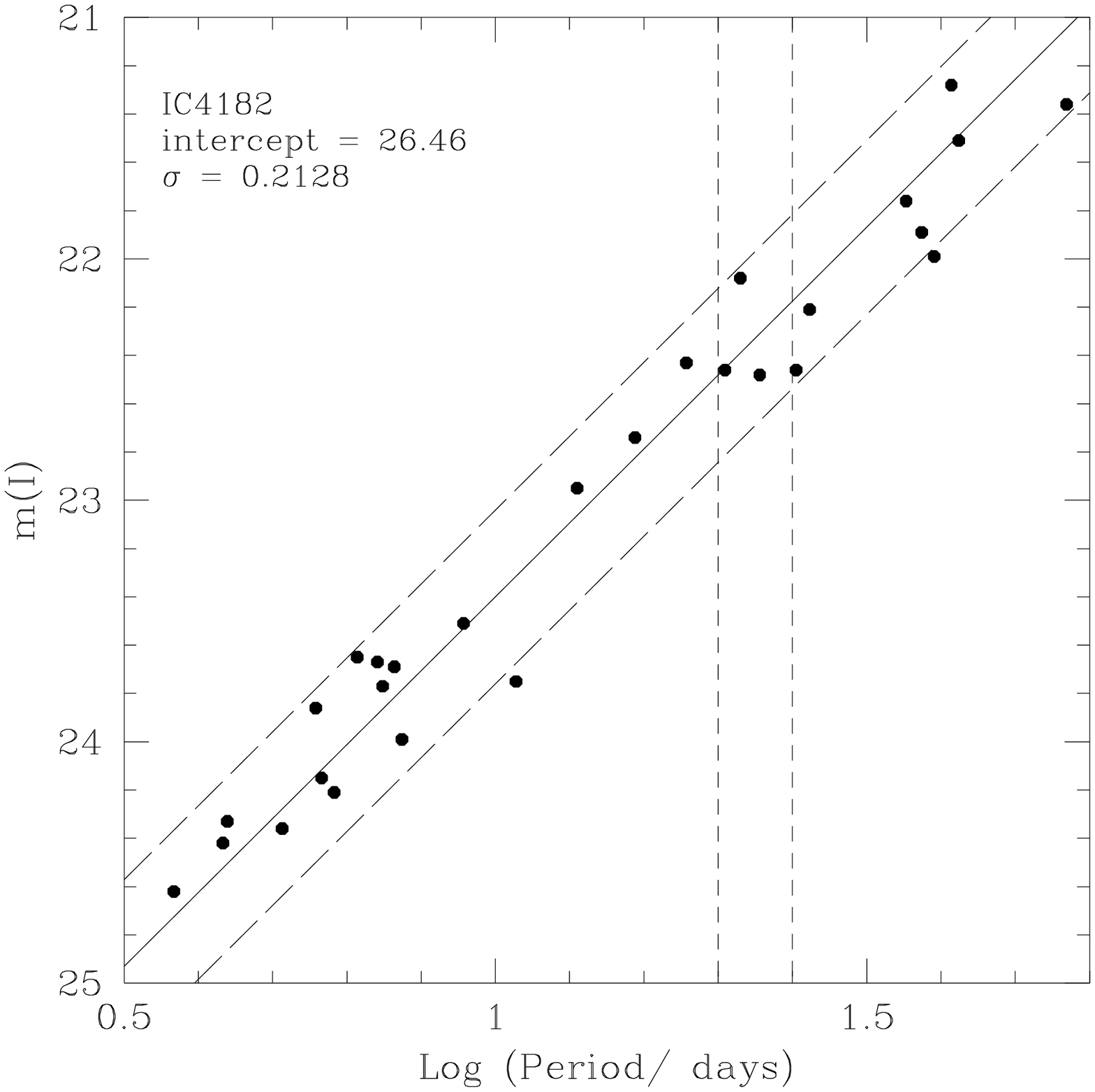}} \\
\end{tabular}
\addtocounter{figure}{-1}
\caption{\emph{continued}}
\label{fig:pl5} 
\end{figure*}

Using these P-L relations, we then calculated distance moduli for the galaxies.
This was done using data as used by the Key Project and ignoring our magnitude
limits. The intercepts found from the best fit to the P-L relations were 
converted to V and I-band distance moduli using the LMC calibration of 
\citeasnoun{mf91}. This was done by adding 1.4 to the V-band intercept and 
1.81 to the I-band intercept (i.e. the galactic zero points from the LMC zero 
point of Madore and Freedman assuming an LMC distance modulus 18.5). The true 
distance modulus was then obtained using the following formula.
\begin{equation}
\mu_{0} = \mu_{AV} - R(\mu_{AV} - \mu_{AI})
\label{eq:distmod}
\end{equation}
where R = 2.45 after \citeasnoun{cardelli}. The V-band modulus is simply
corrected for absorption assuming a galactic reddening law.

\begin{table*}
\begin{tabular}{cccccccc}
\hline
Galaxy & Team & Intercept(V) & Dispersion(V) & Intercept(I) & Dispersion(I) & Distance Modulus & Distance (Mpc)\\ \hline
NGC 925 & H$_{0}$ Key Project & 28.87 & 0.32 & 28.27 & 0.30 & 29.82 $\pm$ 0.08 & 9.29\\
NGC 1326A & H$_{0}$ Key Project & 29.97 & 0.17 & 29.63 & 0.20 & 31.55 $\pm$ 0.07 & 19.32\\
NGC 1365 & H$_{0}$ Key Project & 30.34 & 0.28 & 29.79 & 0.25 & 31.40 $\pm$ 0.10 & 18.28\\
NGC 1425 & H$_{0}$ Key Project & 30.58 & 0.31 & 30.10 & 0.24 & 31.82 $\pm$ 0.06 & 23.01\\
NGC 2090 & H$_{0}$ Key Project & 29.28 & 0.28 & 28.77 & 0.21 & 30.45 $\pm$ 0.08 & 12.30\\
NGC 2541 & H$_{0}$ Key Project & 29.34 & 0.18 & 28.82 & 0.24 & 30.46 $\pm$ 0.08 & 12.42\\
NGC 3031 & H$_{0}$ Key Project & 26.63 & 0.35 & 26.11 & 0.30 & 27.77 $\pm$ 0.08 & 3.63\\
NGC 3198 & H$_{0}$ Key Project & 29.57 & 0.32 & 29.09 & 0.28 & 30.80 $\pm$ 0.06 & 14.45\\
NGC 3319 & H$_{0}$ Key Project & 29.53 & 0.31 & 29.06 & 0.30 & 30.78 $\pm$ 0.10 & 14.32\\
NGC 3351 & H$_{0}$ Key Project & 28.98 & 0.40 & 28.42 & 0.33 & 30.01 $\pm$ 0.08 & 10.05\\
NGC 3621 & H$_{0}$ Key Project & 28.44 & 0.39 & 27.74 & 0.33 & 29.14 $\pm$ 0.11 & 6.70\\
NGC 4321 & H$_{0}$ Key Project & 29.92 & 0.36 & 29.38 & 0.29 & 31.04 $\pm$ 0.09 & 16.14\\
NGC 4414 & H$_{0}$ Key Project & 30.05 & 0.39 & 29.62 & 0.28 & 31.41 $\pm$ 0.10 & 19.14\\
NGC 4535 & H$_{0}$ Key Project & 29.95 & 0.31 & 29.44 & 0.23 & 31.11 $\pm$ 0.07 & 16.60\\
NGC 4548 & H$_{0}$ Key Project & 29.89 & 0.35 & 29.36 & 0.23 & 31.01 $\pm$ 0.08 & 19.14\\
NGC 4725 & H$_{0}$ Key Project & 29.67 & 0.33 & 29.05 & 0.27 & 30.56 $\pm$ 0.08 & 13.00\\
NGC 5457 & H$_{0}$ Key Project & 28.05 & 0.31 & 27.58 & 0.25 & 29.34 $\pm$ 0.10 & 7.38\\
NGC 7331 & H$_{0}$ Key Project & 29.97 & 0.33 & 29.36 & 0.23 & 30.90 $\pm$ 0.10 & 15.07\\
NGC 3368 & Tanvir et al. & 29.13 & 0.25 & 28.56 & 0.21 & 30.20 $\pm$ 0.10 & 11.59\\
NGC 3627 & Sandage et al. & 28.98 & 0.34 & 28.41 & 0.27 & 30.06 $\pm$ 0.17 & 11.07\\
NGC 4639 & Sandage et al. & 30.50 & 0.31 & 30.02 & 0.24 & 31.80 $\pm$ 0.09 & 25.47\\
NGC 4496A & Sandage et al. & 29.70 & 0.36 & 29.23 & 0.29 & 31.02 $\pm$ 0.07 & 16.07\\
NGC 4536 & Sandage et al. & 29.78 & 0.30 & 29.25 & 0.22 & 30.95 $\pm$ 0.07 & 16.60\\
NGC 5253 & Sandage et al. & 26.54 & 0.26 & 25.96 & 0.20 & 27.61 $\pm$ 0.11 & 4.13\\
IC 4182 & Sandage et al. & 26.86 & 0.25 & 26.46 & 0.21 & 28.36 $\pm$ 0.08 & 4.70\\
\hline
\end{tabular}
\caption{Data obtained from P-L fitting. Intercept and dispersion in each waveband, and the calculated distance modulus and distance}
\label{tab:cephdist}
\end{table*}

Table \ref{tab:cephdist} shows the parameters obtained from the P-L fits for 
each galaxy, (intercept and dispersion in the two wavebands, along with the 
calculated distance moduli and the distance in Mpc). The errors quoted on the 
distance moduli are from \citeasnoun{fer}. The error has been split into 
random and systematic components. The random components are photometric 
errors, errors on the P-L fits and errors on the reddening correction. The 
error quoted is the random error. The systematic errors are the errors in the 
LMC calibration, the distance to the LMC, the photometry zero-point and 
metallicity differences. The systematic error would add a further $\pm$ 0.16 
(in quadrature) to the error quoted in the Table. More details can be found in
the relevant papers (see Section \ref{sec:data} for references).

Other parameters relevant to the galaxies are contained in Table
\ref{tab:cephmag}. This includes the metallicity of the galaxy from HII regions
in the region of the Cepheids used in the P-L relations (taken from Table 2 of
\citeasnoun{fer} which is based on the references therein), and the mean
reddening (internal + galactic foreground), calculated from the Cepheids
themselves. It is noted that \citeasnoun{fer} give the metallicity of NGC 3368 as
$12+{\rm log} [O/H]=9.2\pm0.20$ based upon \citeasnoun{oey}. However
\citeasnoun{tanvir99} (and this paper) obtain 8.9$\pm$0.1 based upon
\citeasnoun{oey}. The value of 8.9 is therefore used in the rest of this paper.
Also shown are the exposure times (from the references in Section \ref{sec:data})
and magnitude limits calculated from these. The final column gives the number of
Cepheids used in fitting the P-L relation.
 
\subsubsection*{Comparison with Published Distance Moduli}

With one exception, the distance moduli calculated here (see Table 
\ref{tab:cephdist}) all agree with those published by the Key Project team 
(see Section \ref{sec:data} for references), to within 0.02 mag, and most 
agree exactly. The small discrepancies are thought to be due to simple rounding 
errors. The intercepts from the P-L relations usually agree and in the cases 
where they do not there is only a small change in distance modulus. The only 
exception was NGC 1326A. Here a distance modulus of 31.55 was obtained 
compared with 31.43 from \citeasnoun{fer}. This is one of the galaxies that 
has had data cut at short period, and if all Cepheids are included a distance 
modulus of 31.36 is obtained, in agreement with \citeasnoun{prosser}. After 
checking the right cuts have been made, there seems to be no explanation for 
the discrepancy between the value of 31.55 obtained here and the 31.43 
obtained by \citeasnoun{fer}. The value obtained here is used in the rest of 
this work.

So, using the same methods we confirm the Key Project results and in addition 
obtain the dispersion about the P-L relation (see Table \ref{tab:cephdist}).

\begin{table*}
\begin{tabular}{cccccccc}
\hline
Galaxy & Reddening & Metallicity & Exposure (V) & Exposure (I) & m(V) & m(I) & Number of Cepheids \\ 
 & V-I & 12 + Log(O/H) & s & s & limit & limit & \\ \hline
NGC 925 & 0.18 $\pm$ 0.03 & 8.55 $\pm$ 0.15 & 2200 & 2200 & 25.95 & 24.83 & 75 \\
NGC 1326A & 0.00 $\pm$ 0.01 & 8.50 $\pm$ 0.15 & 3400 & 3600 & 26.19 & 25.10 & 8\\
NGC 1365 & 0.20 $\pm$ 0.04 & 8.96 $\pm$ 0.20 & 5000 & 5000 & 26.40 & 25.28 & 26\\
NGC 1425 & 0.09 $\pm$ 0.04 & 9.00 $\pm$ 0.15 & 3900 & 3900 & 26.26 & 25.14 & 20\\
NGC 2090 & 0.09 $\pm$ 0.01 & 8.80 $\pm$ 0.15 & 2200 & 2200 & 25.95 & 24.83 & 30\\
NGC 2541 & 0.11 $\pm$ 0.07 & 8.50 $\pm$ 0.15 & 2200 & 2600 & 25.95 & 24.92 & 27\\
NGC 3031 & 0.04 & 8.75 $\pm$ 0.15 & 1200 & 1800 & 25.62 & 24.72 & 31 \\
NGC 3198 & 0.07 $\pm$ 0.04 & 8.60 $\pm$ 0.15 & 2200 & 2400 & 25.95 & 24.88 & 52\\
NGC 3319 & 0.05 $\pm$ 0.04 & 8.38 $\pm$ 0.15 & 2200 & 2600 & 25.95 & 24.92 & 28\\
NGC 3351 & 0.15 & 9.24 $\pm$ 0.20 & 2500 & 2400 & 26.02 & 24.88 & 45 \\
NGC 3621 & 0.30 $\pm$ 0.03 & 8.75 $\pm$ 0.15 & 1800 & 1800 & 25.84 & 24.72 & 36\\
NGC 4321 & 0.18 $\pm$ 0.11 & 9.13 $\pm$ 0.20 & 3600 & 3600 & 26.22 & 25.10 & 43\\
NGC 4414 & 0.01 $\pm$ 0.05 & 9.20 $\pm$ 0.15 & 2500 & 2500 & 26.02 & 24.90 & 9\\
NGC 4535 & 0.13 $\pm$ 0.04 & 9.20 $\pm$ 0.15 & 2400 & 2600 & 26.00 & 24.92 & 25\\
NGC 4548 & 0.12 $\pm$ 0.03 & 9.34 $\pm$ 0.15 & 2400 & 2600 & 26.00 & 24.92 & 24\\
NGC 4725 & 0.26 $\pm$ 0.04 & 8.92 $\pm$ 0.15 & 2500 & 2500 & 26.02 & 24.90 & 13\\
NGC 5457 & 0.18 & 9.05 $\pm$ 0.15 & 4200 & 4200 & 26.30 & 25.18 & 29\\
NGC 7331 & 0.19 $\pm$ 0.07 & 8.67 $\pm$ 0.15 & 2800 & 2800 & 26.08 & 24.96 & 13\\
NGC 3368 & 0.19 $\pm$ 0.05 & 8.90 $\pm$ 0.10 & 2400 & 3000 & 26.00 & 25.00 & 7\\
NGC 3627 & 0.19 $\pm$ 0.05 & 9.25 $\pm$ 0.20 & 4900 & 5000 & 26.39 & 25.28 & 17\\
NGC 4639 & 0.07 $\pm$ 0.04 & 9.00 $\pm$ 0.20 & 4900 & 5200 & 26.39 & 25.30 & 17\\
NGC 4496A & 0.05 $\pm$ 0.03 & 8.77 $\pm$ 0.20 & 4000 & 4000 & 26.28 & 25.16 & 51\\
NGC 4536 & 0.13 $\pm$ 0.03 & 8.85 $\pm$ 0.20 & 4000 & 4000 & 26.28 & 25.16 & 27\\
NGC 5253 & 0.19 $\pm$ 0.07 & 8.15 $\pm$ 0.15 & 3600 & 3600 & 26.22 & 25.10 & 7\\
IC 4182 & -0.04 $\pm$ 0.04 & 8.40 $\pm$ 0.20 & 4200 & 4200 & 26.30 & 25.18 & 28\\ \hline
\end{tabular}
\caption{Some Cepheid galaxy parameters. Total mean reddenings for Cepheids in
each galaxy, metallicity of region where Cepheids are found (from
\protect\citeasnoun{fer}), exposure times (from papers listed in section
\ref{sec:data}) along with calculated magnitude limits and the number of Cepheids
used in fitting the P-L relation.}
\label{tab:cephmag}
\end{table*}

\section{Relationship between Metallicity and Dispersion}
\label{sec:metsig}

One of the aims of this paper is to try and detect any signature of the effect
of metallicity on Cepheids within a galaxy. In the first instance, our aim is
to plot   P-L  dispersion  against metallicity. The motivation is that
high metallicity galaxies  may have a wider intrinsic metallicity 
distribution  than  low metallicity galaxies and that this may result
in a broader P-L relation for high metallicity galaxies.

\subsection{V-Band Data}
\label{sec:vband}

\begin{figure}
{\epsfxsize=8.5truecm \epsfysize=8.5truecm
\epsfbox[17 143 565 695]{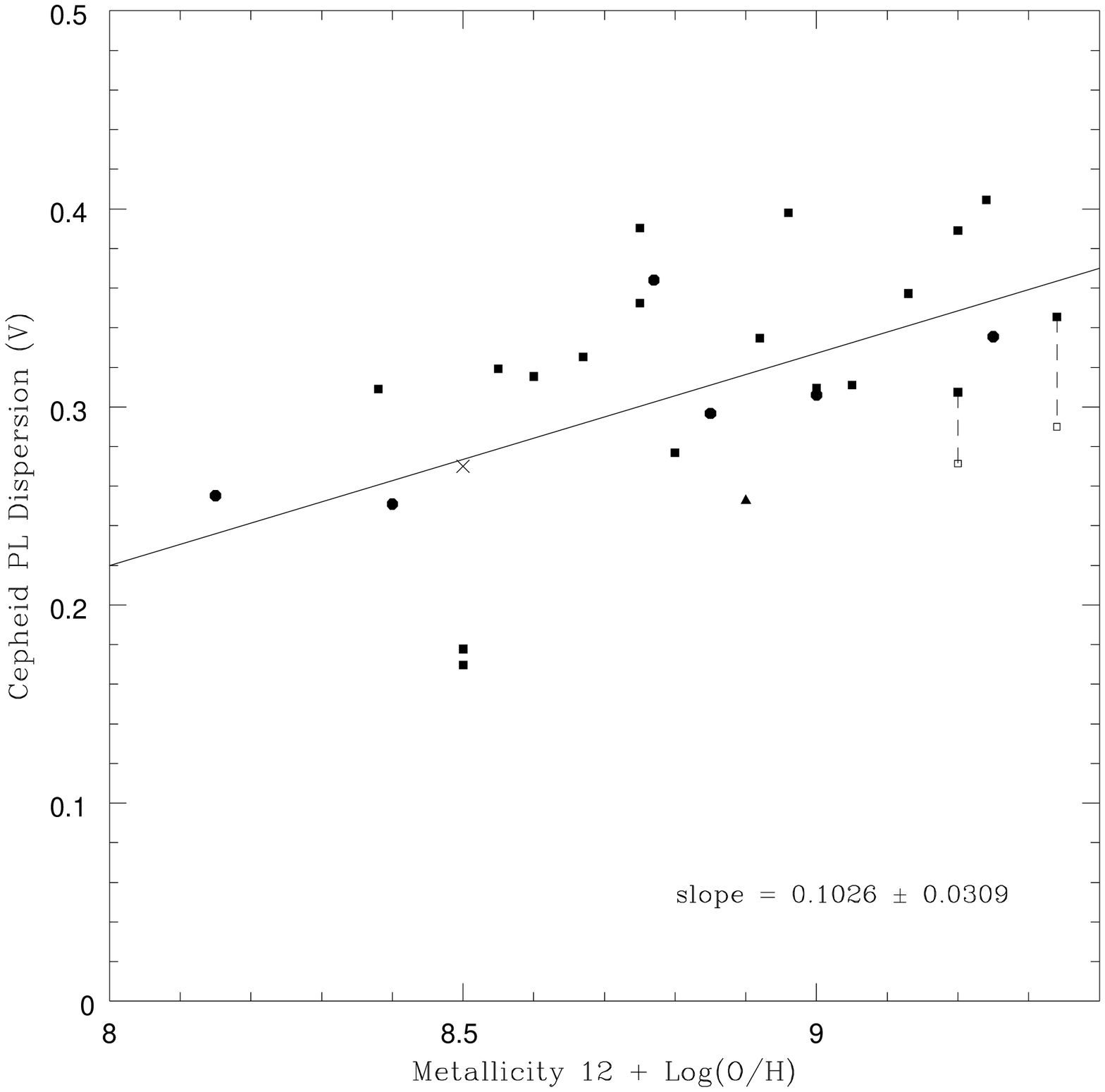}}
\caption{Relationship between mean r.m.s. dispersion (V-band) about the Cepheid P-L
relation and metallicity of HII regions in the vicinity of the Cepheids. Data
from the H$_{0}$ Key Project team is shown as squares. Sandage et al. data is
shown as circles and the Tanvir et al data as a triangle (although in both these
cases the Key Project photometry was used). Finally, the LMC is shown as a cross.
The result of a least squares fit to the data is also shown.}
\label{fig:metsig1}
\end{figure}

The V-band dispersion about the best fit to the P-L relation (Figure 
\ref{fig:pl1} and Table \ref{tab:cephdist}), was plotted against the 
metallicity (Table \ref{tab:cephmag}). The results are shown in Figure 
\ref{fig:metsig1}.

A least squares fit was performed on the data in Figure \ref{fig:metsig1} as 
described in the previous section except the slope was not constrained to any 
particular value. A slope of 0.103 $\pm$ 0.031 mag dex$^{-1}$ was obtained, 
which is significant at the 3.3$\sigma$ level. The error is the standard error
on the slope.  

\begin{figure}
{\epsfxsize=8.5truecm \epsfysize=8.5truecm
\epsfbox[17 143 565 695]{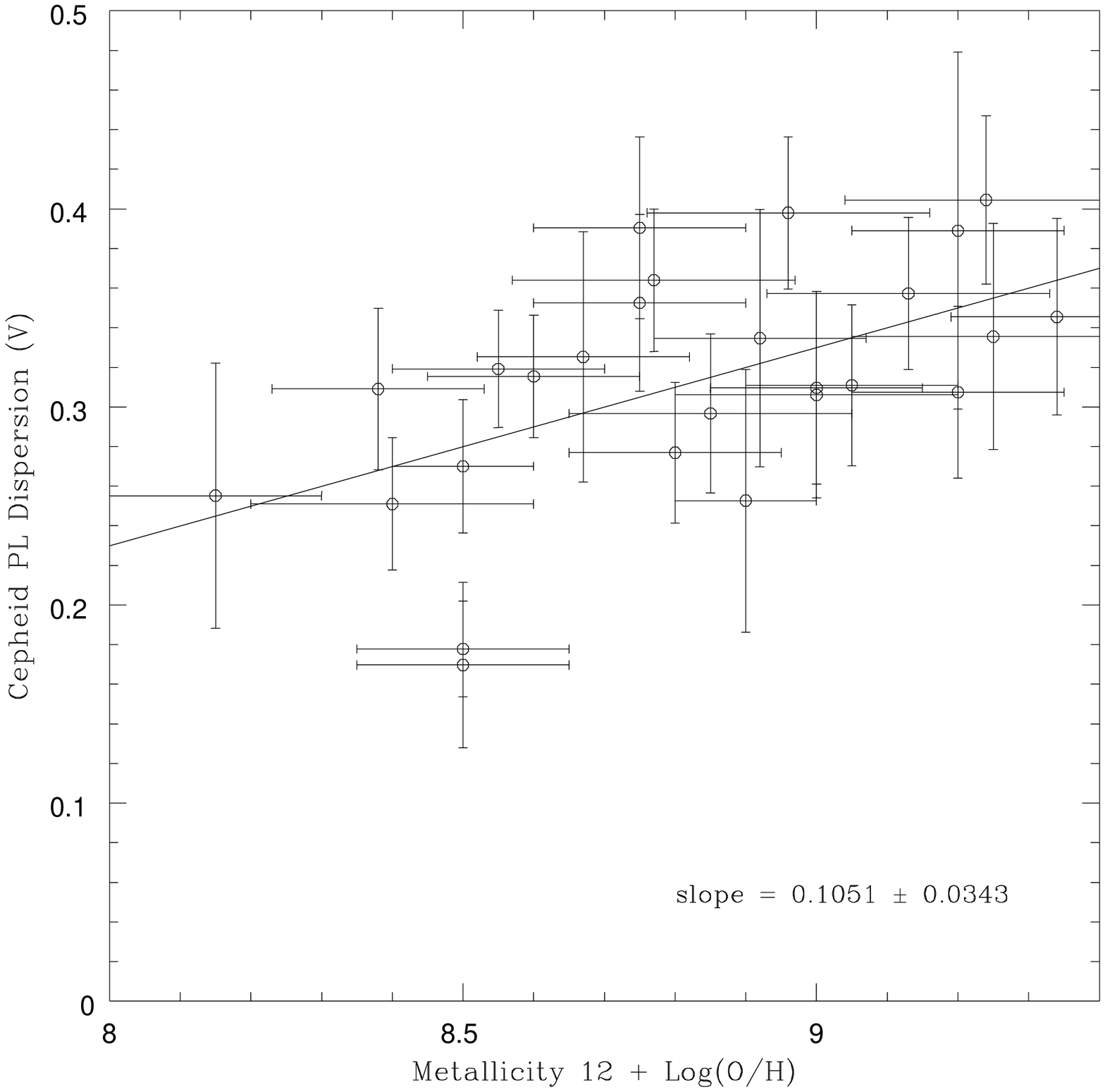}}
\caption{Relationship between mean r.m.s. dispersion (V-band) about the Cepheid 
P-L relation and metallicity of HII regions in the vicinity of the Cepheids, now showing
the errors on both axes.  The same unweighted least squares fit to the data is shown as in Fig. 2.}
\label{fig:err}
\end{figure}

The least squares fit is unweighted as the errors on the metallicities and 
dispersions are comparable. The error bars are shown in Figure \ref{fig:err}. 
One possible issue here is how the Cepheids are spatially distributed in the 
galaxies. Since there is a metallicity gradient in galaxies, Cepheids that are
taken from a larger area of a galaxy should have a larger dispersion. It is 
assumed here that the errors on the metallicities take this into account. 
Unfortunately it is not clear from \citeasnoun{fer} if this is the case, 
although it appears that measurement errors dominate in any case, so this 
effect is likely to be small.  

\begin{figure*} 
\begin{tabular}{ccc} 
{\epsfxsize=5.5truecm \epsfysize=5.5truecm \epsfbox[17 144 590 715]{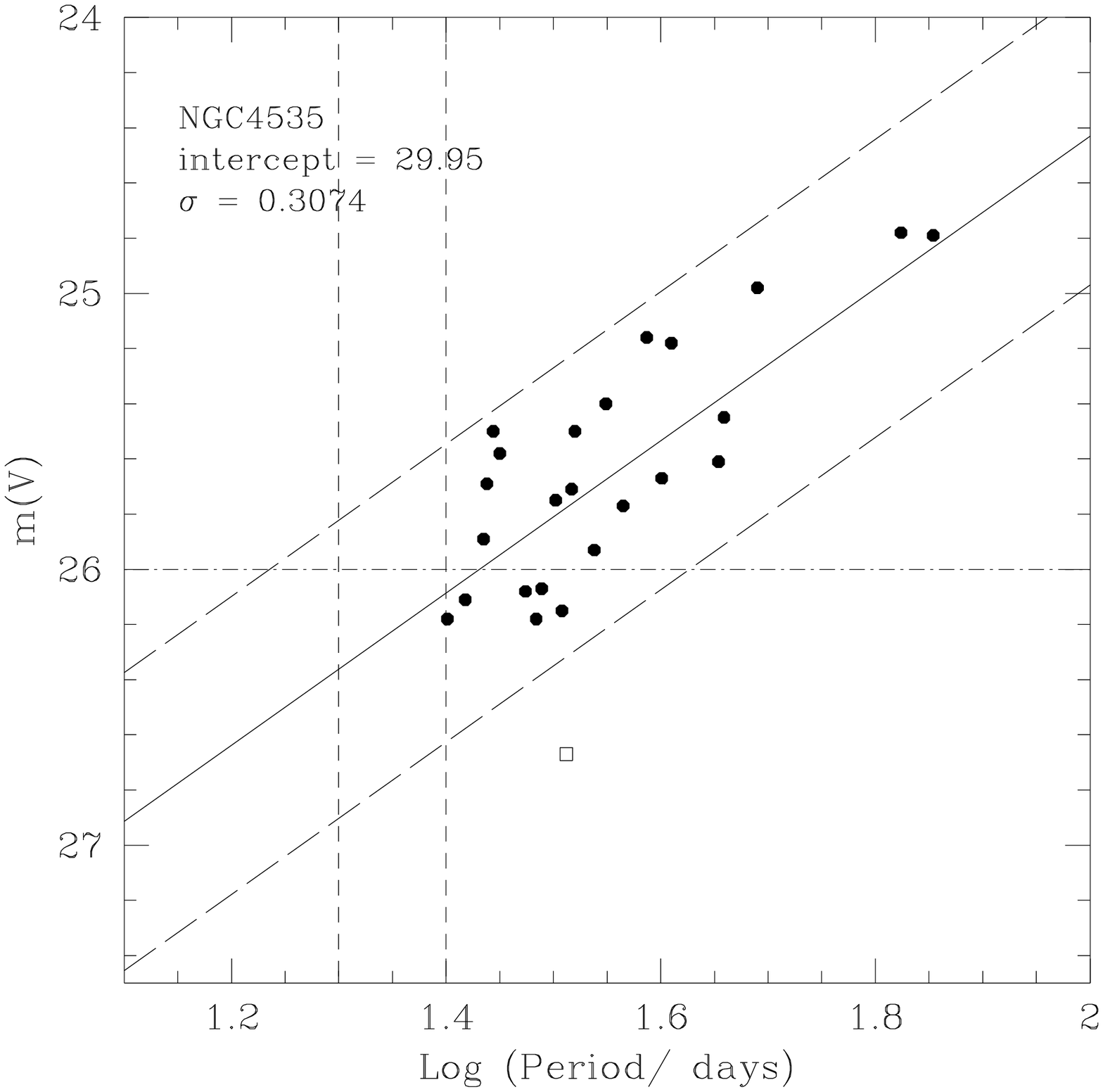}} &
{\epsfxsize=5.5truecm \epsfysize=5.5truecm \epsfbox[17 144 590 715]{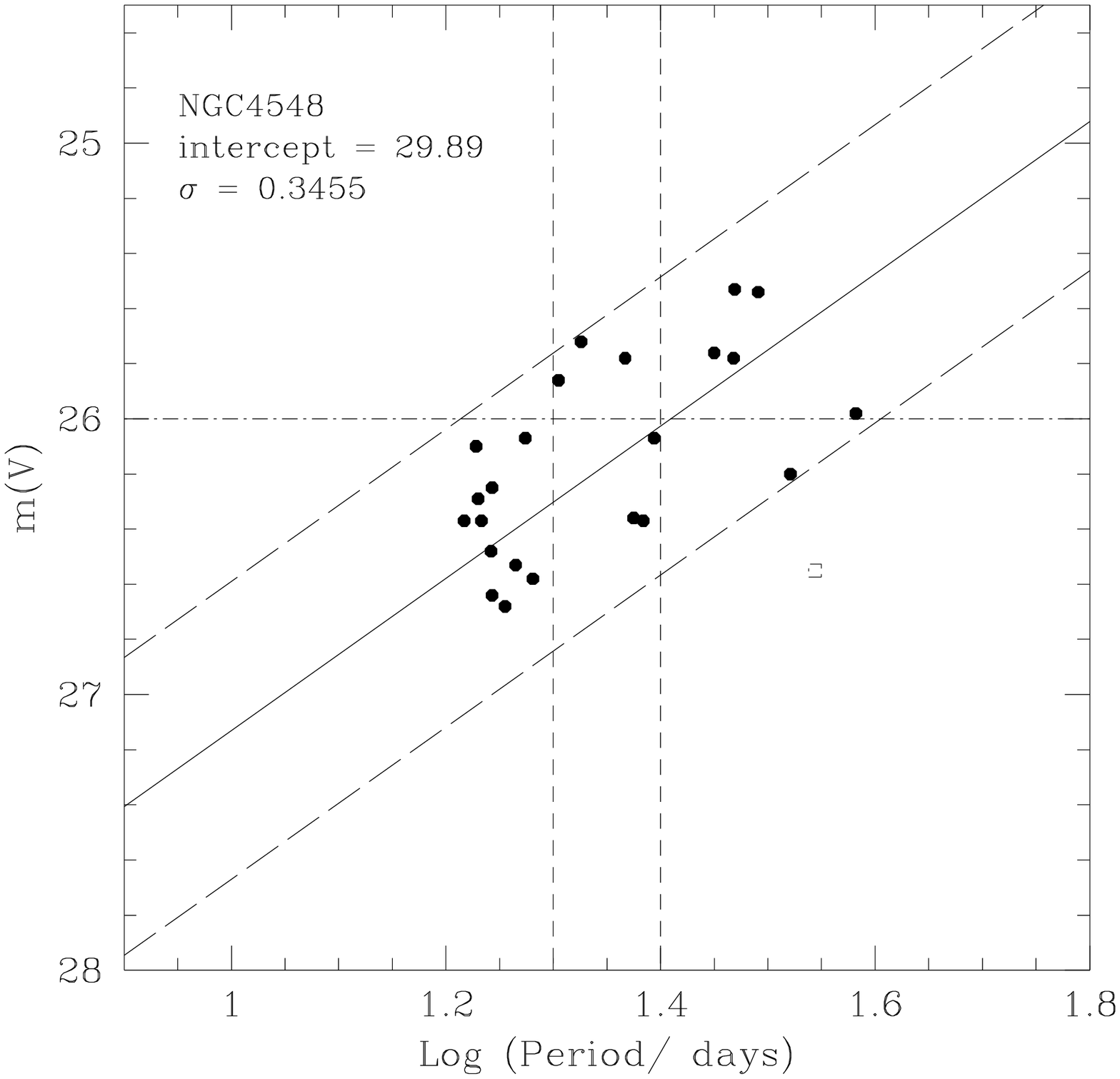}} \\
\end{tabular}
\caption{P-L (V) relations for NGC 4535 and NGC 4548. Each has a possible outlying point 
(shown as an empty square rather than a filled circle). This reduces the dispersion from 
0.3074 to 0.2714, and from 0.3455 to 0.2900 respectively}
\label{fig:plout} 
\end{figure*}

An analysis of the P-L relations shows that two of the high dispersion/high 
metallicity galaxies contain outlying points. The galaxies are NGC 4535 and 
NGC 4548. The questionable points are shown as open squares rather than filled
circles in the P-L relations in Figure \ref{fig:plout}. Consequently, the 
dispersion decreases for both the galaxies. NGC 4535 decreases from 0.3074 to 
0.2714 and NGC 4548 from 0.3455 to 0.2900. On the dispersion/metallicity graph
(Figure \ref{fig:metsig1}) these alternative points are shown as open squares.
Using these new points, a best fit to Figure \ref{fig:metsig1} gives a slope 
of 0.085 $\pm$ 0.033 mag dex$^{-1}$, which is significant at the 2.6$\sigma$
level. The Key Project Team included these points in their work, and their 
Cepheid selection procedure is fairly conservative. It is unlikely that the 
points are not normal Cepheids, (they are also not outliers in the I-band data), 
and so they are included in this work, but the effect of their removal is 
noted.

There was some concern over incompleteness at short periods due to a magnitude
limit or crowding at short periods. To investigate the effects of this, the 
data was cut in two ways. Using a cut in period at P = 25 days and a cut in 
period at the point where the lower LMC 2$\sigma$ line crosses the magnitude 
limits calculated in Table \ref{tab:cephmag}. The results are shown for the 
V-band in Table \ref{tab:cuts}. In many cases the number of Cepheids that 
could be used in the P-L fit reduces dramatically. 

A period cut at P = 25 days was applied to all the P-L relations. This either 
has little effect, or causes the number of Cepheids to be so small that the 
measured dispersion is not reliable. Also, the Key Project team (Ferrarese et 
al (1999)) has already examined the effects of cutting data at P = 25 days. 
We therefore revert to the Key Project reccomendations for period cuts at P = 
25 days (\citeasnoun{fer} and \citeasnoun{gibson}) in the rest of this paper.

Cuts in period where the lower LMC 2$\sigma$ line crosses the magnitude limit 
can also mean that much of the data is lost, and in some cases it was not even
possible to construct a P-L relation because only one Cepheid was left. In many
cases there was also little change in dispersion or intercept. It was decided 
to consider changing those galaxies where there were 7 or more remaining 
Cepheids, and the dispersion had changed by more than 0.03 mag. Table 
\ref{tab:cuts} shows all these changes, with the last column showing the final
selection. 

\begin{figure}
{\epsfxsize=8.5truecm \epsfysize=8.5truecm
\epsfbox[17 143 565 695]{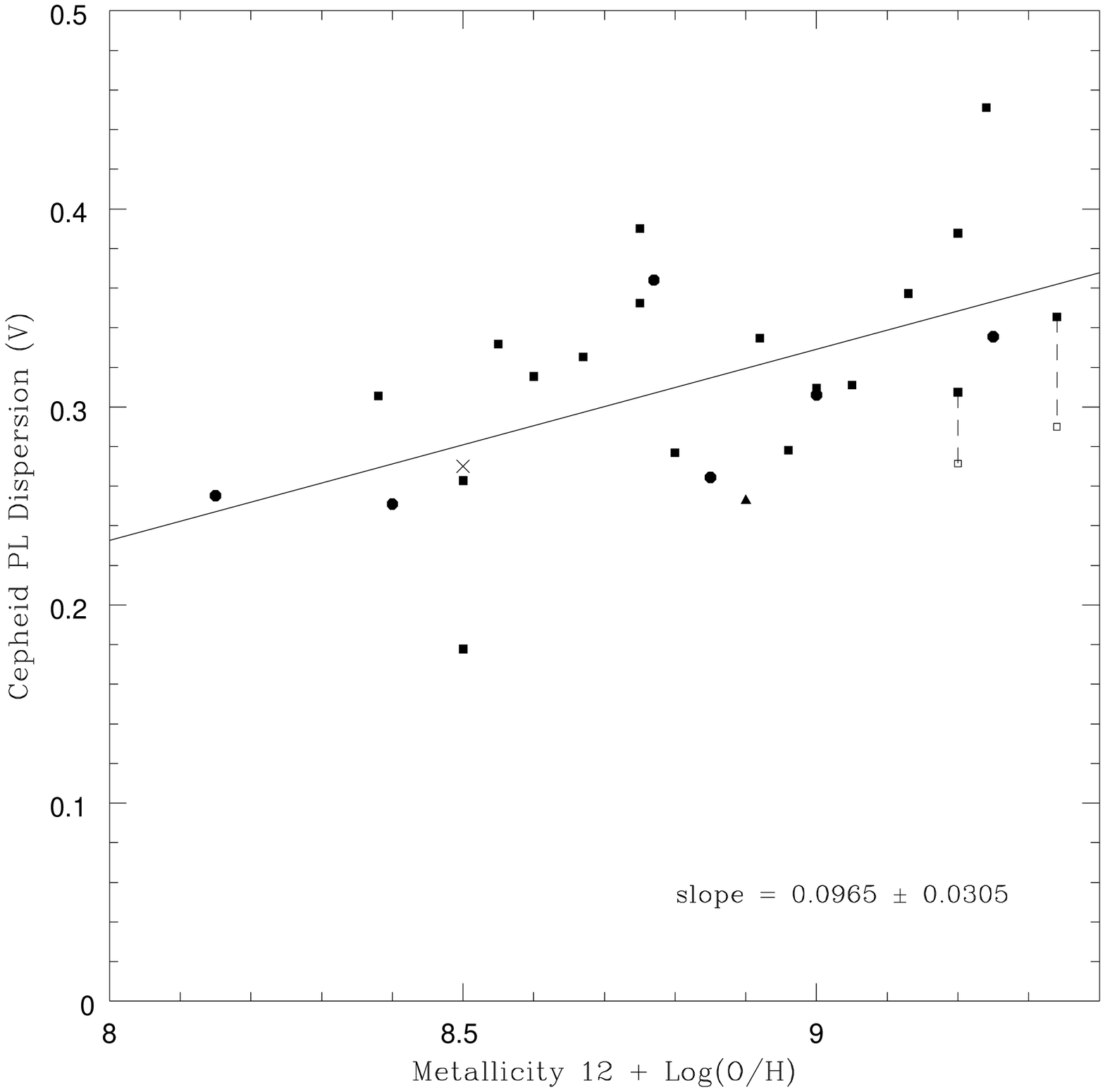}}
\caption{Relationship between mean r.m.s. dispersion (V-band) about the Cepheid P-L 
relation and metallicity of HII regions in the vicinity of the Cepheids. 
Magnitude cuts have been applied to data for the galaxies NGC 925, NGC 
1326A, NGC 3351 and NGC 4536. Data from the H$_{0}$ Key Project team are 
shown as squares. Sandage et al data are shown as circles and the Tanvir 
et al data as a triangle (although in both these cases the Key Project 
photometry was used). Finally, the LMC is shown as a cross. The result of a 
least squares fit to the data is also shown.}
\label{fig:metsig2}
\end{figure}

Figure \ref{fig:metsig2} shows the relationship between metallicity and data 
that has been selectively cut in period at the point where the 2$\sigma$ LMC 
dispersion lines cross the magnitude limit. Cuts have been applied to NGC 925,
NGC 1326A, NGC 3351, and NGC 4536. The best fit to the data is shown on the 
graph. A slope of 0.096 $\pm$ 0.030 is obtained which is still significant 
at the 3$\sigma$ level. If the galaxies NGC 4535 and NGC 4548 have their 
dispersion changed due to outlying points, as described in the previous 
section, the slope reduces to 0.079 $\pm$ 0.033 mag dex$^{-1}$. This is 
significant at the 2.4$\sigma$ level, and not much different than the value 
quoted in Figure \ref{fig:metsig1}.   

Having considered the effects of systematically applying period cuts to the 
Cepheid data, we note that this can only be done in a small number of cases. 
It should also be noted that the changes to the metallicity/dispersion 
relation are small. For these reasons we use the original values based
upon the Key Project period cutting recommendations given in Figure 
\ref{tab:cephdist} in the rest of this paper.

\begin{table}
\begin{tabular}{ccccc}
\hline
Galaxy & Dispersion & Dispersion & Dispersion & Change\\
 & (published) & P= 25 days & P (mag) & y/n\\ \hline
NGC 925 & 0.32 & 0.37 & 0.33 & y \\
NGC 1326A & 0.17 & 0.18 & 0.26 & y \\
NGC 1365 & 0.28 & 0.27 & N/A & n \\
NGC 1425 & 0.31 & 0.31 & N/A & n \\
NGC 2090 & 0.28 & 0.29 & 0.29 & n \\
NGC 2541 & 0.18 & 0.15 & 0.16 & n \\
NGC 3031 & 0.35 & 0.23 & 0.35 & n \\
NGC 3198 & 0.32 & 0.38 & 0.29 & n \\
NGC 3319 & 0.31 & 0.36 & N/A & n \\
NGC 3351 & 0.40 & 0.47 & 0.45 & y \\
NGC 3621 & 0.39 & 0.45 & 0.41 & n \\
NGC 4321 & 0.36 & 0.38 & 0.36 & n \\
NGC 4414 & 0.39 & 0.45 & N/A & n \\
NGC 4535 & 0.31 & 0.31 & N/A & n \\
NGC 4548 & 0.35 & 0.46 & N/A & n \\
NGC 4725 & 0.33 & 0.33 & N/A & n \\
NGC 5457 & 0.31 & 0.32 & 0.31 & n \\
NGC 7331 & 0.33 & 0.38 & 0.35 & n \\
NGC 3368 & 0.25 & 0.24 & 0.24 & n \\
NGC 3627 & 0.34 & 0.34 & 0.34 & n \\
NGC 4639 & 0.31 & 0.31 & N/A & n \\
NGC 4496A & 0.36 & 0.36 & 0.36 & n \\
NGC 4536 & 0.30 & 0.28 & 0.26 & y \\
NGC 5253 & 0.25 & N/A & 0.25 & n \\
IC 4182 & 0.25 & 0.29 & 0.25 & n \\ \hline
\end{tabular}
\caption{V-band P-L relation dispersion using data as published and with period 
cut at P = 25 days and in period where the 2$\sigma$ dispersion lines 
cross the magnitude limits}
\label{tab:cuts}
\end{table}

\subsection{I-Band Data}
\label{sec:iband}

\begin{figure}
{\epsfxsize=8.5truecm \epsfysize=8.5truecm
\epsfbox[17 143 565 695]{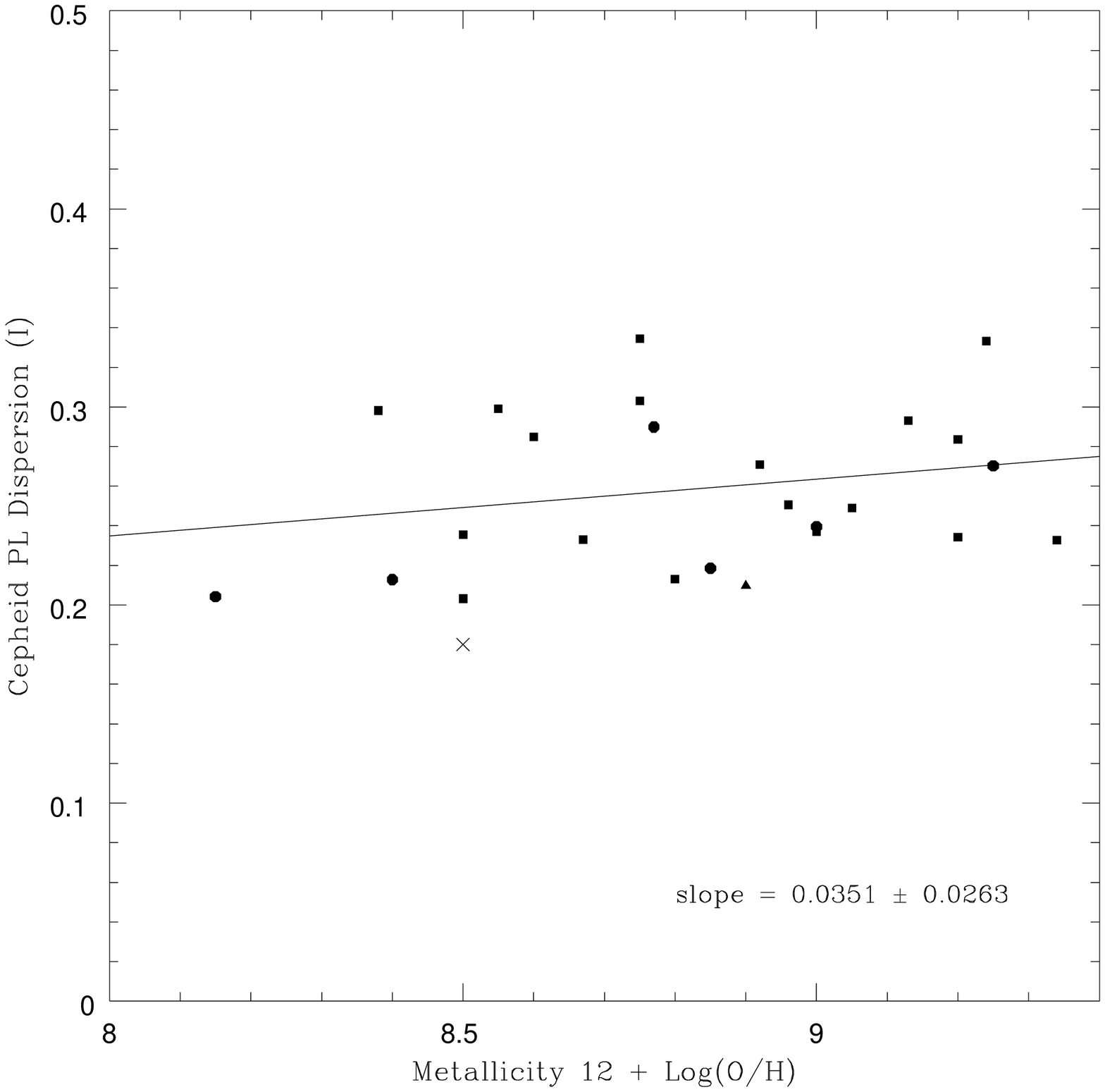}}
\caption{Relationship between mean r.m.s. dispersion (I-band) about the Cepheid P-L
relation and metallicity of HII regions in the vicinity of the Cepheids. The same
symbols are used as in Fig. 5. The result of a least squares fit to the data is
also shown.}
\label{fig:metsigi}
\end{figure}

As with the V-band data, the dispersions from Table \ref{tab:cephdist} were 
plotted against the metallicities in Table \ref{tab:cephmag}. The result is 
shown in Figure \ref{fig:metsigi}. The best fit to this data gives a slope of 
0.035 $\pm$ 0.026. Although the relationship is in the same direction as the
V-band data the result is less significant.  It is therefore not possible to 
infer a relationship between the I-band dispersion and metallicity from this 
data.

The galaxies NGC 4535 and NGC 4548 do not have outlying points in the I-band
data and so no attempt is made to recalculate their dispersions. Any magnitude
cut that could be made either leaves the dispersion unchanged (by more than 
0.03 mag), or means that too much data is lost, and it is not possible to 
construct an accurate P-L relation.

\subsection{Why is there a relationship in V but not in I?}
\label{sec:whyV}

One reason for expecting a relationship in I as well as V is that if Main
Sequence fitted distance moduli to NGC7790 are overestimates due to low
metallicity as claimed by \citeasnoun{hoyle}, then the effect of metallicity  on
Cepheid luminosity is expected to be broadly independent of passband. This is
because  a compensating and similar effect of metallicity on the three NGC7790
Cepheids is then required to maintain their tight position in the Galactic P-L
relations from B through to K. Since a relationship between dispersion and
metallicity can be inferred in  V at the 3 $\sigma$ level but not in  I, possible
reasons for  this result are now considered.

\subsubsection*{Reddening}
\label{sec:red}

\begin{figure}
{\epsfxsize=8.5truecm \epsfysize=8.5truecm
\epsfbox[17 143 570 695]{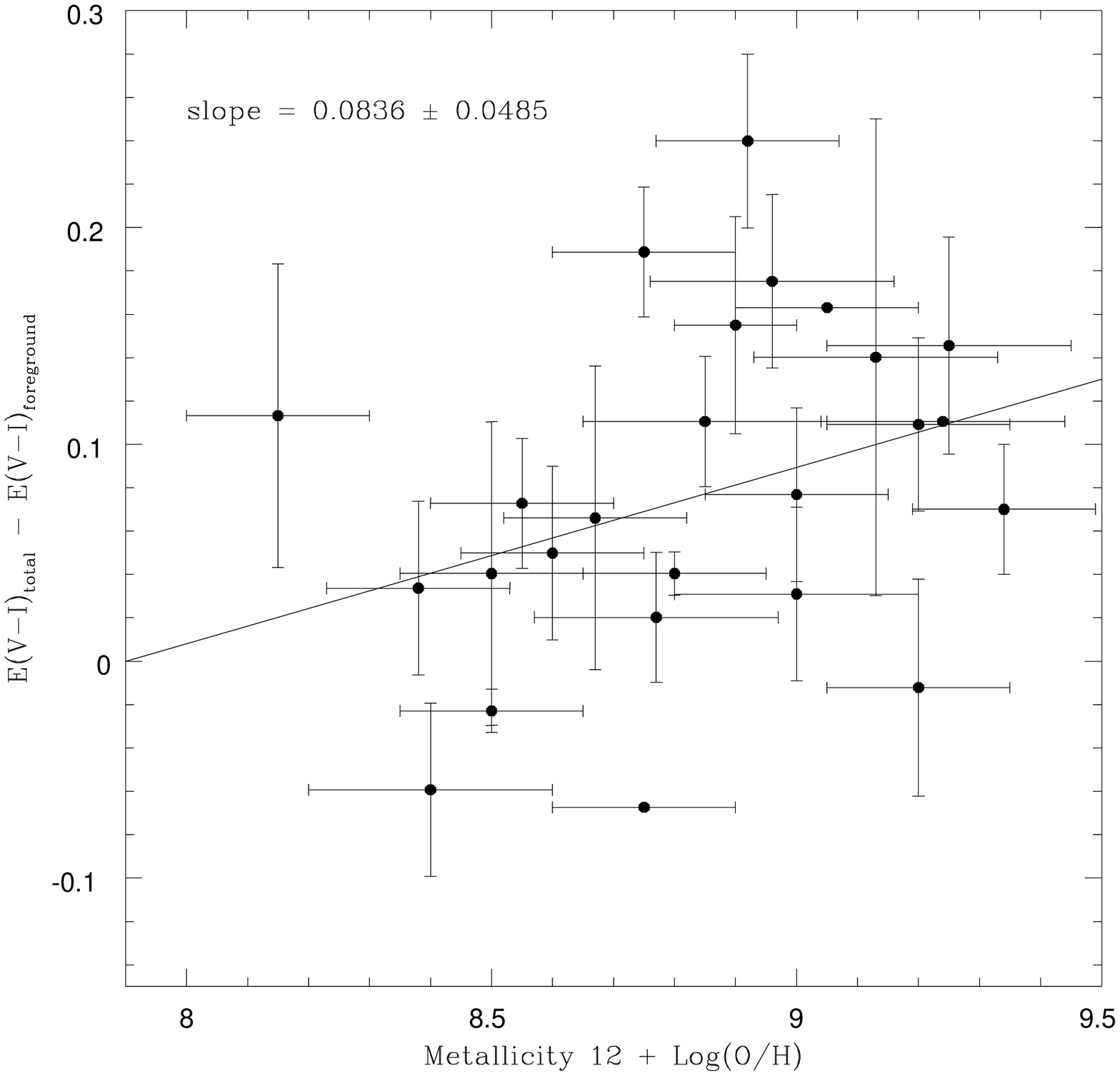}}
\caption{Relationship between metallicity and intrinsic reddening of the 25 
HST Cepheid galaxies. The reddenings of the Cepheids have been corrected for 
galactic foreground reddening. The best fit to the data and its gradient is 
shown on the graph.}
\label{fig:redmet}
\end{figure}

Absorption due to dust is  more significant in the V-band than the I-band. If a
galaxy is particularly dusty (i.e. with a high reddening), then an uneven
distribution of this dust would mean that the light from Cepheids is absorbed by
differing amounts. In turn this could cause more variation in magnitude and so
increase the dispersion about the P-L relation. Moreover, galaxies with high
metallicity are more likely to be highly evolved and hence more dusty. This means
that reddening due to absorption from dust could be the real cause of the
relationship between metallicity and V-band dispersion. The relationship between
reddening and metallicity is shown in Figure \ref{fig:redmet}, which is an update
of similar graphs in \citeasnoun{kochanek} and \citeasnoun{kennicutt}. The graph
shows that again that there is a weak correlation between the amount of reddening
toward a galaxy and the metallicities. The reddenings have the Galactic
foreground reddening subtracted. The foreground reddenings were taken from the
maps of \citeasnoun{schlegel} via the NASA/IPAC extragalactic database (NED)
website (http://nedwww.ipac.caltech.edu).

\begin{figure}
{\epsfxsize=8.5truecm \epsfysize=8.5truecm
\epsfbox[17 143 565 695]{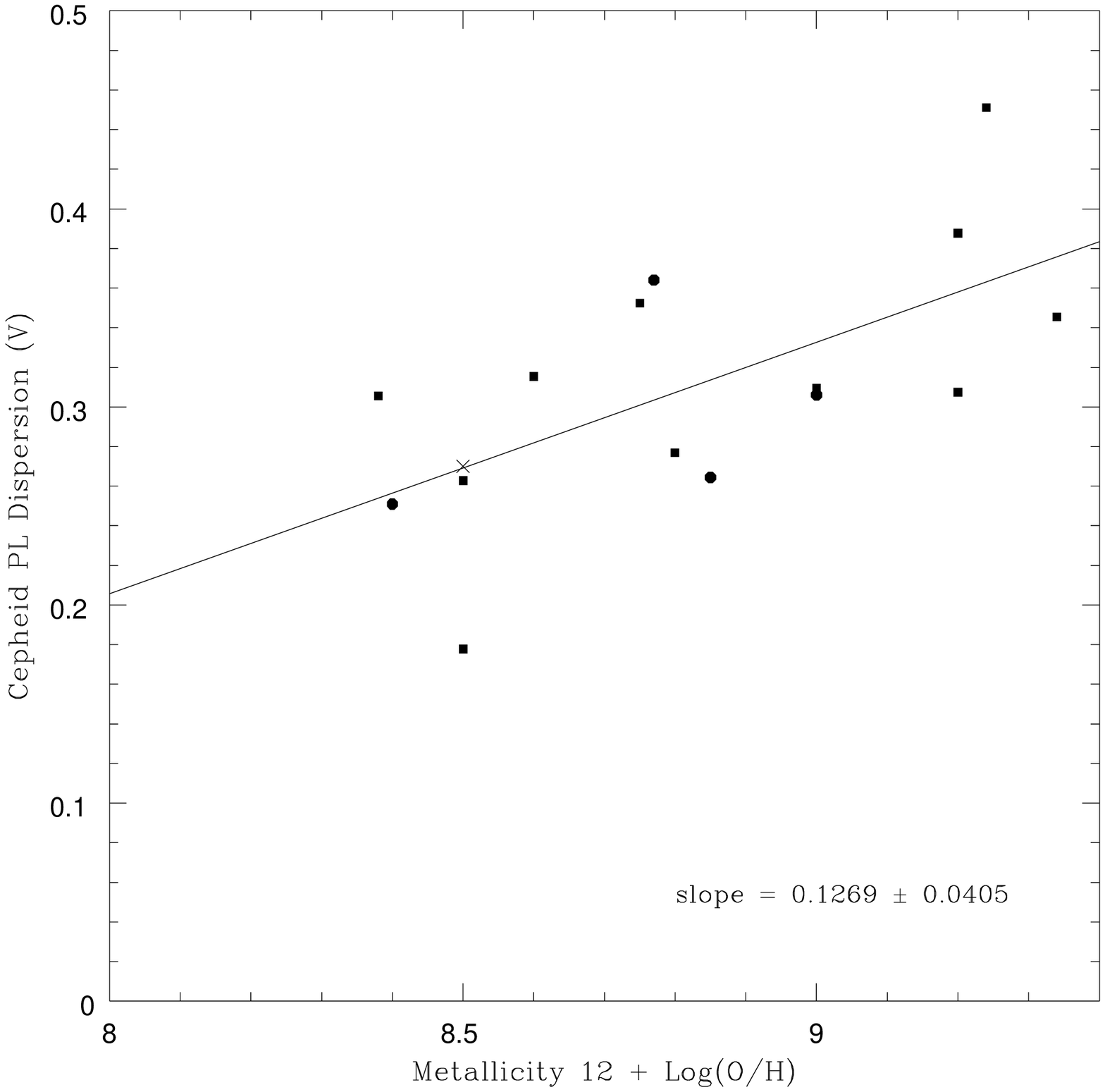}}
\caption{Relationship between mean r.m.s. dispersion (V-band) about the Cepheid P-L
relation and metallicity of HII regions in the vicinity of the Cepheids, now
using only the 10 least reddened galaxies. The same symbols are used as in Fig.
5. The result of a least squares fit to the data is also shown.} 
\label{fig:metsig3}
\end{figure}

The best fit to the unweighted points in Figure \ref{fig:redmet} gives a slope of
0.084 $\pm$ 0.049 mag dex$^{-1}$. The significance level is 1.7$\sigma$,
suggesting that reddening is unlikely to be the full cause of the observed
relationship. Despite this, to fully verify that reddening does not cause the
dispersion/ metallicity relationship, dispersion is plotted against metallicity
in Figure \ref{fig:metsig3} for those galaxies with low reddenings (i.e.
$E_{V-I}<0.13$mag). Here, the 10 most reddened galaxies are excluded from the
sample of galaxies used in the fit to the dispersion/metallicity data. The
dispersion data uses those galaxies with data cut due to possible magnitude
incompleteness. The galaxies NGC 925, NGC 1365, NGC 3621, NGC 4321, NGC 4725, NGC
5457, NGC 7331, NGC 3368, NGC 3627 and NGC 5253 are excluded. A best fit to this
data gives a slope of 0.13 $\pm$ 0.040 mag dex$^{-1}$, a result significant at
the 3$\sigma$ level. Removal of outlier points in NGC 4535 and NGC 4548 (see
Figure \ref{fig:plout}) reduces this slope to 0.10 $\pm$ 0.049 mag dex$^{-1}$,
which is significant at the 2$\sigma$ level. Both these results are comparable to
those obtained in the previous section. Removal of highly reddened galaxies has
little effect on the observed correlation and if anything makes the relationship
slightly stronger, which again suggests that reddening is unlikely to be the
cause of the relationship between V-band dispersion and metallicity.

\begin{figure}
{\epsfxsize=8.5truecm \epsfysize=8.5truecm
\epsfbox[17 143 565 695]{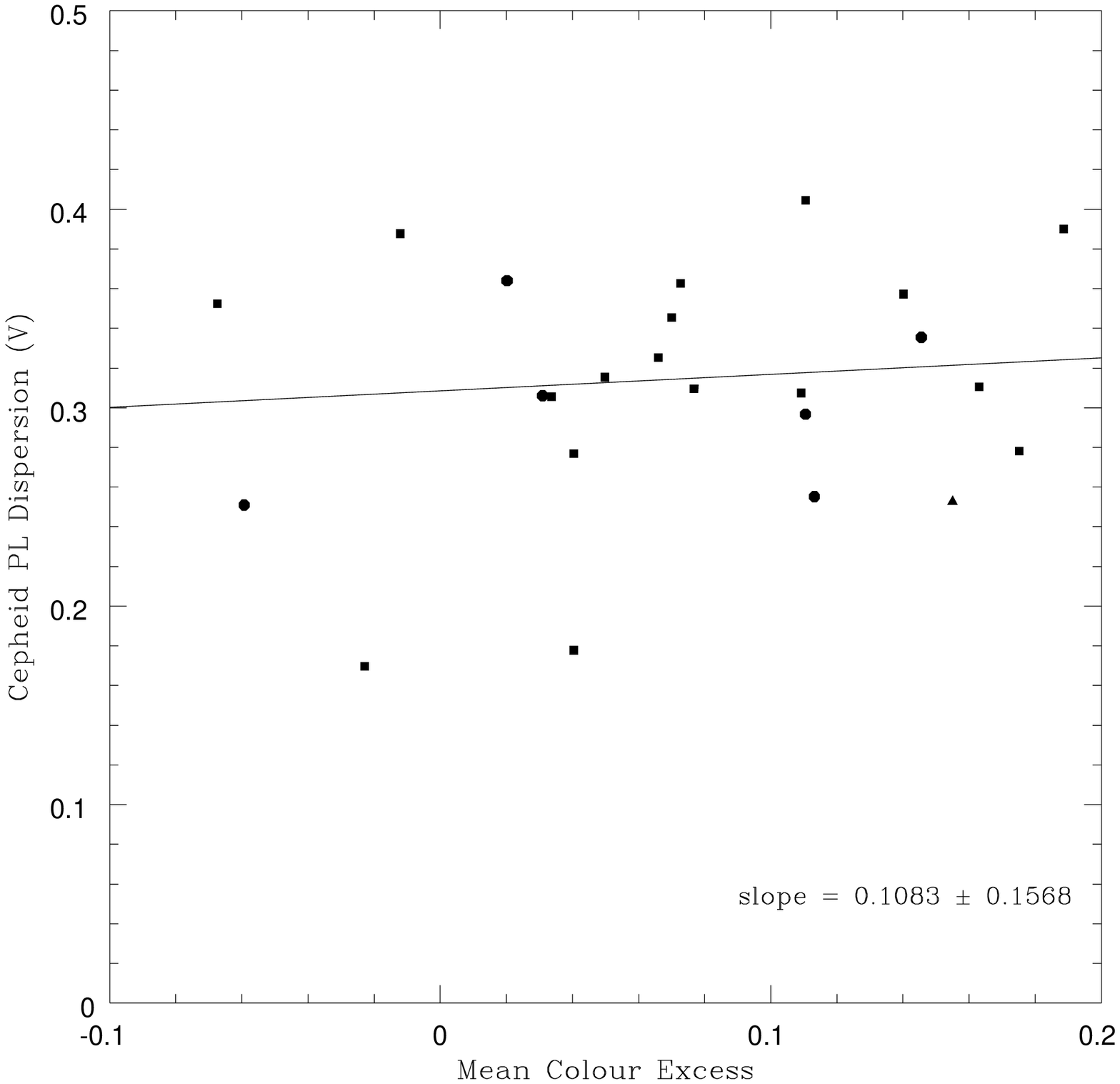}}
\caption{Relationship between mean r.m.s. dispersion (V-band) about the Cepheid P-L
relation and mean colour excess of the Cepheids. The same symbols are used as in
Fig. 5. The result of a least squares fit to the data is also shown.}
\label{fig:redsig}
\end{figure}

Finally, in Figure \ref{fig:redsig}, dispersion is plotted against reddening. No
correlation is observed with a best fit to the data giving a slope of 0.083 
$\pm$ 0.16. On the basis of this data and these arguments it seems very
unlikely that reddening is the cause of the high V-band dispersion at high
metallicity.

\subsubsection*{Photometric Errors}
\label{sec:photerr}

The main possible causes of dispersion in the Cepheid P-L relation are variation
in Cepheid temperature across the instability strip, reddening, metallicity and
measurement errors, particularly in photometry. Unfortunately, photometry errors
were only published for 15 of the galaxies, which are listed in Table
\ref{tab:phot}. The photometry errors quoted by the Key Project team are very
small (these were obtained from the relevant papers given in section
\ref{sec:data}) and are essentially based on epoch-epoch reproducibility for
non-variable stars. In all cases the quoted I-band error is significantly larger
than the V-band error (mean V-band photometry error = 0.045 mag c.f. mean I-band
photometry error = 0.070mag). This is because the V-band light curves are
constructed from observations at 12-15 epochs, whilst the I-band light curves are
constructed from observations at 3-5 epochs. If the mean photometric errors are
subtracted from the total dispersion in quadrature (see Table \ref{tab:phot}),
then in both cases the mean change in dispersion is small. For the V-band this is
0.0049 mag, and for the I-band, 0.012 mag. These changes are shown in Table
\ref{tab:phot}.

\begin{table*}
\begin{tabular}{ccccccc}
\hline
Galaxy & Photometry Error (V) & Phototmetry Error (I) & Corrected Dispersion (V) & Corrected Dispersion (I) & $\Delta$$\sigma$(V) & $\Delta$$\sigma$(I) \\ \hline
NGC 1365 & 0.10  & 0.11  & 0.26 & 0.22 & 0.020 & 0.026 \\
NGC 3198 & 0.11  & 0.13  & 0.30 & 0.25 & 0.019 & 0.031 \\
NGC 3368 & 0.031 & 0.037 & 0.25 & 0.21 & 0.0020 & 0.0033 \\
NGC 3621 & 0.041 & 0.059 & 0.39 & 0.33 & 0.0022 & 0.0052 \\
NGC 3627 & 0.025 & 0.033 & 0.33 & 0.27 & 0.0009 & 0.0020 \\
IC 4182  & 0.018 & 0.098 & 0.25 & 0.19 & 0.0006 & 0.024 \\
NGC 4414 & 0.040 & 0.069 & 0.39 & 0.28 & 0.0021 & 0.0085 \\
NGC 4535 & 0.029 & 0.039 & 0.31 & 0.23 & 0.0014 & 0.0032 \\
NGC 4536 & 0.026 & 0.041 & 0.30 & 0.21 & 0.0011 & 0.0038 \\
NGC 4639 & 0.032 & 0.063 & 0.30 & 0.23 & 0.0017 & 0.0084 \\
NGC 4725 & 0.024 & 0.041 & 0.33 & 0.27 & 0.0009 & 0.0031 \\
NGC 5253 & 0.023 & 0.060 & 0.25 & 0.20 & 0.0010 & 0.0090 \\
NGC 5457 & 0.090 & 0.11  & 0.30 & 0.23 & 0.013 & 0.024 \\
NGC 7331 & 0.065 & 0.12  & 0.32 & 0.20 & 0.0066 & 0.035 \\
NGC 4496A & 0.020 & 0.039 & 0.36 & 0.29 & 0.0006 & 0.0027 \\ \hline
\end{tabular}
\caption{Columns 4,5 contain the dispersions corrected for the photometry errors 
given in Columns 2,3. The resulting differences in the dispersions are given
in Columns 6,7.}
\label{tab:phot}
\end{table*}

Since the change to the dispersion is small, little effect is made on the
metallicity/dispersion relations, especially in the V-band. The best fitting
V-band slope is now 0.0750 $\pm$ 0.035 mag dex$^{-1}$, which although significant
at the 2$\sigma$ level, is shallower and less significant than previous results.
However, 10 data points are missing and these seem to be those galaxies with more
extreme dispersions. It should also be noted that V-band photometry corrections
make little difference to those points that have photometry errors available so it
is very likely that this would also be the case for the remaining 10 points. It
is therefore expected that photometry corrections would make little difference to
the relationship between metallicity and V-band dispersion implied in this paper.
In the I-band the changes are slightly more significant with the new best fitting
slope 0.049 $\pm$ 0.032 mag dex$^{-1}$. Although this is slightly steeper and
more significant than Figure \ref{fig:metsigi}, it is still not possible to infer
any relationship between metallicity and I-band dispersion. It appears that the
photometric errors quoted by the KP are unlikely to have much effect on the
measured dispersions about the P-L relations. They are small because they are
based on observations  at several epochs, even in the I-band.

\subsubsection*{Intrinsic Scatter}
\label{sec:intscatt}

Intrinsic scatter in the P-L relation cause by Cepheid temperature differences at
fixed period, which drive the Cepheid P-L-C relation, are significantly bigger
than photometric errors. In the low-metallicity LMC, the intrinsic scatter in V
is $\sigma_V=0.27$ mag and in I it is $\sigma_I=0.18$ mag. From Figs.
\ref{fig:metsig1}, \ref{fig:metsigi} it is clear that intrinsic scatter plus
photometric errors could plausibly dominate any metallicity induced P-L
dispersion in HST galaxies at low metallicity ($12+{\rm log}(O/H)\approx8.5$). At
high metallicity, ($12+{\rm log}(O/H)\approx9.4$), it is also clear that if the
difference of $\sigma_{diff}=\sqrt(0.27^2-0.18^2)=0.20$ between the I and V band
intrinsic dispersions is added to the observed HST I band dispersion of
$\sigma_I\approx0.28$ then a total dispersion of $\sigma_I=0.34$ results, which
is comparable to the observed $\sigma_V=0.35$ at high metallicity in Fig.
\ref{fig:metsig1}. Thus the  data in Figs. \ref{fig:metsig1}, \ref{fig:metsigi} at
least at high metallicity is not inconsistent with the idea that the effect of
metallicity on the P-L dispersions at I and V might be the same.

However, the dispersion in the lowest metallicity galaxies is  similar in both I
and V at $\approx\pm0.25$mag, so here there is no chance of arguing that the I
dispersion drops with similar slope to low metallicity as in V, when the  smaller
I intrinsic dispersion is taken into account. Indeed, combining the LMC I
dispersion of $\approx\pm0.18$  and the average photometric error in I of
$\pm0.07$mag only accounts for $\pm0.19$mag, leaving a residual r.m.s. dispersion of
$\pm0.16$mag of the observed $\approx\pm0.25$mag. The same problem does not exist
in the V band where the average of the low metallicity galaxies' dispersion is
similar to that of the LMC (see Fig. \ref{fig:metsig1}). This makes it unlikely
that the cause of the excess dispersion at low metallicity in I is reddening
which would be expected to cause a bigger dispersion at V than in I.

\subsubsection*{Discussion}
\label{sec:discussion}

In the V-band it appears likely the observed increase in dispersion with
metallicity is indeed caused by metallicity.  Both reddening and photometry are
likely to have little effect on this relationship. As pointed out in previous
sections, there have been several recent empirical attempts to measure the
effects of metallicity on Cepheids (\citeasnoun{gould},\citeasnoun{kennicutt}, 
\citeasnoun{hoyle}). These all seem to indicate that high metallicity Cepheids are
brighter than low metallicity ones. A high metallicity galaxy is likely to be
more highly evolved and complex, and therefore could contain Cepheids with
varying metal content, including some low metallicity Cepheids. In a low
metallicity, perhaps younger, galaxy, all the Cepheids are likely to be of low
metallicity whereas in a high metallicity galaxy there is more likely to be a
wider range of Cepheid metallicity and hence luminosity, causing increased
dispersion about the P-L relation.

In the I-band there are a number of reasons why the dispersion-metallicity
relation appears less strong. First, the effects of metallicity could be less
significant in the I-band. This is suggested by \citeasnoun{kennicutt} when
referring to their version of Figure \ref{fig:redmet}. However, as we have
considered above it is also possible that because the I-band P-L relation is
intrinsically tighter and the data more erroneous, the tight dispersion expected
in the HST galaxy P-L relations at low metallicity  may be masked by increased
error. The difficulty for this explanation is that the quoted photometric errors
are not enough to explain why the HST I dispersions are so much larger than the
LMC dispersion.  One possibility is that the effects of crowding are mostly
neglected in the KP estimate of the photometric error (as summarised in Table
\ref {tab:phot}), which is essentially a measure of how well a magnitude
reproduces epoch-epoch. The error caused by two unresolved stars being merged is
not included in the above estimate. Using tests where artificial stars are added
to real HST frames, \citeasnoun {fer00b} claim crowding does not systematically
bias HST Cepheid photometry bright by more than 0.07mag and they use this to
rebut claims of a larger effect by \citeasnoun {moch} and \citeasnoun {saha00}.
However, \citeasnoun {fer00b} do not report the increased r.m.s. error caused by
crowding. Also these authors sub-select the artificial stars that deviate by
$>3\sigma$ at one epoch against their mean over 4 epochs and this would seem to
exclude  the stars with crowding errors because here  the underlying star will
contribute at all epochs and so bias the mean magnitude. The fact that the HST
P-L relations at LMC metallicity generally have significantly larger dispersions
than observed in the LMC may thus argue that crowding makes a significant
contribution of $\approx\pm$0.16mag in I to the real r.m.s. photometric error. In
turn, this may also mean that the {\it systematic} error due to crowding is
higher than claimed by \citeasnoun {fer00b} and closer to their average single
frame offset for NGC1365 of $\approx$0.13mag.

We conclude that a combination of crowding  together with other smaller
contributions from other effects such as reddening  may increase the P-L
dispersion and mask the existence of a similar  metallicity-dispersion correlation
in I as at at V. Certainly, if the LMC dispersion is taken to be a more accurate
measure of the I dispersion at low metallicity,  the dispersion rises from
$\pm$0.18 at $12+{\rm log}(O/H)\approx8.5$ to $\pm$0.25 at $12+{\rm log}(O/H)\approx9.4$ and
this is of the same order as the differential effect of metallicity on the V
dispersion.

\section{Consequences of a Relationship Between Dispersion and Metallicity}
\label{sec:conseq}

\subsection{Application of a Global Metallicity Correction}
\label{sec:global}

Several recent studies have attempted to measure the effects of metallicity on
Cepheid magnitudes. Although there are now several authors claiming that high
metallicity Cepheids are brighter than low metallicity ones e.g.
\citeasnoun{gould}, \citeasnoun{sasselov}, \citeasnoun{kennicutt},
\citeasnoun{hoyle}, the Key Project team do not yet apply any correction to their
distances. This is because the errors associated with the metallicity correction
are  high and at best the relationships are significant at the 2$\sigma$ level.
For example, \citeasnoun{kennicutt} measured $\Delta
M/[O/H]\approx-0.24\pm0.16$mag whereas \citeasnoun{hoyle} claimed $\Delta
M/[O/H]\approx-0.66$ which is only consistent with the relation of
\citeasnoun{kennicutt} at the 2.6$\sigma$ level. However, this work, whilst not 
measuring the size of  the global metallicity effect very accurately, does imply
with a reasonable significance that Cepheid magnitudes are affected by
metallicity.

In fact, if we take the observed error in the  logarithmic metallicity of the
highest metallicity galaxies' as $\approx\pm$0.2 then the excess r.m.s. error in
their V P-L  dispersion  of $\pm$0.25mag converts to $|\Delta
M|/[O/H]\approx0.25/0.2\approx1$. The excess in the I P-L dispersion for the high
metallicity galaxies is smaller at $\pm$0.13mag, converting to $|\Delta
M|/[O/H]\approx0.13/0.2\approx0.65$. Both these numbers are close to the claimed
$\Delta M/[O/H]\approx-0.66$ metallicity correction of \citeasnoun{hoyle}. If the
smaller -0.24 mag dex$^{-1}$ of \citeasnoun{kennicutt} applied, then the expected
increase in dispersion is only $\pm0.2\times0.24\approx\pm0.05$mag which is too
small to explain even the weaker I band correlation.  We therefore take this as
supporting  evidence that a global metallicity correction should be applied to
Cepheid distances. As a result of this, the effects of the -0.66 mag dex$^{-1}$
metallicity correction of \citeasnoun{hoyle} are now examined.

\subsubsection*{The Distance to the LMC}
\label{sec:lmcdist}

The distance to the LMC is likely to be one of the largest possible sources
of systematic error in the Cepheid distance scale. One of the effects of
applying a metallicity correction would be to reduce the Cepheid distance to
the LMC, changing the zero point of the Cepheid P-L relation and cancelling
the effect of the increased extragalactic distances. However, there is
general disagreement between authors, over what the Cepheid distance to the
LMC actually is (e.g. \citeasnoun{hoyle} obtain a distance modulus of 18.5,
whilst \citeasnoun{feastcatch} obtain 18.7). \citeasnoun{tanvir98} (after 
\citeasnoun{mf91}) cautions that it is perhaps unwise to rely on Cepheid 
distances alone for this step of the distance ladder, and recommends that a 
distance of 18.5 should be adopted as an average of different methods 
\cite{mf91}. These can include Cepheids, RR Lyrae and various techniques 
involving Supernova 1987A. As \citeasnoun{tanvir98} points out, the Cepheid 
distances are generally on the high side in LMC distance determinations
(i.e. $>$ 18.5), and application of a metallicity correction may bring the 
Cepheid distances into line with those obtained using other methods.

\citeasnoun{hoyle} obtain an LMC distance modulus of 18.5$\pm$0.12. Application
of a metallicity correction to this (assuming [O/H] for the LMC = -0.4) reduces
the distance modulus to 18.24$\pm$0.12. However, this is clearly low compared
with the independent distances obtained using RR Lyrae  and SN1987A. It should
also be noted that the distance modulus obtained by \citeasnoun{hoyle}  has a
relatively large error associated with it.  As a result of these uncertainties
over the LMC distance the effects of a global metallicity correction are looked
at in two different cases, taking the LMC absolute distance modulus to be 18.5
and 18.24, although 18.5 is the preferred distance modulus.

\subsubsection*{Cepheid Distances with a Global Metallicity Correction}
\label{sec:hoyledistances}

The result of applying a -0.66 mag dex$^{-1}$ metallicity correction is shown
in Table \ref{tab:hoyledist}, which list the modified distance moduli for LMC
distance moduli of 18.5, and 18.24. The change to the published values is 
also noted along with their mean change. For LMC modulus = 18.24 there is 
little overall change (i.e. a mean of -0.039). However, for LMC modulus = 
18.5, it can be seen that the published distances are too short by a mean 
value of 0.22$\pm$0.17, where the error estimate comes from the 
dispersion of the data. Whilst this result could have a significant effect, 
it should be noted that the error estimate is quite large.

\begin{table*}
\begin{tabular}{ccccccc}
\hline
Galaxy & Number of Cepheids & Distance Modulus & Distance Modulus & $\Delta$$\mu$ & Distance Modulus & $\Delta$$\mu$\\
 & & (As Published) & (Corrected)  & &  (Corrected)  & \\
 & & & $\mu$$_{LMC}$ = 18.5 & & $\mu$$_{LMC}$ = 18.24 & \\ \hline
NGC 925 & 75 & 29.82 & 29.85 & 0.03 & 29.59 & -0.23\\
NGC 1326A & 8 & 31.55 & 31.55 & 0 & 31.29 & -0.26\\
NGC 1365 & 26 & 31.40 & 31.71 & 0.30 & 31.45 & 0.05\\
NGC 1425 & 20 & 31.82 & 32.15 & 0.33 & 31.89 & 0.07\\
NGC 2090 & 30 & 30.45 & 30.65 & 0.20 & 30.39 & -0.06\\
NGC 2541 & 27 & 30.46 & 30.46 & 0 & 30.20 & -0.26\\
NGC 3031 & 25 & 27.77 & 27.94 & 0.17 & 27.68 & -0.09\\
NGC 3198 & 52 & 30.80 & 30.87 & 0.07 & 30.61 & -0.19\\
NGC 3319 & 28 & 30.78 & 30.70 & -0.08 & 30.44 & -0.34\\
NGC 3351 & 45 & 30.01 & 30.50 & 0.49 & 30.24 & 0.23\\
NGC 3627 & 36 & 29.14 & 29.30 & 0.16 & 29.04 & -0.10\\
NGC 4321 & 43 & 31.04 & 31.46 & 0.42 & 31.20 & 0.16\\
NGC 4414 & 9 & 31.41 & 31.87 & 0.46 & 31.61 & 0.20\\
NGC 4535 & 25 & 31.11 & 31.57 & 0.46 & 31.31 & 0.20\\
NGC 4548 & 24 & 31.01 & 31.56 & 0.55 & 31.30 & 0.29\\
NGC 4725 & 13 & 30.56 & 30.84 & 0.28 & 30.58 & 0.02\\
NGC 5457 & 29 & 29.34 & 29.70 & 0.36 & 29.44 & 0.10\\
NGC 7331 & 13 & 30.90 & 31.01 & 0.11 & 30.75 & -0.15\\
NGC 3368 & 7 & 30.20 & 30.47 & 0.27 & 30.21 & 0.01\\
NGC 3627 & 17 & 30.06 & 30.56 & 0.50 & 30.30 & 0.24\\
NGC 4639 & 17 & 31.80 & 32.13 & 0.33 & 31.87 & 0.07\\
NGC 4496A & 51 & 31.02 & 31.20 & 0.18 & 30.94 & -0.08\\
NGC 4536 & 27 & 30.95 & 31.18 & 0.23 & 30.92 & -0.03\\
NGC 5253 & 7 & 27.61 & 27.38 & -0.23 & 27.12 & -0.49\\
IC 4182 & 28 & 28.36 & 28.30 & -0.06 & 28.04 & -0.32\\ \hline
\end{tabular}
\caption{The effects of metallicity on Cepheid distances and the change
in distance, taking absolute distance moduli $\mu$$_{LMC}$ = 18.50 and $\mu$$_{LMC}$ = 18.24.}
\label{tab:hoyledist}
\end{table*}
 
\begin{figure}
{\epsfxsize=8.5truecm \epsfysize=8.5truecm
\epsfbox[17 143 565 695]{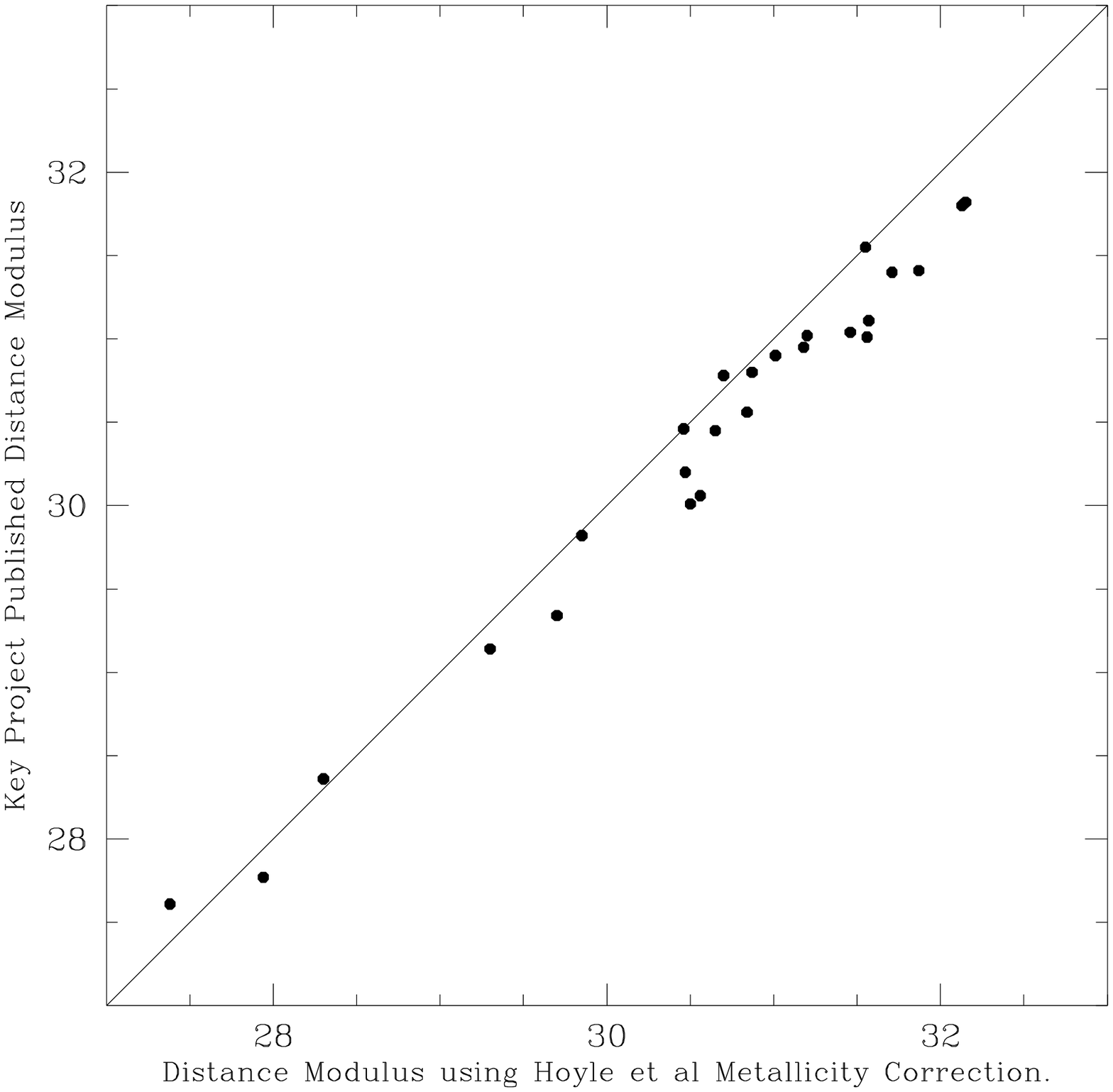}}
\caption{Relationship between published Cepheid distance modulus and metallicity
corrected distance modulus. The solid line shows the line of equal distances. It
can be seen that the published distances are generally shorter than the
corrected ones. This assumes an absolute distance modulus of $\mu$$_{LMC}$ = 18.50.}
\label{fig:hoyledist}
\end{figure}

The relationship between published Cepheid distance and corrected Cepheid
distance is shown in Figure \ref{fig:hoyledist}. It can be seen here more clearly
that most of the original distances are too short. The black line is the line of
equal distances and  most of the points lie below this line.

\subsection{Bias due to incompleteness at faint magnitudes}
\label{sec:incomplete}

There is another consequence of the observed relationship between dispersion and
metallicity. An increase in dispersion means that any bias due to incompleteness
at faint magnitudes that is present, would become more significant. The
implication is that high metallicity galaxies could be even further away than
they are with just a global metallicity correction applied, because their
distances are also biased too short. Of course, even if there were no relation
with metallicity, the bigger than expected P-L dispersions observed in the HST 
sample would still produce a bias towards distances which are too low.

There are two  ways to deal with this problem. The first method is to only
use data in regions of the P-L relation where there is confidence that most of 
the data is above the magnitude limit. The best way of doing this is by 
calculating the period where the 2$\sigma$ dispersion lines cross the magnitude
limit, and then cutting all data with $\emph{periods}$ less than this value. 
This technique has already been used to check the effects of magnitude 
incompleteness on dispersion in section \ref{sec:vband}. However, this method 
is fairly unsatisfactory for what is required here (i.e. accurate distance 
determination). This is because except in the small number of cases where the 
data is well clear of the magnitude limit (and no data has to be cut), the cuts
applied are usually too drastic. In many cases the range in period left to use 
in the P-L relation is less than 0.3 (log (days)). The number of remaining 
Cepheids is also usually very small. This means that the accuracy of the 
distances obtained is questionable. After analysing the V-band data, it was 
decided that this method was unsuitable.

\subsubsection*{The Maximum Likelihood Method}
\label{sec:maxlike}

The other way to calculate distances under a possible (magnitude limited)
incompleteness bias is to use a maximum likelihood technique. We assume that the
probability of a point being a certain distance from the line is given by a
Gaussian, with the added restriction that the probability that a point exists
below the magnitude limit is zero. The Gaussian must therefore be
normalised to account for this. For the V-band this is

\begin{equation}
p_{i} = \frac{exp(\frac{-(m_{i} - 2.76Logp_{i} + a)}{2\sigma^{2}})}{\int_{m_{0}}^{\infty}exp(\frac{-(m_{i} - 2.76Logp_{i} + a)}{2\sigma^{2}})}
\end{equation}

\noindent where p$_{i}$ is the probability, m$_{i}$ the magnitude, $\sigma$ the 
dispersion, a the intercept and m$_{0}$ is the calculated magnitude limit.

The total probability for the whole data set is simply a product of these 
probabilities.

\begin{equation}
p = \prod_{i}p_{i}(a)
\end{equation}

This is usually converted to a logarithm so that the product symbol can be
exchanged for a summation.

\begin{equation}
Logp = \sum_{i}Logp_{i}(a)
\label{eq:maxlike}
\end{equation}

This likelihood can be calculated using all data points that lie above the
magnitude limit, with no restriction on period. The value of the intercept, $a$,
which maximises this function can then be found. The number of points lost 
using this technique is minimised and the range in period available for the 
P-L relation is not greatly restricted. The maximum likelihood technique 
therefore allows distances to be determined accounting for magnitude limits.

The maximum likelihood intercept for all 25 galaxies in both V and I bands
was calculated using a FORTRAN programme written to find the value of the
intercept, $a$, which maximises equation \ref{eq:maxlike}. From these 
intercepts new distances were calculated and these are shown in Table \ref{tab:maxlike}.

\begin{table}
\begin{tabular}{cccc}
\hline
Galaxy & $\mu_0$ & $\mu_0$ & $\Delta$$\mu_0$ \\
 & (as published) & (maximum likelihood) & \\ \hline
NGC 925 & 29.82 & 29.88 & 0.06 \\
NGC 1326A & 31.55 & 31.42 & -0.13 \\
NGC 1365 & 31.40 & 31.48 & 0.08 \\
NGC 1425 & 31.82 & 32.22 & 0.40 \\
NGC 2090 & 30.45 & 30.52 & 0.07 \\
NGC 2541 & 30.46 & 30.48 & 0.02 \\
NGC 3031 & 27.77 & 27.77 & 0.00 \\
NGC 3198 & 30.80 & 30.93 & 0.13 \\
NGC 3319 & 30.78 & 30.80 & 0.02 \\
NGC 3351 & 30.01 & 29.95 &-0.06 \\
NGC 3621 & 29.14 & 29.15 & 0.01 \\
NGC 4321 & 31.04 & 31.03 &-0.01 \\
NGC 4414 & 31.41 & 31.99 & 0.58 \\
NGC 4535 & 31.11 & 31.18 & 0.07 \\
NGC 4548 & 31.01 & 31.24 & 0.23 \\
NGC 4725 & 30.56 & 30.56 & 0.00 \\
NGC 5457 & 29.34 & 29.35 & 0.01 \\
NGC 7331 & 30.90 & 31.28 & 0.38 \\
NGC 3368 & 30.20 & 30.21 & 0.01 \\
NGC 3627 & 30.06 & 30.06 & 0.00 \\
NGC 4639 & 31.80 & 31.92 & 0.12 \\
NGC 4496A& 31.02 & 31.02 & 0.00 \\
NGC 4536 & 30.95 & 31.08 & 0.13 \\
NGC 5253 & 27.61 & 27.61 & 0.00 \\
IC 4182 & 28.36 & 28.36 & 0.00 \\ \hline
\end{tabular}
\caption{Published distance moduli and distance moduli using maximum likelihood 
method.}
\label{tab:maxlike}
\end{table}

\begin{figure}
{\epsfxsize=8.5truecm \epsfysize=8.5truecm
\epsfbox[17 143 565 695]{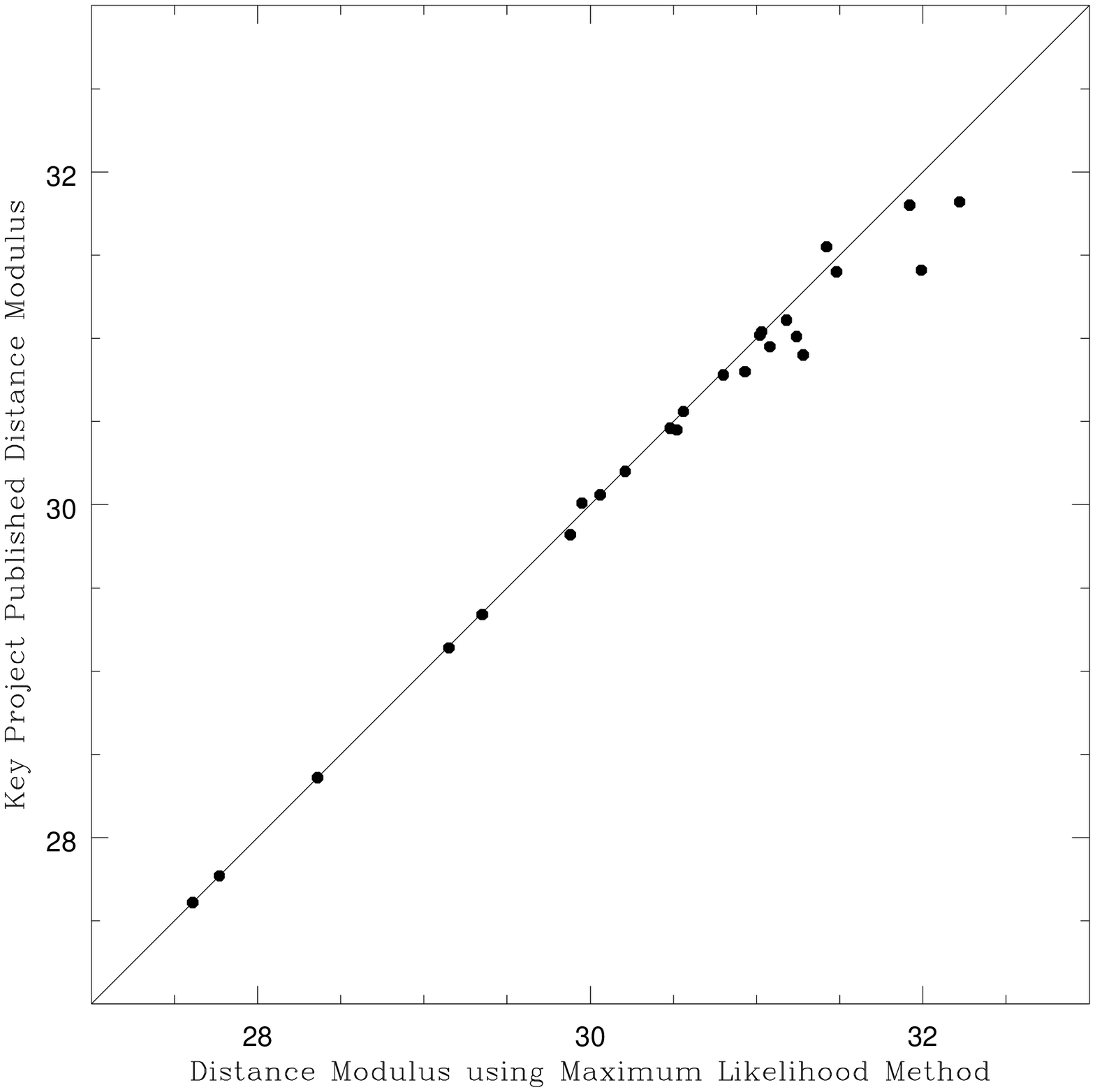}}
\caption{Relationship between published Cepheid distance modulus 
and distance modulus using maximum likelihood method. 
The solid line shows the line of  equal distances.}
\label{fig:distml}
\end{figure}

Table \ref{tab:maxlike} shows that in general, the distance moduli are increased by a
small amount after maximum likelihood fitting. An average increase of 0.085$\pm$0.03mag
is obtained, although many of the values are close to zero. Plotting maximum
likelihood distance against the previous distances (Figure \ref{fig:distml}) 
reveals that most of the significant differences are at large distances. This 
is likely to be because these are the furthest galaxies, and so more prone to
magnitude incompleteness and bias.

\subsection{Total Metallicity Effects}
\label{sec:totmet}

\begin{figure}
{\epsfxsize=8.5truecm \epsfysize=8.5truecm
\epsfbox[17 143 565 695]{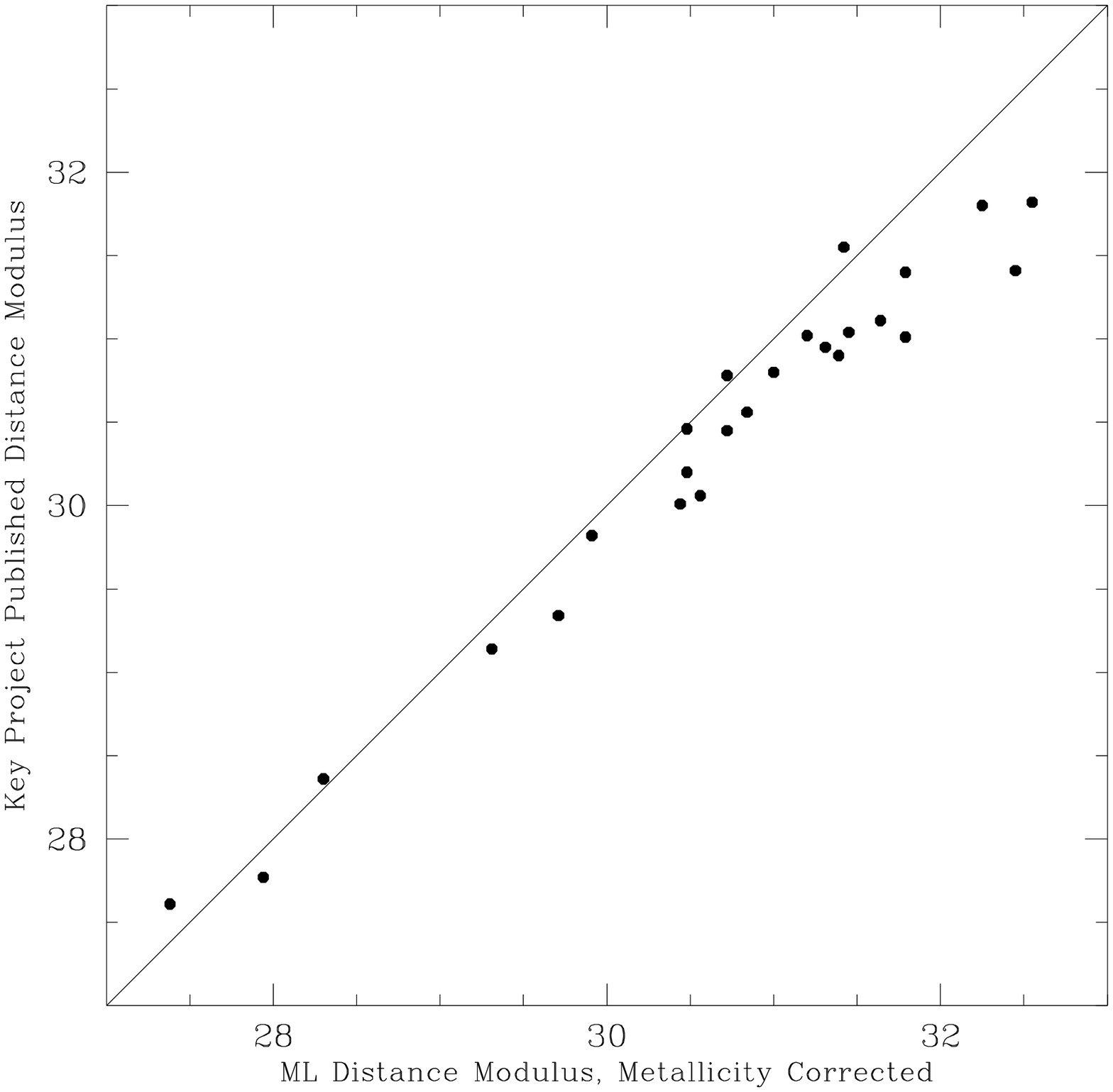}}
\caption{Relationship between published Cepheid distance modulus and distance
modulus using maximum likelihood method $\emph{and}$ the Hoyle et al. metallicity
correction. The solid line shows the line of equal distances.}
\label{fig:both}
\end{figure}

The maximum effect of the \citeasnoun{hoyle} global metallicity correction (i.e.
taking LMC modulus = 18.5) and the magnitude incompleteness (maximum likelihood)
effect are combined in Table \ref{tab:pubmet} and Figure \ref{fig:both}. The mean
difference in distance modulus between metallicity corrected and non-metallicity
corrected data, is now 0.31$\pm$0.06mag. There is also some evidence here of a
scale error. The best fit to the data gives a slope of 0.085$\pm$0.036, differing
from unity by 4.2$\sigma$. In the HST sample, there is a tendency for the highest
redshift galaxies to be brighter and have high metallicity and the combined
effect of the resulting high P-L dispersion and the global metallicity
correction, means that there is a scale error, rather than just a simple offset,
between the corrected and uncorrected distances.

\begin{table}
\begin{tabular}{cccccc}
\hline
Galaxy   & KP     &  Hoyle & ML     & ML     & ML \\
     &(as pub.)&      &     &$+$Hoyle& (mean disp.) \\  \hline 
NGC925   & 29.82  & 29.85  & 29.88  & 29.91  & 29.81 \\
NGC1326A & 31.55  & 31.55  & 31.42  & 31.42  & 31.44 \\
NGC1365  & 31.40  & 31.71  & 31.48  & 31.79  & 31.45 \\
NGC1425  & 31.82  & 32.15  & 32.22  & 32.55  & 32.32 \\
NGC2090  & 30.45  & 30.65  & 30.52  & 30.72  & 30.53 \\
NGC2541  & 30.46  & 30.46  & 30.48  & 30.48  & 30.39 \\
NGC3031  & 27.77  & 27.94  & 27.77  & 27.94  & 27.77 \\
NGC3198  & 30.80  & 30.87  & 30.93  & 31.00  & 30.96 \\
NGC3319  & 30.78  & 30.70  & 30.80  & 30.72  & 30.73 \\
NGC3351  & 30.01  & 30.50  & 29.95  & 30.44  & 29.94 \\
NGC3621  & 29.14  & 29.30  & 29.15  & 29.31  & 29.14 \\
NGC4321  & 31.04  & 31.46  & 31.03  & 31.45  & 31.03 \\
NGC4414  & 31.41  & 31.87  & 31.99  & 32.45  & 32.04 \\
NGC4535  & 31.11  & 31.57  & 31.18  & 31.64  & 31.21 \\
NGC4548  & 31.01  & 31.56  & 31.24  & 31.79  & 31.55 \\
NGC4725  & 30.56  & 30.84  & 30.56  & 30.84  & 30.57 \\
NGC5457  & 29.34  & 29.70  & 29.35  & 29.71  & 29.35 \\
NGC7331  & 30.90  & 31.01  & 31.28  & 31.39  & 31.41 \\
NGC3368  & 30.20  & 30.47  & 30.21  & 30.48  & 30.22 \\
NGC3627  & 30.06  & 30.56  & 30.06  & 30.56  & 30.06 \\
NGC4639  & 31.80  & 32.13  & 31.92  & 32.25  & 31.95 \\
NGC4496a & 31.02  & 31.20  & 31.02  & 31.20  & 31.01 \\
NGC4536  & 30.95  & 31.18  & 31.08  & 31.31  & 31.11 \\
NGC5253  & 27.61  & 27.38  & 27.61  & 27.38  & 27.61 \\
NGC4182  & 28.36  & 28.30  & 28.36  & 28.30  & 28.36 \\\hline
\end{tabular}
\caption{Distance moduli as published by Key Project (2), corrected for global 
metallicity correction of Hoyle et al. (2000) (3),  using maximum 
likelihood method (4), combined maximum likelihood and 
global correction (5) and just maximum likelihood with no metallicity correction (6).}
\label{tab:pubmet}
\end{table}

\section{Potential Implications for the Distance Scale}

The dispersion-metallicity correlation supports the idea that the metallicity
dependence of Cepheid luminosities may be stronger than expected and is
consistent with the $\frac{\Delta M}{[O/H]}$ =-~0.66 coefficient for the effect on
the P-L relation suggested by \citeasnoun{hoyle}. Further, the increased
dispersion observed in HST Cepheid galaxy P-L relations over that seen in the LMC
suggests that the effects of magnitude incompleteness is larger than expected,
particularly for the most distant galaxies. This latter effect would be present
even if there is assumed to be no relation between P-L dispersion and
metallicity, since the HST Cepheid P-L relations have a much higher dispersion
than the LMC P-L relations.

\begin{figure}
{\epsfxsize=8.truecm \epsfysize=16.0truecm
\epsfbox{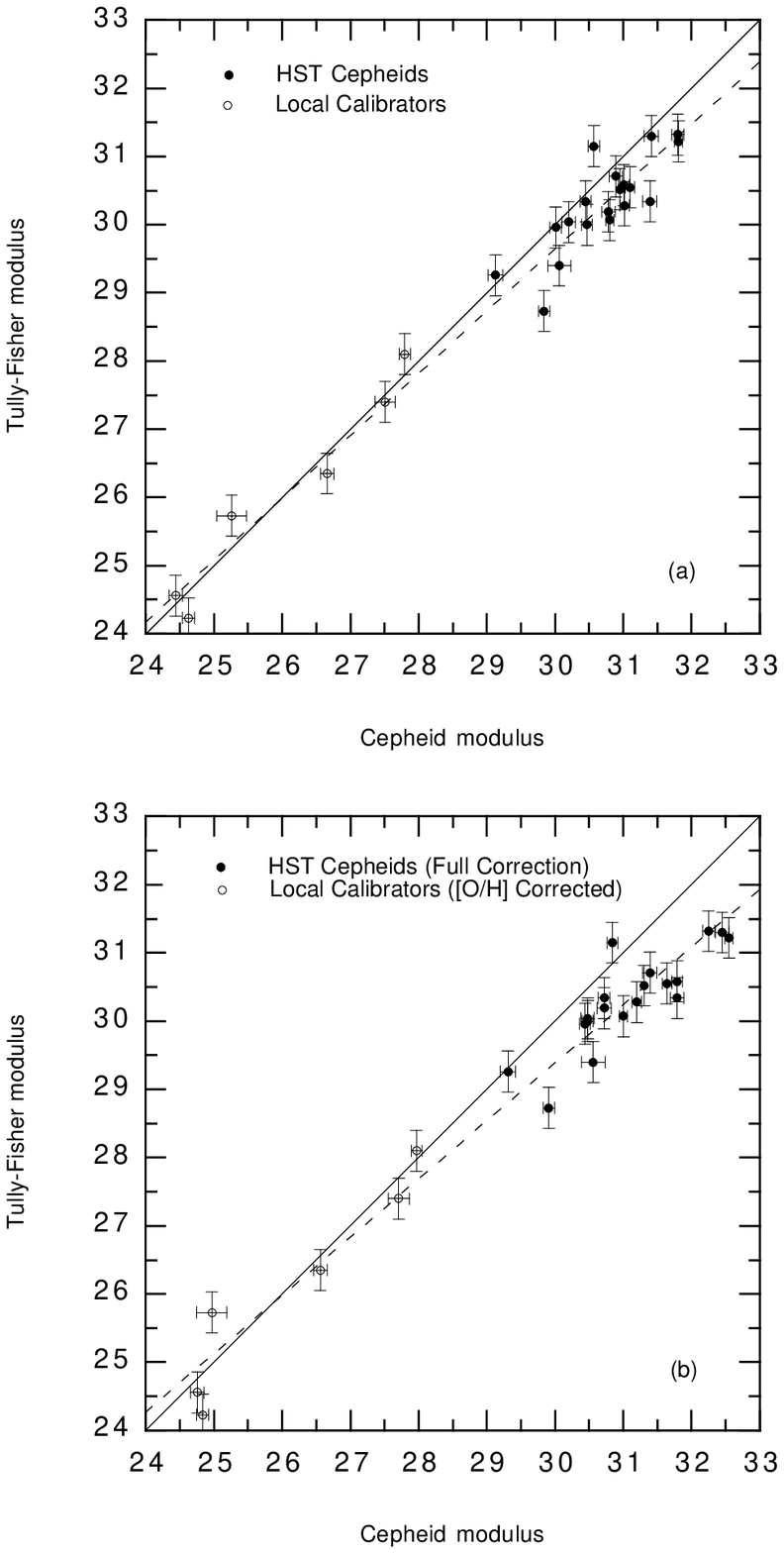}} 
\caption{(a) A comparison between HST Cepheid and TF distances which suggests that
TF distances show a significant scale-error with the TF distance to galaxies at the 
distance of the Virgo cluster being underestimated by 22$\pm$5.2\%. The solid line
shows the best fit with $(m-M)_{TF}= 0.915\pm0.036\times(m-M)_{Ceph}+2.204$.
(b)The same as (a) with the Cepheid distances now having been fully corrected
for metallicity from Table \ref {tab:pubmet}. An even stronger scale error is now
seen, with the TF distances to galaxies at the distance of the Virgo cluster now 
shown to be  underestimates by 46$\pm$6.7\%. The solid line shows the best fit with
$(m-M)_{TF}= 0.85\pm0.036\times(m-M)_{Ceph}+3.847$.}
\label{fig:2v9ab}
\end{figure}

\subsection{Corrected distances to the Virgo, Fornax and Leo Galaxy Clusters}

We first  illustrate the possible effect on the distance scale and Hubble's
Constant by  comparing the new metallicity corrected  average of the four
galaxies in the Virgo cluster core, defined as within a radius of 6 degrees of M87
(NGC4321, NGC4414, NGC4535, NGC4548) with the published average (from Table
\ref{tab:pubmet}, cols. 2, 5). The Virgo distance modulus increases from
$(m-M)_0=31.24\pm0.19$ to $(m-M)_0=31.78\pm0.17$  or from 17.7$\pm$1.6Mpc to
22.7$\pm$1.8Mpc. Including the further two galaxies which are within  a radius of
10 degrees of M87, NGC4496A and NGC4536, the Virgo distance modulus  now
increases from $(m-M)_0=31.15\pm0.13$ to $(m-M)_0=31.61\pm0.16$ or from
17.0$\pm$1.0Mpc to 21.0$\pm$1.6Mpc. The errors in the Virgo distance are large
because the Virgo spirals  are thought to have significant line-of-sight depth as
evidence by their almost flat velocity distribution. Below we shall use corrected
TF distances to obtain an alternative estimate of the Virgo distance which is
based on more Virgo spirals. The Leo distance modulus increases from
30.09$\pm$0.06 to 30.49$\pm$0.04 or from 10.4$\pm$0.3Mpc to 12.5$\pm$0.3Mpc based
on the 3 galaxies NGC3351, NGC3368 and  NGC3627. The Fornax distance modulus
increases from 31.59$\pm$0.12 to 31.92$\pm$0.33 or from 20.8$\pm$1.2Mpc to
24.2$\pm$4.0Mpc based on the galaxies NGC1326a, NGC1365 and NGC1425. The Fornax
cluster forms the Eastern wall of the Sculptor Void and  the dispersion in the
Fornax spirals may be expected to be large \cite{arat}. As for Virgo, below we
shall therefore also use corrected  TF distances to obtain an alternative
estimate of the Fornax distance based on 21 Fornax spirals.

\begin{table*}
\begin{tabular}{ccccccccccc}
\hline
Name    & TF Distance&  Inc&log ${W^R}_I$&$I_T$&$I_{corr}$&$B-I_{corr}$&KP&Corrected&KP-TF&Corr-TF \\ 
\hline
NGC0925 & 28.73$\pm$0.3$^1$ & 57.0 & 2.349 &9.30 & 9.11 & 1.34 & 29.84$\pm$0.08 & 29.91 & 1.11 & 1.18 \\
NGC1365 & 30.34$\pm$0.3$^3$ & 57.0 & 2.634 &8.31 & 8.23 & 1.91 & 31.39$\pm$0.10 & 31.79 & 1.05 & 1.45 \\
NGC1425 & 31.22$\pm$0.3$^3$ & 63.5 & 2.566 &9.82 & 9.70 & 1.32 & 31.81$\pm$0.06 & 32.55 & 0.59 & 1.33 \\
NGC2090 & 30.34$\pm$0.3$^5$ & 61.0 & 2.475 &9.72 & 9.62 & 2.14 & 30.45$\pm$0.08 & 30.72 & 0.11 & 0.38 \\
NGC2541 & 30.00$\pm$0.3$^7$ & 59.9 & 2.296 &11.02&10.84 & 1.02 & 30.47$\pm$0.08 & 30.48 & 0.47 & 0.48 \\
NGC3198 & 30.07$\pm$0.3$^7$ & 67.1 & 2.484 &9.41 & 9.27 & 1.28 & 30.80$\pm$0.06 & 31.00 & 0.73 & 0.93 \\
NGC3319 & 30.19$\pm$0.3$^6$ & 56.7 & 2.348 &10.66&10.58 & 0.72 & 30.78$\pm$0.10 & 30.72 & 0.59 & 0.53 \\
NGC3351 & 29.96$\pm$0.3$^2$ & 47.5 & 2.511 &  -  & 8.92 & 1.45 & 30.01$\pm$0.08 & 30.44 & 0.05 & 0.48 \\
NGC3368 & 30.04$\pm$0.3$^2$ & 46.2 & 2.641 &  -  & 7.88 & 2.08 & 30.20$\pm$0.10 & 30.48 & 0.16 & 0.44 \\
NGC3621 & 29.26$\pm$0.3$^4$ & 54.9 & 2.471 &8.83 & 8.57 & 1.12 & 29.13$\pm$0.11 & 29.31 &-0.13 & 0.05 \\
NGC3627 & 29.40$\pm$0.3$^1$ & 58.0 & 2.603 &7.67 & 7.56 & 1.77 & 30.06$\pm$0.17 & 30.56 & 0.66 & 1.16 \\
NGC4414 & 31.30$\pm$0.3$^1$ & 50.0 & 2.696 & -   &  -   &  -   & 31.41$\pm$0.10 & 32.45 & 0.11 & 1.15 \\
NGC4496a& 30.28$\pm$0.3$^1$ & 43.0 & 2.328 &10.88&10.84 & 1.47 & 31.02$\pm$0.07 & 31.20 & 0.74 & 0.92 \\
NGC4535 & 30.55$\pm$0.3$^7$ & 44.9 & 2.559 &9.13 & 9.09 & 1.41 & 31.10$\pm$0.07 & 31.64 & 0.55 & 1.09 \\
NGC4536 & 30.52$\pm$0.3$^1$ & 70.0 & 2.525 &9.53 & 9.36 & 1.69 & 30.95$\pm$0.07 & 31.31 & 0.43 & 0.79 \\
NGC4548 & 30.58$\pm$0.3$^7$ & 37.4 & 2.586 &8.95 & 8.89 & 1.94 & 31.01$\pm$0.08 & 31.79 & 0.43 & 1.21 \\
NGC4639 & 31.32$\pm$0.3$^1$ & 55.0 & 2.510 &10.39&10.29 & 1.66 & 31.80$\pm$0.09 & 32.25 & 0.48 & 0.93 \\
NGC4725 & 31.15$\pm$0.3$^7$ & 44.9 & 2.724 &8.31 & 8.25 & 1.73 & 30.57$\pm$0.08 & 30.84 &-0.58 &-0.31 \\
NGC7331 & 30.71$\pm$0.3$^7$ & 69.0 & 2.712 &8.23 & 7.92 & 1.72 & 30.89$\pm$0.10 & 31.39 & 0.18 & 0.68 \\

\hline
\end{tabular}
\caption{The Tully-Fisher parameters for HST Cepheid galaxies with inclinations,
$i>35$deg. These galaxies' I band TF parameters have been obtained from the work
of  (1) Pierce (1994), (2) Pierce (priv. comm.) reported by Ciardullo et al
(1989), (3) Bureau, Mould \& Staveley-Smith (1996) (4) the TF distance for NGC3621
is based on an $I_T$  derived  from de Vaucouleurs \& Longo (1988) using
$(V-I)_{Johnson}=1.3(V-I)_{KC}$ to convert $I_{Johnson}$ into $I_{KC}$ and  an
aperture correction of 0.35 mag, giving $I_T$=8.83mag. (5) The $I_T$ magnitude
is from Mathewson \& Ford (1996) (6) The $I_T$ magnitude is derived from Sakai
et al.(1999) (7) Where the TF distances were unavailable, they have been derived
using  line-widths,  inclinations and V mags from The Third Reference Catalogue
(de Vaucouleurs et al., 1991) and V-I mags from Buta \& Williams (1995), using
the precepts of Tully \& Fouque (1985) and Pierce \& Tully (1992).}
\label{tab:tf}
\end{table*}

\begin{table*}
\begin{tabular}{cccccccccc}
\hline
Name  & TF Distance&$B-I_{corr}$&log ${W^R}_I$&$I_{corr}$&KP &KP-TF &Corrected&Corr-TF& 12+logO/H \\ 
\hline
M31     & 24.56$\pm$0.3 & 1.69 & 2.712 & 1.77 & 24.44$\pm$0.10 & -0.12 & 24.76 & 0.20 & 8.98$\pm$0.15 \\
M33     & 24.23$\pm$0.3 & 1.15 & 2.322 & 4.84 & 24.63$\pm$0.09 &  0.40 & 24.84 & 0.61 & 8.82$\pm$0.15 \\
NGC2403 & 27.40$\pm$0.3 & 1.16 & 2.411 & 7.24 & 27.51$\pm$0.15 &  0.11 & 27.71 & 0.31 & 8.80$\pm$0.10 \\
M81     & 28.10$\pm$0.3 & 1.82 & 2.685 & 5.54 & 27.80$\pm$0.08 & -0.30 & 27.97 &-0.13 & 8.75$\pm$0.15 \\
NGC3109 & 25.73$\pm$0.3 & 0.88 & 2.032 & 8.87 & 25.26$\pm$0.22 & -0.47 & 24.97 &-0.76 & 8.06$\pm$0.15 \\
NGC300  & 26.35$\pm$0.3 & 0.94 & 2.284 & 7.30 & 26.66$\pm$0.10 &  0.31 & 26.56 & 0.21 & 8.35$\pm$0.15 \\
\hline
\end{tabular}
\caption{The TF parameters for the local calibrators are taken from Pierce \& Tully (1992).
The KP distances come from  Table 3 of Ferrarese et al (2000).}
\label{tab:loctf}
\end{table*}

\subsection{New Tests of Tully-Fisher Distances}

We next illustrate the possible effect on the distance scale and Hubble's
Constant by using the example of the Tully-Fisher relation, since a significant
number of the HST Cepheid galaxies have Tully-Fisher distances. In Figure
\ref{fig:2v9ab}(a) we therefore show the comparison of I-band Tully-Fisher
distances, calculated on the same premises as \citeasnoun{pt92} and with our
assumed TF parameters given in Table \ref{tab:tf}, with the published KP Cepheid
distances from Table \ref{tab:pubmet}. Note that the average difference between
our $I_T$ magnitudes and  those published by \citeasnoun{macri00} is only on
average $0.03\pm0.04$mag. Also shown are the previous 6 local TF calibrators, M31,
M33, M81, NGC2403, NGC300, NGC3109. This is similar to Fig. 2 of Shanks (1999),
updated from the 13 HST Cepheid/TF galaxies available then to the 19 such
galaxies available now, taking a lower limit for galaxy inclination of
$i>35$deg.  For the 18 HST Cepheid galaxies beyond (m-M)$_0>$29.5, we find
that the TF distance moduli underestimate the Cepheid distance moduli by
0.44$\pm$0.093mag, a 4.7$\sigma$ effect. Thus TF distances are confirmed to have
been underestimated by 22$\pm$5\%. Fitting a least squares line to all the
galaxies in Figure \ref{fig:2v9ab}(a) gives a slope of 0.915$\pm$0.036 which is
2.4$\sigma$ evidence that it is a real scale-error rather than a simple off-set
causing the effect.

\begin{figure}
{\epsfxsize=8.truecm \epsfysize=16.0truecm
\epsfbox{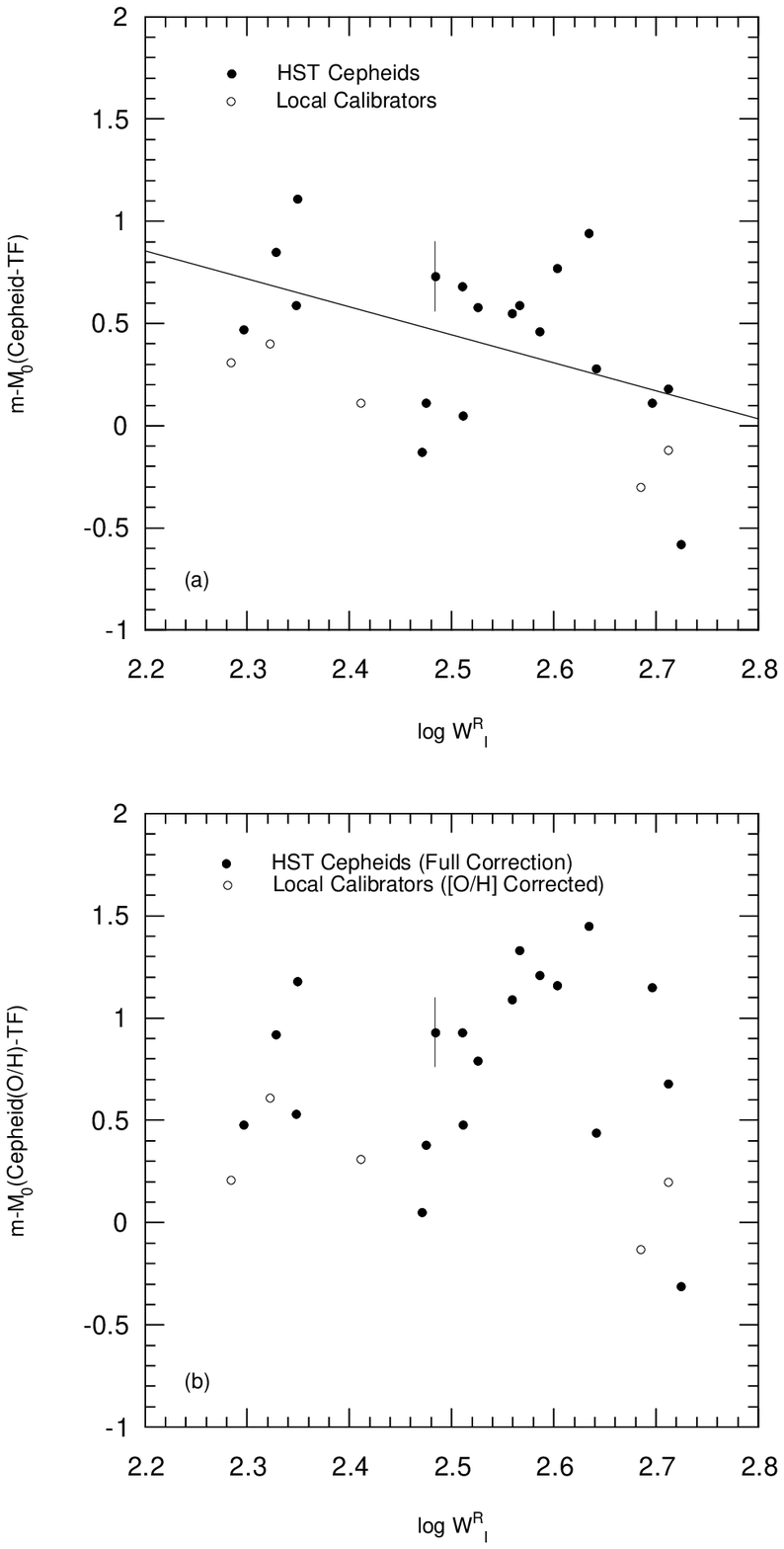}}
\caption{(a)Cepheid-TF residuals plotted against linewidth. A significant
correlation is seen in the sense that low linewidth galaxies have bigger
residuals. The least-squares line, Cepheid-TF distance modulus difference  =
-1.38($\pm$0.68)logW + 3.88, is also shown. (b) The same as (a) but now the  metallicity
corrected Cepheid distances are being used. The correlation is much reduced.}
\label{fig:malm8ab}
\end{figure}

In Figure \ref{fig:malm8ab}(a) we show the Cepheid-TF distance modulus
residuals plotted against galaxy linewidth. The  19 HST Cepheid  galaxies  show a 2$\sigma$
correlation in the sense that the lower linewidth, lower luminosity galaxies are
fainter. \citeasnoun {shanks99} previously suggested this was a possible
signature of Malmquist bias (see \citeasnoun{sandage94}) in the TF relation.
The  Local Calibrators in Figure \ref{fig:malm8ab}(a)show the same slope but at 
approximately zero offset.

\begin{figure}
{\epsfxsize=8.5truecm \epsfysize=8.5truecm
\epsfbox{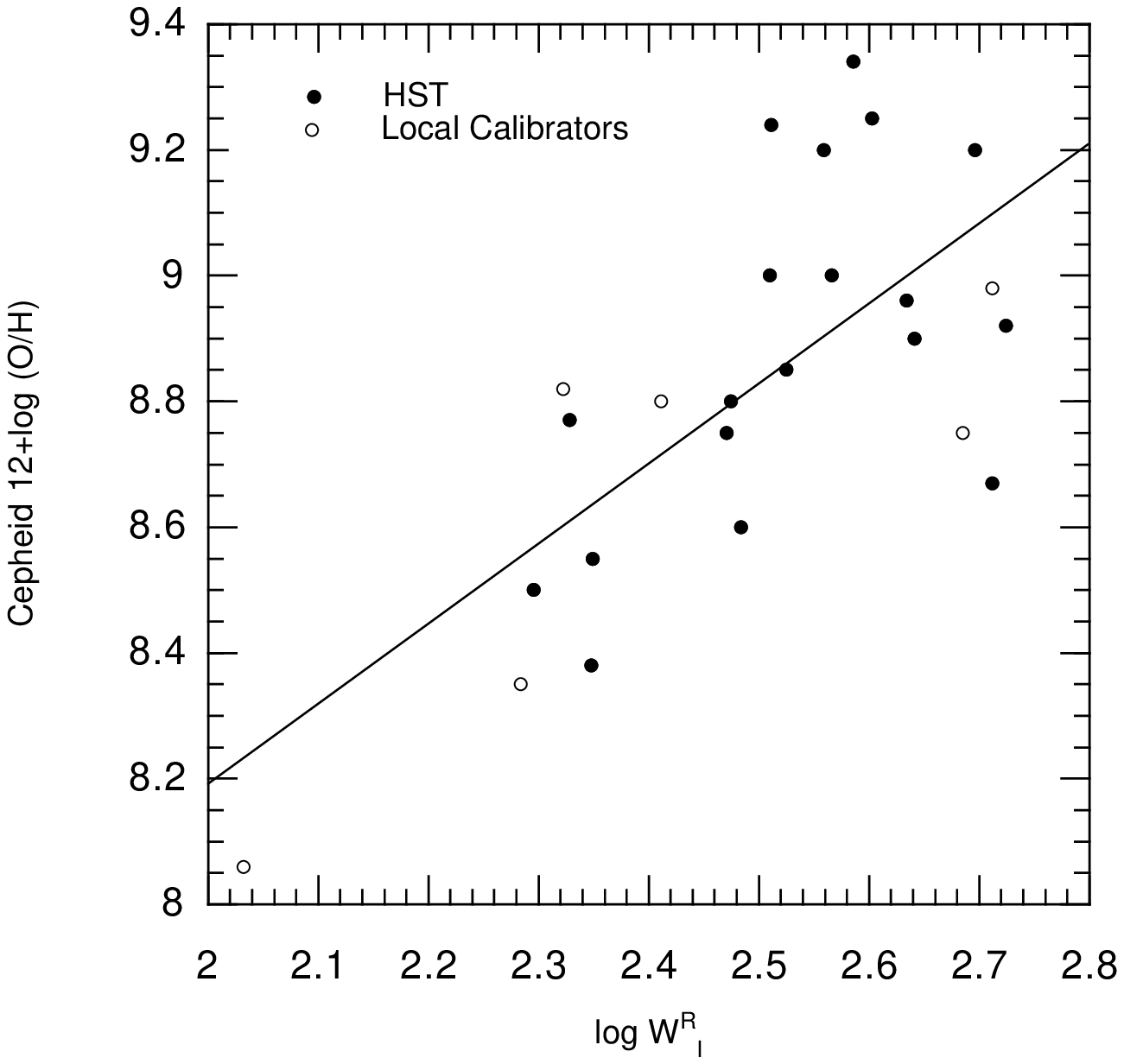}}
\caption{The metallicity-linewidth correlation for the HST Cepheid 
galaxies. The least-squares fitted line, $12+log(O/H)=1.274(\pm0.268)logW+5.643$,
is also shown.}
\label{fig:lwmet}
\end{figure}

Now in Fig. \ref{fig:2v9ab}(b) we show the same TF galaxy data as in Figure
\ref{fig:2v9ab}(a) but with Cepheid moduli now fully corrected for metallicity in
the case of the HST galaxies  (i.e. from col. 5 of Table \ref{tab:pubmet}) and in
the case of the Local Calibrators just for the global metallicity correction
(i.e. from col. 8 of Table \ref {tab:loctf}). Note  that in both cases this means
we have normalised the metallicity to the LMC value of $12+{\rm log}[O/H]=8.5$
rather than the Galactic value of $12+{\rm log}[O/H]=8.9$, effectively assuming
that the LMC distance is known independently of the Cepheids. The Local
Calibrators overall have a slightly smaller average Cepheid distance because of
the metallicity corrections, with the Cepheid-TF distance modulus difference
moving to 0.073$\pm$0.19 from -0.01$\pm$0.14; this is principally due to the low
metallicities of NGC3109 and NGC300. It should also be noted that the TF
linewidth of NGC3109 is uniquely small at $log {W^R}_I=2.032$ (see Figure
\ref{fig:lwmet}) and therefore the TF distance may be unreliable. The previously
published Cepheid distance to this galaxy is also the subject of discussion \cite
{fer,cap}.

For the HST Cepheid galaxies,  it can be seen  that TF distances 
beyond distance modulus (m-M)$_0$=29.5 are now found to underestimate the HST
Cepheid distance moduli by 0.82$\pm$0.10 mag or by 46$\pm$6.7\% in distance.
The least squares fit to the TF:Cepheid data shown as the dashed line
in Fig. \ref{fig:2v9ab}(b) has slope 0.85$\pm$0.036 which represents  4.2$\sigma$
evidence for a scale error in the TF distances rather than a simple offset.

In Fig. \ref{fig:malm8ab}(b) we show the residual-linewidth plot as in Fig.
\ref{fig:malm8ab}(a) but for the corrected Cepheid data. The metallicity
corrections have left the low linewidth points  approximately unchanged in this
diagram but have made the residuals for the higher linewidth points  bigger and
comparable to those for the lower linewidth points. If the corrected data are
more accurate, then it would suggest that the linewidth correlation in Fig.
\ref{fig:malm8ab}(a) was more likely due to the higher linewidth galaxies having
higher metallicity rather than the alternative Malmquist bias explanation
previously considered. In this view, the TF distances have a constant offset with
the HST Cepheid distances uncorrelated with linewidth. 

We now argue that the continuing offset between the HST Cepheid galaxies and the Local
Calibrators in Fig. \ref{fig:malm8ab}(b) may be evidence that the TF distances also
require a metallicity correction. The average Cepheid metallicity of the 19 HST
galaxies with TF distances is 8.89$\pm$0.28 whereas the average Cepheid
metallicity of the 6 Local Calibrators is 8.63$\pm$0.14; only one (M31) has a
higher Cepheid metallicity (8.98$\pm$0.15) than the HST average. And almost half
(7/19) of the HST galaxies have a Cepheid metallicity which is higher than that
of Cepheids in the highest metallicity local calibrator, M31. Thus generally, the
HST Cepheid galaxies extend to much higher metallicities than the Local Calibrators. In
addition, as shown in Fig. \ref{fig:lwmet}, there is a strong 4.7$\sigma$
correlation between TF linewidth and metallicity in the HST/local calibrator  TF
sample. Even when the lowest linewidth point (NGC3109) is removed the correlation
is still a 3.4$\sigma$ effect. In the HST and Local samples taken individually 
there is a correlation at the 3$\sigma$ significance level in both cases. This
strong metallicity-linewidth correlation in the 25 galaxy Cepheid-TF sample means
that the linewidth-TF residual  correlation seen before could actually be caused
by a metallicity dependence in the TF relation. The correlations of TF-Cepheid
residuals are less strong with metallicity than with linewidth but it is clear
that metallicity could plausibly be acting as a third parameter for the TF
relation. For example, when Local Calibrators have the same linewidth and
metallicity as an HST galaxy then they tend to act consistently and lie in the
same part of Fig. \ref{fig:malm8ab}(b). e.g. M31 and NGC4725 are similar and both
lie to the bottom right and the 3 local calibrators and the 4 HST galaxies on the
left also have similar metallicities.

Of course, it might not be unexpected that TF distances show a correlation with
metallicity. It could be argued that a spiral galaxy with a  metallicity which is
a tenth of solar will be populated by a main sequence of sub-dwarfs which lies
some 0.7mag fainter than a solar metallicity main sequence. On the crude
assumption that the same number of stars are formed per unit mass,  low
metallicity spirals would then form a TF relation which lay parallel but
$\approx$0.7mag fainter than for solar metallicity spirals. This may be the
underlying explanation of the effects seen in Fig. \ref{fig:malm8ab}(a,b).

Even though it may have a scale error, the TF route may supply a better estimate
of the distance to the Virgo and Fornax cluster mean spiral distances (and hence
$H_0$) because it is thought that the Virgo/Fornax spirals  are extended in the
line of sight and therefore the average distance of only a few Cepheid galaxies
in these cases may not be very accurate. The \citeasnoun{pt} TF distance to Virgo
is 15.6$\pm$1.5Mpc and the \citeasnoun{bureau} TF distance to Fornax is
15.4$\pm$2.3Mpc. Given that TF underestimates corrected Cepheid distances by
46$\pm$6.7\% at the distance of Virgo/Fornax, this means that the corrected TF
distance to Virgo should be 22.8$\pm$2.2Mpc,  similar to  the average of the four
Virgo core member HST Cepheid distances found above. The corrected TF distance to
Fornax is 22.5$\pm$3.4Mpc, in statistical agreement with the average of three
Fornax HST Cepheid distances found earlier. After \citeasnoun{shanks99}, we also
include the Ursa Major cluster which has a TF distance of 15.5$\pm$1.2Mpc
\cite{pt} which is similar to the TF distances for Virgo and Fornax and so within
the HST Cepheid range although as yet without actual HST Cepheid observations.
The corrected TF distance to Ursa Major is then 22.6$\pm$1.8Mpc. The heliocentric
velocities of these clusters are 1016$\pm$42kms$^{-1}$ for Virgo\cite{pt},
1450$\pm$34kms$^{-1}$ for Fornax \cite {bureau} and 967$\pm$20kms$^{-1}$ for Ursa
Major \cite{pt}. Ignoring infall and simply taking the ratio of heliocentric
velocity to corrected distance results in  values of  $H_0$=45, 64, 43
kms$^{-1}$Mpc$^{-1}$ from Virgo, Fornax and Ursa Major respectively. The mean of
these results gives $H_0=51\pm7$kms$^{-1}$Mpc$^{-1}$. Alternatively, assuming an
infall model, \citeasnoun{pt}  derived a value of
$H_0=85\pm10$kms$^{-1}$Mpc$^{-1}$ from the Virgo and Ursa Major TF distances.
Therefore, under the same assumptions, our corrected TF distances would give
$H_0=58\pm7$kms$^{-1}$Mpc$^{-1}$. However, caution is again required because the
continuing  large uncertainties surrounding the infall model translate directly
into uncertainty in H$_0$. All we take from this discussion is that values of
H$_0$ around 50km s$^{-1}$Mpc$^{-1}$ or even lower cannot yet be ruled out by
considerations of the Cepheid or TF distance scales.

\subsection{A Metallicity Dependence for Type Ia Supernovae?}

\begin{figure}/cos/h/pallen/plplots/
{\epsfxsize=8.0truecm \epsfysize=16.0truecm
\epsfbox{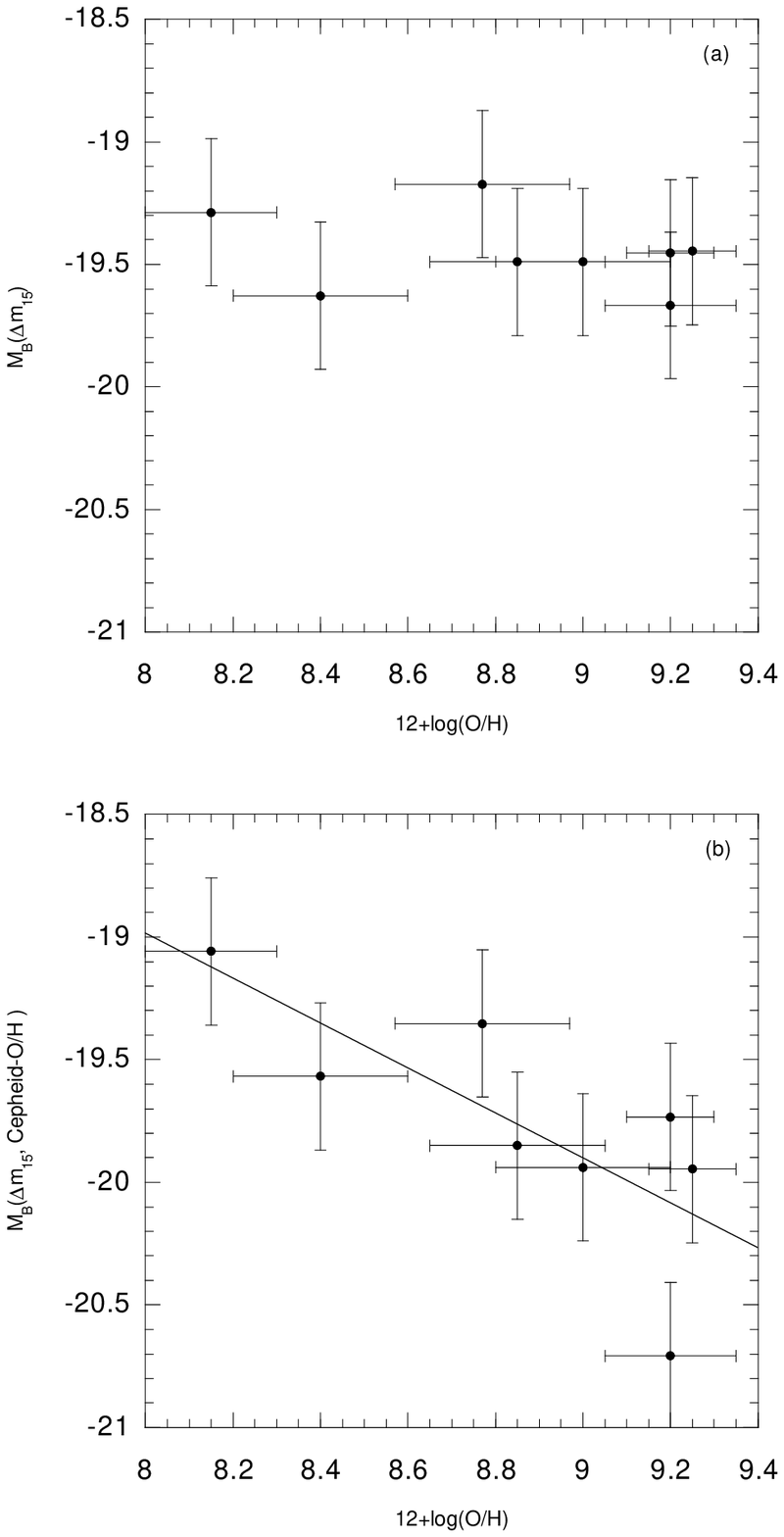}}
\caption{(a) The SNIa absolute magnitude-metallicity relation using the SNIa peak
magnitudes of Gibson et al, corrected for $\Delta m_{15}$ as in Table \ref{tab:sn1a}. 
(b) The SNIa absolute magnitude-metallicity relation using the SNIa
peak magnitudes of Gibson et al. (2000), now corrected for $\Delta m_{15}$  and
metallicity as in Table \ref {tab:sn1a}. The least-squares fitted line, 
$M_B=-0.92(\pm0.33)logW-11.65$, is also shown.}
\label{fig:sn1ab2}
\end{figure}

Finally, we show in Fig \ref{fig:sn1ab2} the effect that a Cepheid metallicity
term of the size described here would have on the SNIa `standard candle'. There
are 8 galaxies with both SNIa and Cepheid distances. Fig. \ref{fig:sn1ab2}(a)
shows the plot of SNIa absolute B magnitude vs host galaxy metallicity assuming
the apparent magnitudes, Cepheid distance moduli and reddenings of
\citeasnoun{gibson} as summarised in Table \ref{tab:sn1a}. Little correlation
with metallicity is seen. Fig. \ref{fig:sn1ab2}(b) shows the same graph now with
the same data except that the Cepheid distance moduli are corrected for the
metallicity effect of \citeasnoun {hoyle}. A  correlation with metallicity is now
seen. A least squares fit gives the slope as -0.92$\pm$0.33 in the sense that
higher metallicity galaxies have brighter SNIa.

\begin{table*}
\begin{tabular}{ccccccc}
\hline
Name &  $M_B$ (Gibson)&$\Delta m_{15}$ & $\Delta M(\Delta m_{15}$)&$M_B$($\Delta m_{15}$)&Corr-KP&$M_B$($\Delta m_{15}$) \\
     &                &      &      &        &      & -(Corr-KP) \\ \hline
NGC4639 & -19.51 & 1.07 & 0.02 & -19.49 & 0.45 & -19.94 \\
NGC4536 & -19.49 & 1.10 & 0.00 & -19.49 & 0.36 & -19.85 \\
NGC3627 & -19.25 & 1.31 &-0.20 & -19.45 & 0.50 & -19.95 \\
NGC3368 & -19.51 & 1.01 & 0.06 & -19.45 & 0.28 & -19.73 \\
NGC5253 & -19.40 & 0.87 & 0.11 & -19.29 &-0.23 & -19.06 \\
IC4182  & -19.74 & 0.87 & 0.11 & -19.63 &-0.06 & -19.57 \\
NGC4496 & -19.20 & 1.06 & 0.03 & -19.17 & 0.18 & -19.35 \\
NGC4414 & -19.66 & 1.11 &-0.01 & -19.67 & 1.04 & -20.71 \\
\hline
\end{tabular}
\caption{The SN1a peak absolute B magnitude $M_B$ and the  $\Delta m_{15}$ parameters
are taken from Gibson et al. (2000). $\Delta M(\Delta m_{15}$) is the correction for
$\Delta m_{15}$  from equation (1) of Gibson et al.(2000), ignoring the  reddening correction
to $\Delta m_{15}$ which is small.}
\label{tab:sn1a}
\end{table*}

In fact, we note that there is a correlation between $\Delta m_{15}$ and 
metallicity amongst these eight galaxies, in the sense that higher metallicity
galaxies have higher $\Delta m_{15}$. A least squares fit to this relation gives a
slope of 0.284$\pm$0.086. Since the peak-luminosity decline-rate correlation
implies that higher $\Delta m_{15}$ means fainter SNIa luminosities it is clear
that if this correlation is real the use of the peak-luminosity decline-rate
correlation invokes a bigger metallicity effect in the SNIa luminosities than if it
is ignored. When the metallicity correction of \citeasnoun{hoyle} is applied without
the $\Delta m_{15}$ correction, the slope of the peak luminosity-metallicity
correlation reduces from -0.92$\pm$0.33 to -0.75$\pm$0.34.

Now it might be argued that the fact that the $\Delta m_{15}$  correction works
only to accentuate the effect of metallicity  on Type Ia luminosity means that
the empirically determined decline rate correlation is evidence against a strong
metallicity effect for SNIa. However, the decline rate correction is much smaller
than the proposed metallicity correction which would be dominant. Also, the
selection effects induced by the wide dispersion in SNIa luminosity would have to
be carefully taken into account before coming to any conclusion. For example,
SNIa surveys at intermediate redshifts would preferentially detect the brighter
SNIa in higher metallicity galaxies, leading to SNIa which were systematically
too bright relative to local SNIa and with an artificially too small dispersion.
There would also be a potentially serious impact on the SNIa Hubble Diagram determinations
of q$_0$; if the average metallicity of galaxies was a factor of 3 lower at $z>0.5$ 
than observed locally then the SNIa would on average be $\approx$0.3 mag fainter
than expected and this is order of the amount needed to explain away the 
positive detection of the cosmological constant by the Supernova Cosmology 
teams \cite {perlmutter,schmidt}.

\section{Conclusions}\label{conc}

After a detailed analysis of the relationship between the dispersion in Cepheid
P-L relations and metallicity, it can be inferred that there is a relationship in
the V-band at the level 0.10 $\pm$ 0.03 mag dex$^{-1}$, in the sense that the
dispersion about the P-L relation increases with metallicity. This is a good fit
and significant at the 3$\sigma$ level and may  arise because high metallicity
galaxies contain Cepheids with a wider range of metallicities and hence
luminosities. The reason that a less strong relationship is observed in the
I-band is unclear. It could be that metallicity effects are less significant in
the I-band or that a relationship with the same slope but a lower amplitude is
masked in I  because of the effect of photometry/crowding errors on  the
intrinsically tighter dispersion. The size of the dispersion metallicity correlation
that we find is roughly consistent with the global P-L metallicity dependence of
$\Delta M/[O/H]=-0.66$ mag dex$^{-1}$ claimed by \citeasnoun{hoyle}.

The first consequence of this is  that there is now more evidence that a global
metallicity correction should be applied to Cepheid distances. The second
consequence is that the effects of incompleteness bias may also be important,
especially for high metallicity galaxies at large distances. These results combined imply that
Cepheid distance moduli are too short on the average by  0.29$\pm$0.21mag and by
up to $\approx$0.5mag in the case of the distant high metallicity galaxies. 

We have described the effects on the distance scale if the metallicity
corrections to Cepheid distances suggested here prove accurate. We have shown
that the distances to the Virgo galaxy cluster increase from 16.6 to 20.9 Mpc. We
have shown that there would be  a significant scale error in the TF scale which
would mean that the TF relation underestimates distances by 46\% at the distance
of Virgo. We have suggested thatthis may be due to the high metallicity of the 
HST Cepheid galaxies compared to the  Local Calibrators and that metallicity
may act as a significant second parameter in the TF relation.

We have shown by correcting TF distances to the Virgo, Fornax and Ursa Major 
clusters that this leads to  a range of values of $H_0$ between 43-64
kms$^{-1}$Mpc$^{-1}$ if no infall model is assumed and a value of
H$_0$=58$\pm$7km s$^{-1}$Mpc$^{-1}$ if an infall model is assumed, although the
infall model itself is highly uncertain.  We have also shown
that if the Cepheid luminosities are metallicity dependent then so are the peak
luminosities of Type Ia supernovae. New information on the metallicity of Type Ia
supernovae at both intermediate and high redshifts will be needed before
conclusions on supernova estimates of H$_0$ and q$_0$ can be drawn.

Of course, Cepheids with or without metallicity corrections provide no
direct check on the secondary distance indicators that are purely
associated with early-type galaxies such as SBF, PNLF or fundamental
plane indicators. Therefore we simply note that the early-type route
to H$_0$  through the Leo-I Group \cite{tanvir95,tanvir99},
which is the most nearby galaxy group containg both spirals with HST
Cepheids and early-type galaxies is that our corrected distance to
Leo-I of 12.5$\pm$0.3 Mpc from the 11.2$\pm$1.0Mpc quoted by
\citeasnoun {tanvir99} is that the derived value of H$_0$ will
therefore decrease from H$_0$=67$\pm$7kms$^{-1}$Mpc$^{-1}$ to
H$_0$=60$\pm$6kms$^{-1}$Mpc$^{-1}$.  But this estimate depends on
using secondary distance indicators to obtain the Coma-Leo relative
distance and it would be surprising if the metallicity corrections etc.
were only found to be important for the spiral distance indicators!

We therefore believe that there  is increasing evidence that  Cepheid distances
require significant corrections for the effects of metallicity and incompleteness
bias with important potential consequences for the distance scale,  Hubble's
Constant and cosmology. Further tests of the metallicity dependence of the
Cepheid P-L relation include a study of the metallicity of the main-sequence
stars in the open cluster NGC7790\cite{hs2001}. We are also using  our
metallicity-corrected Cepheid distances to make new checks of dynamical infall
models into Virgo and Fornax (Shanks and Warfield in prep.). Ultimately, a full
understanding of the metallicity effects on Cepheids and so the value of Hubble's
Constant may require parallaxes to Galactic Cepheids outside the Solar
Neighbourhood from astrometry satellites such as GAIA and a larger sample of
Cepheids in galaxies with a wide range of metallicities from instruments such as
the HST Advanced Camera and the NGST fitted with a  Visible Camera.

\section*{Acknowledgments} We thank the HST Distance Scale Key Project for
making their Cepheid data publically available. PDA thanks the University of Durham
for support.

\end{document}